\documentclass[onecolumn,notitlepage,floats,floatfix,amssymb,aps,prd,showpacs,superscriptaddress,groupedaddress,nofootinbib]{revtex4-2}  

\usepackage[T1]{fontenc}
\usepackage{lmodern} 
\usepackage{graphicx}  
\usepackage{dcolumn}   
\usepackage{bm}        
\usepackage{amssymb}   
\usepackage{amsmath}
\usepackage{bbold}
\usepackage{array}
\usepackage{makecell}
\usepackage{braket}
\usepackage{longtable}
\usepackage{supertabular,booktabs}

\usepackage{titlesec} 
\usepackage[colorlinks,citecolor=blue,urlcolor=blue,hypertexnames=true]{hyperref}
\usepackage{subfigure}

\usepackage[usenames,dvipsnames,svgnames,table]{xcolor} 

\renewcommand{\arraystretch}{2}

\newcommand{\be}{\begin{equation}}
\newcommand{\ee}{\end{equation}}
\newcommand{\bea}{\begin{eqnarray}}
\newcommand{\eea}{\end{eqnarray}}
\newcommand{\bml}{\begin{subequations}}
\newcommand{\eml}{\end{subequations}}
\newcommand{\bfig}{\begin{figure}}
\newcommand{\efig}{\end{figure}}

\newcommand{\del}{\delta}

\newcommand{\mbf}{\mathbf}

\newcommand{\slashed}{\hspace{-1.1ex}/}

\newcommand{\bmat}{\begin{pmatrix}}
\newcommand{\emat}{\end{pmatrix}}
\usepackage{graphicx,slashed,booktabs,xcolor,multirow,float,
amsfonts,bbold,mathtools,sidecap,tikz,bm,enumitem}
\usepackage{multirow}
\usepackage{bbding}
\usepackage{titlesec}
\usepackage{hyperref}
\usepackage{wasysym}
\usepackage{amssymb}
\usepackage{pifont}
\newcommand{\grad}{\nabla}

\renewcommand{\leq}{\leqslant}
\renewcommand{\geq}{\geqslant}

\usepackage[dvipsnames, usenames]{xcolor}

\definecolor{linkcolor}{rgb}{0.55, 0.13, .32}

\definecolor{oucrimsonred}{rgb}{0.6, 0.0, 0.0}
\definecolor{persianblue}{rgb}{0.11, 0.22, 0.73}
\definecolor{forestgreen}{rgb}{0.13,0.35,0.13}
\definecolor{lightgray}{rgb}{0.83, 0.83, 0.83}
 \hypersetup{colorlinks, citecolor=oucrimsonred, linkcolor=persianblue, urlcolor=oucrimsonred}
\definecolor{cornellred}{rgb}{0.7, 0.11, 0.11}
\definecolor{navyblue}{rgb}{0.0, 0.0, 0.5}
\definecolor{amethyst}{rgb}{0.6, 0.4, 0.8}
\definecolor{yellow}{rgb}{1.0, 1.0, 0.0}
\definecolor{firebrick}{rgb}{0.7, 0.13, 0.13}
\definecolor{tangerineyellow}{rgb}{1.0, 0.8, 0.0}
\definecolor{deepfuchsia}{rgb}{0.76, 0.33, 0.76}
\definecolor{amber}{rgb}{1.0, 0.75, 0.0}
\definecolor{VioletRed4}{rgb}{0.55, 0.13, .32}
\definecolor{indiagreen}{rgb}{0.07, 0.53, 0.03}
\definecolor{VioletRed4}{rgb}{0.55, 0.13, .32}

\usepackage{hyperref}

\usepackage{graphics,appendix,afterpage,makecell}

\usepackage{bbold}
\usepackage{tikz}
\usepackage{adjustbox}
\usepackage{tikz-feynman}
\usepackage{tcolorbox}

\def\abs#1{\left| #1\right|}

\definecolor{oucrimsonred}{rgb}{0.6, 0.0, 0.0}
\definecolor{persianblue}{rgb}{0.11, 0.22, 0.73}
\definecolor{forestgreen}{rgb}{0.13,0.35,0.13}
\definecolor{lightgray}{rgb}{0.83, 0.83, 0.83}
 \hypersetup{colorlinks, citecolor=oucrimsonred, linkcolor=persianblue, urlcolor=oucrimsonred}
\definecolor{cornellred}{rgb}{0.7, 0.11, 0.11}
\definecolor{navyblue}{rgb}{0.0, 0.0, 0.5}
\definecolor{amethyst}{rgb}{0.6, 0.4, 0.8}
\definecolor{yellow}{rgb}{1.0, 1.0, 0.0}
\definecolor{firebrick}{rgb}{0.7, 0.13, 0.13}
\definecolor{tangerineyellow}{rgb}{1.0, 0.8, 0.0}
\definecolor{deepfuchsia}{rgb}{0.76, 0.33, 0.76}
\definecolor{amber}{rgb}{1.0, 0.75, 0.0}
\definecolor{VioletRed4}{rgb}{0.55, 0.13, .32}
\definecolor{indiagreen}{rgb}{0.07, 0.53, 0.03}
\definecolor{VioletRed4}{rgb}{0.55, 0.13, .32}

\definecolor{oucrimsonred}{rgb}{0.6, 0.0, 0.0}
\newcommand\vertarrowbox[3][6ex]{%
  \begin{array}[t]{@{}c@{}} #2 \\
  \left\uparrow\vcenter{\hrule height #1}\right.\kern-\nulldelimiterspace\\
  \makebox[0pt]{\scriptsize#3}
  \end{array}%
}

\definecolor{mtcolor}{rgb}{.8,.3,.1}

\definecolor{violachiaro}{rgb}{1,0.6,1}

\definecolor{gbcolor}{rgb}{.43,.22,.12}
 
\definecolor{gbcolor2}{rgb}{.9,.2,.6}
\definecolor{gbcolor3}{rgb}{.3,.2,.6}

\definecolor{verdechiaro}{rgb}{0.6,1,0.6}
\definecolor{giallochiaro}{rgb}{1,1,0.6}
\definecolor{bluscuro}{rgb}{0.15, 0.2, 0.9}
\definecolor{verdes}{rgb}{0.1, 0.5, 0.1}%
\definecolor{tangerineyellow}{rgb}{1.0, 0.8, 0.0}
\definecolor{smokyblack}{rgb}{0.06, 0.05, 0.03}

\definecolor{americanrose}{rgb}{1.0, 0.01, 0.24}
\definecolor{cobalt}{rgb}{0.0, 0.28, 0.67}
\definecolor{brandeisblue}{rgb}{0.0, 0.44, 1.0}
\definecolor{mycolor}{rgb}{0.0, 0.0, 0.5}
\definecolor{oxfordblue}{rgb}{0.0, 0.13, 0.28}
\definecolor{azure}{rgb}{0.0, 0.5, 1.0}
\definecolor{turquoiseblue}{rgb}{0.0, 1.0, 0.94}
\newtcolorbox{mynewbox}[1]{colback=white!5!white,colframe=azure!75!black,fonttitle=\bfseries,title=#1}
\newtcolorbox{mybox}{colback=mycolor!5!white,colframe=azure!75!black}
\newtcolorbox{mynamedbox}[1]{colback=mycolor!5!white,colframe=azure!75!black,title=#1}
\definecolor{venetianred}{rgb}{0.78, 0.03, 0.08}
\newtcolorbox{mynamedbox1}[1]{colback=venetianred!5!white,colframe=venetianred!80!black,title=#1}
\newtcolorbox{mynamedbox2}[1]{colback=azure!5!white,colframe=azure!80!black,title=#1}

\definecolor{rossocorsa}{rgb}{0.83, 0.0, 0.0}

\tikzset{->-/.style={decoration={
  markings,
  mark=at position #1 with {\arrow{>}}},postaction={decorate}}}
\tikzset{-<-/.style={decoration={
  markings,
  mark=at position #1 with {\arrow{<}}},postaction={decorate}}} 

\def\be{\begin{equation}}
\def\ee{\end{equation}}
\def\ba{\begin{eqnarray}}
\def\ea{\end{eqnarray}}

\def\L*{{\cal L}_*}
\def\L{\mathcal{L}}
\def\({\left(}
\def\){\right)}

\def\<{\langle}
\def\>{\rangle}

\def\cs2{c_{s}^{2}}

 \def\be   {\begin{equation}}   \def\ee   {\end{equation}}
 \def\ba   {\begin{array}}      \def\ea   {\end{array}}
 \def\bea  {\begin{eqnarray}}   \def\eea  {\end{eqnarray}}
 \def\bean {\begin{eqnarray*}}  \def\eean {\end{eqnarray*}}

\titleclass{\subsubsubsection}{straight}[\subsection]

\newcounter{subsubsubsection}[subsubsection]
\renewcommand\thesubsubsubsection{\thesubsubsection.\arabic{subsubsubsection}}

\titleformat{\subsubsubsection}
  {\normalfont\normalsize\bfseries}{\thesubsubsubsection}{1em}{}
\titlespacing*{\subsubsubsection}
{0pt}{3.25ex plus 1ex minus .2ex}{1.5ex plus .2ex}

\makeatletter
\renewcommand\paragraph{\@startsection{paragraph}{5}{\z@}%
  {3.25ex \@plus1ex \@minus.2ex}%
  {-1em}%
  {\normalfont\normalsize\bfseries}}
\renewcommand\subparagraph{\@startsection{subparagraph}{6}{\parindent}%
  {3.25ex \@plus1ex \@minus .2ex}%
  {-1em}%
  {\normalfont\normalsize\bfseries}}
\def\toclevel@subsubsubsection{4}
\def\toclevel@paragraph{5}
\def\toclevel@paragraph{6}
\def\l@subsubsubsection{\@dottedtocline{4}{7em}{4em}}
\def\l@paragraph{\@dottedtocline{5}{10em}{5em}}
\def\l@subparagraph{\@dottedtocline{6}{14em}{6em}}
\makeatother

\begin{document}


\definecolor{lime}{HTML}{A6CE39}
\DeclareRobustCommand{\orcidicon}{\hspace{-2.1mm}
\begin{tikzpicture}
\draw[lime,fill=lime] (0,0.0) circle [radius=0.13] node[white] {{\fontfamily{qag}\selectfont \tiny \,ID}}; \draw[white, fill=white] (-0.0525,0.095) circle [radius=0.007]; 
\end{tikzpicture} \hspace{-3.7mm} }
\foreach \x in {A, ..., Z}{\expandafter\xdef\csname orcid\x\endcsname{\noexpand\href{https://orcid.org/\csname orcidauthor\x\endcsname} {\noexpand\orcidicon}}}
\newcommand{\orcidauthorA}{0000-0002-0459-3873}
\newcommand{\orcidauthorC}{0009-0003-9227-8615}
\newcommand{\orcidauthorD}{0000-0001-9434-0505}
\newcommand{\orcidauthorE}{0000-0003-1081-0632}


\title{\textcolor{Sepia}{\textbf \huge\Large\LARGE  
Realisation of the ultra-slow roll phase in Galileon inflation and 
PBH overproduction
}}


\author{{\large  Sayantan Choudhury\orcidA{}${}^{1}$}}
\email{sayantan\_ccsp@sgtuniversity.org,  \\ sayanphysicsisi@gmail.com (Corresponding author)}
\author{{\large  Ahaskar Karde\orcidC{}${}^{1}$}}
\email{kardeahaskar@gmail.com}
\author{\large Sudhakar~Panda\orcidD{}${}^{1,2}$}
\email{panda@niser.ac.in}
\author{ \large M.~Sami\orcidE{}${}^{1,3,4}$}
\email{ sami\_ccsp@sgtuniversity.org,  samijamia@gmail.com}

\affiliation{ ${}^{1}$Centre For Cosmology and Science Popularization (CCSP),\\
        SGT University, Gurugram, Delhi- NCR, Haryana- 122505, India.}
\affiliation{${}^{2}$School of Physical Sciences,  National Institute of Science Education and Research, Bhubaneswar, Odisha - 752050, India.}
\affiliation{${}^{3}$Center for Theoretical Physics, Eurasian National University, Astana 010008, Kazakhstan.}
	\affiliation{${}^{4}$Chinese Academy of Sciences,52 Sanlihe Rd, Xicheng District, Beijing.}

\begin{abstract}

We demonstrate the explicit realisation of the ultra-slow roll phase in the framework of the effective field theory of single-field Galileon inflation. The pulsar timing array (PTA) collaboration hints at the scalar-induced gravity waves (SIGW) from the early universe as an explanation for the origin of the observed signal, which, however, leads to an enhancement in the amplitude of the scalar power spectrum giving rise to the overproduction of primordial black holes (PBHs). In the setup under consideration, we examine the generation of SIGW consistent with PTA (NANOGrav15 and EPTA) data, in addition to which we also consider the impact from QCD crossover at the nHz frequencies and address the PBH overproduction issue assuming linear approximations for the over-density without incorporating  non-Gaussian effects from the comoving curvature perturbation. The framework is shown to give rise to SIGWs well consistent with the PTA signal with comfortable PBH abundance, $10^{-3} \lesssim f_{\rm PBH} < 1$, of near solar-mass black holes.

\end{abstract}

\maketitle
\tableofcontents
\newpage

\section{Introduction}

Primordial black holes (PBHs) have recently caught enormous attention as potential candidates for dark matter and for their connection with the induced gravitational waves \cite{Zeldovich:1967lct,Hawking:1974rv,Carr:1974nx,Carr:1975qj,Chapline:1975ojl,Carr:1993aq,Choudhury:2012whm,Choudhury:2012yh,Choudhury:2013woa,Yokoyama:1998pt,Kawasaki:1998vx,Rubin:2001yw,Khlopov:2002yi,Khlopov:2004sc,Saito:2008em,Khlopov:2008qy,Carr:2009jm,Choudhury:2011jt,Lyth:2011kj,Drees:2011yz,Drees:2011hb,Ezquiaga:2017fvi,Kannike:2017bxn,Hertzberg:2017dkh,Pi:2017gih,Gao:2018pvq,Dalianis:2018frf,Cicoli:2018asa,Ozsoy:2018flq,Byrnes:2018txb,Ballesteros:2018wlw,Belotsky:2018wph,Martin:2019nuw,Ezquiaga:2019ftu,Motohashi:2019rhu,Fu:2019ttf,Ashoorioon:2019xqc,Auclair:2020csm,Vennin:2020kng,Nanopoulos:2020nnh,Inomata:2021uqj,Stamou:2021qdk,Ng:2021hll,Wang:2021kbh,Kawai:2021edk,Solbi:2021rse,Ballesteros:2021fsp,Rigopoulos:2021nhv,Animali:2022otk,Correa:2022ngq,Frolovsky:2022ewg,Escriva:2022duf,Ozsoy:2023ryl,Ivanov:1994pa,Afshordi:2003zb,Frampton:2010sw,Carr:2016drx,Kawasaki:2016pql,Inomata:2017okj,Espinosa:2017sgp,Ballesteros:2017fsr,Sasaki:2018dmp,Ballesteros:2019hus,Dalianis:2019asr,Cheong:2019vzl,Green:2020jor,Carr:2020xqk,Ballesteros:2020qam,Carr:2020gox,Ozsoy:2020kat,Baumann:2007zm,Saito:2008jc,Saito:2009jt,Choudhury:2013woa,Sasaki:2016jop,Raidal:2017mfl,Papanikolaou:2020qtd,Ali-Haimoud:2017rtz,Di:2017ndc,Raidal:2018bbj,Cheng:2018yyr,Vaskonen:2019jpv,Drees:2019xpp,Hall:2020daa,Ballesteros:2020qam,Carr:2020gox,Ozsoy:2020kat,Ashoorioon:2020hln,Papanikolaou:2020qtd,Wu:2021zta,Kimura:2021sqz,Solbi:2021wbo,Teimoori:2021pte,Cicoli:2022sih,Ashoorioon:2022raz,Papanikolaou:2022chm,Papanikolaou:2023crz,Wang:2022nml,ZhengRuiFeng:2021zoz,Cohen:2022clv,Arya:2019wck,Correa:2022ngq,Cicoli:2022sih,Brown:2017osf,Palma:2020ejf,Geller:2022nkr,Braglia:2022phb,Frolovsky:2023xid,Aldabergenov:2023yrk,Aoki:2022bvj,Frolovsky:2022qpg,Aldabergenov:2022rfc,Ishikawa:2021xya,Gundhi:2020kzm,Aldabergenov:2020bpt,Cai:2018dig,Cheng:2021lif,Balaji:2022rsy,Qin:2023lgo,Riotto:2023hoz,Ragavendra:2020vud,Ragavendra:2021qdu,Ragavendra:2020sop,Gangopadhyay:2021kmf,Papanikolaou:2022did,Harada:2013epa,Harada:2017fjm,Kokubu:2018fxy,Gu:2023mmd,Saburov:2023buy,Stamou:2023vxu,Libanore:2023ovr,Friedlander:2023qmc,Chen:2023lou,Cai:2023uhc,Karam:2023haj,Iacconi:2023slv,Gehrman:2023esa,Padilla:2023lbv,Xie:2023cwi,Meng:2022low,Qiu:2022klm,Mu:2022dku,Fu:2022ypp,Davies:2023hhn,Choudhury:2024ybk,Choudhury:2024jlz,Choudhury:2024dei}. The latest confirmation of a stochastic gravitational wave background (SGWB) by the pulsar timing array collaborations (PTA), which includes the NANOGrav \cite{NANOGrav:2023gor, NANOGrav:2023hde, NANOGrav:2023ctt, NANOGrav:2023hvm, NANOGrav:2023hfp, NANOGrav:2023tcn, NANOGrav:2023pdq, NANOGrav:2023icp, Inomata:2023zup}, EPTA \cite{EPTA:2023fyk, EPTA:2023sfo, EPTA:2023akd, EPTA:2023gyr, EPTA:2023xxk, EPTA:2023xiy, Lozanov:2023rcd}, PPTA \cite{Reardon:2023gzh, Reardon:2023zen, Zic:2023gta}, and CPTA \cite{Xu:2023wog} have accelerated investigations  for the search of possible sources of the observed signal.
Numerous possible cosmological sources have gained interest since the release of the data, some of which include first-order phase transitions, cosmic strings, domain walls, and inflation \cite{Choudhury:2023hfm,Bhattacharya:2023ysp,Franciolini:2023pbf,Inomata:2023zup,Wang:2023ost,Balaji:2023ehk,HosseiniMansoori:2023mqh,Gorji:2023sil,DeLuca:2023tun,Choudhury:2023kam,Yi:2023mbm,Cai:2023dls,Cai:2023uhc,Huang:2023chx,Vagnozzi:2023lwo,Frosina:2023nxu,Zhu:2023faa,Jiang:2023gfe,Cheung:2023ihl,Oikonomou:2023qfz,Liu:2023pau,Liu:2023ymk,Wang:2023len,Zu:2023olm, Abe:2023yrw, Gouttenoire:2023bqy,Salvio:2023ynn, Xue:2021gyq, Nakai:2020oit, Athron:2023mer,Ben-Dayan:2023lwd, Madge:2023cak,Kitajima:2023cek, Babichev:2023pbf, Zhang:2023nrs, Zeng:2023jut, Ferreira:2022zzo, An:2023idh, Li:2023tdx,Blanco-Pillado:2021ygr,Buchmuller:2021mbb,Ellis:2020ena,Buchmuller:2020lbh,Blasi:2020mfx, Madge:2023cak, Liu:2023pau, Yi:2023npi,Gangopadhyay:2023qjr,Vagnozzi:2020gtf,Benetti:2021uea,Inomata:2023drn,Lozanov:2023rcd,Basilakos:2023jvp,Basilakos:2023xof,Li:2023xtl,Domenech:2021ztg,Yuan:2021qgz,Chen:2019xse,Cang:2023ysz,Cang:2022jyc,Konoplya:2023fmh,Huang:2023chx,Ellis:2023oxs,Yu:2023jrs,Nassiri-Rad:2023asg,Heydari:2023rmq,Li:2023xtl,Chang:2023aba,Bernardo:2023jhs,Choi:2023tun,Elizalde:2023rds,Chen:2023bms,Nojiri:2023mbo,Domenech:2023jve,Liu:2023hpw,Huang:2023zvs,Oikonomou:2023bli,Cyr:2023pgw,Fu:2023aab,Kawai:2023nqs,Kawasaki:2023rfx,Maji:2023fhv,Bhaumik:2023wmw,He:2023ado,An:2023jxf,Zhu:2023lbf,Das:2023nmm,Roshan:2024qnv,Chen:2024fir}. However, the formation of PBHs from enhanced curvature perturbations in the very early universe raises concerns about their subsequent overproduction \cite{Franciolini:2023pbf,Franciolini:2023wun,Inomata:2023zup,LISACosmologyWorkingGroup:2023njw, Inui:2023qsd,Chang:2023aba, Gorji:2023ziy,Li:2023xtl, Li:2023qua,Firouzjahi:2023xke,Gorji:2023sil,Ota:2022xni,Raatikainen:2023bzk,Choudhury:2023fwk,Choudhury:2023fjs}.

We choose Galileon Inflation as the underlying framework for our analysis. The Galileon action comes with a Galilean shift symmetry, and a mild breaking of the said symmetry allows to successfully validate inflation. For the setup of interest in this paper, where an ultra-slow roll (USR) phase is sandwiched between two slow roll (SR) phase, we explicitly justify the applicability and validity of such a 
construction using the effective field theory (EFT) framework. 
An important point in our construction is the use of sharp transitions before and after the USR phase.
The use of sharp transitions, in general, in single field inflation, severely constraints the masses of PBHs produced in the SR/USR/SR setup \cite{Choudhury:2023vuj, Choudhury:2023jlt, Choudhury:2023rks,Choudhury:2023hvf,Choudhury:2024ybk,Choudhury:2024jlz}. However, these constraints on the allowed PBH masses are evaded in the case of Galileon inflation \cite{Choudhury:2023hvf}, in the presence of the sharp transition. Another possibility to evade PBH mass constraints and incorporating the sharp transitions is examined using a modified setup containing multiple sharp transitions in single field inflation \cite{Bhattacharya:2023ysp, Choudhury:2023fjs}. Alternatively, there are approaches that use the smooth transitions  \cite{Riotto:2023hoz,Riotto:2023gpm,Firouzjahi:2023ahg,Firouzjahi:2023aum} or a bump or dip-like feature \cite{Mishra:2019pzq} during inflation. 
An important property of the Galileon theory is the noteworthy non-renormalization theorem. The theorem effectively prevents radiative corrections from affecting the computation process of the interested correlators, thereby circumventing the need to invoke any rigorous renormalization or further resummation procedures. The Galileon interactions remain un-renormalized and, therefore, remain radiatively stable. This fact helps in building  one-loop corrected scalar power spectrum \footnote{For more details on the one-loop corrected power spectrum, see refs. \cite{Adshead:2008gk,Senatore:2009cf,Senatore:2012nq,Pimentel:2012tw,Sloth:2006az,Seery:2007we,Seery:2007wf,Bartolo:2007ti,Seery:2010kh,Bartolo:2010bu,Senatore:2012ya,Chen:2016nrs,Markkanen:2017rvi,Higuchi:2017sgj,Syu:2019uwx,Rendell:2019jnn,Cohen:2020php,Green:2022ovz,Premkumar:2022bkm}. }.

Due to the non-renormalization theorem, the effects of the loop quantum corrections are absent, and as a result, the PBH mass constraints get evaded in the Galileon inflationary framework \cite{Choudhury:2023hvf}. 
In standard practice, we can begin our analysis by considering a narrow Dirac delta-like power spectrum or with a log-normal power spectrum to study the PBH formation and SIGW production. \textcolor{black}{A Dirac delta-like power spectrum describes the simplest cosmological scenario where most of the power is coming from a single wavenumber, and the log-normal spectrum allows us to consider scenarios of having a broad spectrum, which quantitatively also have intimate similarities to a Gaussian power spectrum. It can also help when dealing with the presence of fluctuations. These two models have been used in many studies to significantly ease the analysis of various modes' contribution to the PBH collapse. In a more realistic scenario, when we expect new physics to affect the contribution of modes to the power spectrum, we may not necessarily have a high-peak power spectrum or a Gaussian, which, while a good approximation, can have deviations that become important to characterize. Thus,} although it is perfectly reasonable to consider such shapes for the spectrum in one analysis, obtaining such a form may not always be realizable considering the early Universe dynamics. Thus, we have used a power spectrum derived from the Galilean inflationary setup within the effective field theory (EFT) \cite{Weinberg:2008hq,Cheung:2007st,Choudhury:2017glj,Naskar:2017ekm,Delacretaz:2016nhw}.

Generally, the standard picture involves the collapse of the large primordial curvature perturbations exceeding the threshold value during their re-entry into the radiation-dominated (RD) ($w=1/3$) epoch to form PBHs. 
However, it might be interesting 
to formally extend the analysis to a background evolution with an arbitrary EoS parameter  
Recently, many authors have focused on an arbitrary EoS-dependent background for the Universe right before BBN and during the end of inflation \cite{Altavista:2023zhw,Liu:2023pau,Liu:2023hpw,Balaji:2023ehk,Domenech:2021ztg,Domenech:2020ers,Domenech:2019quo}. 
The basis for the present work is that we integrate the EoS parameter into the framework of Galileon inflation to analyze the production of Scalar Induced Gravitational Waves (SIGWs) and the abundance of Primordial Black Holes (PBHs). Here, we work within the linear regime of the cosmological perturbation theory, which does not account for the effects of non-Gaussianity at super-Hubble scales. Moreover, we adopt the Press-Schechter formalism in this paper to derive the PBH mass fraction. We show that choosing the general EoS formalism to avoid PBH overproduction works well only when its values lie near the RD epoch and under the assumptions of linearity for the overdensity and resulting Gaussian statistics. Including non-linear effects in the overdensity on the super-Hubble scales becomes the source of non-Gaussianity \cite{Young:2019yug,DeLuca:2019qsy}, which demands a separate analysis to tackle overproduction issue \cite{Choudhury:2023fwk}. We shall provide a detailed comparative analysis between these two approaches.

We shall, in particular, focus on the formation of PBHs within a specific mass range predicted by the observed NANOGrav15 signal and generate the spectrum of the GW energy density $\Omega_{\rm GW}h^{2}$ which can together remain consistent with the recent data PTA. It was earlier noted \cite{Choudhury:2023hfm,Choudhury:2023fwk} that the SIGWs generated from Galileon can be consistent with the NANOGrav15 data, but the analysis was carried out primarily in the RD era. To strengthen the choice of the EoS and investigate possible new features we now incorporate the EoS as the new element for our analysis. We shall discuss the production of PBH by incorporating an SR/USR/SR-like setup, where the USR regime provides for the necessary enhancement in the curvature perturbations facilitating PBH formation\cite{Kristiano:2022maq,Riotto:2023hoz,Choudhury:2023vuj,Choudhury:2023jlt,Choudhury:2023rks,Choudhury:2023hvf,Choudhury:2023kdb,Choudhury:2023hfm,Bhattacharya:2023ysp,Choudhury:2023fjs,Banerjee:2021lqu,Kristiano:2023scm,Riotto:2023gpm,Firouzjahi:2023ahg,Firouzjahi:2023aum,Firouzjahi:2023bkt,Franciolini:2023lgy,Cheng:2023ikq,Tasinato:2023ukp,Tasinato:2023ioq,Motohashi:2023syh,Jackson:2023obv,Davies:2023hhn,Iacconi:2023ggt,Mu:2023wdt,Domenech:2023dxx,Ahmadi:2023qcw,Dalianis:2023pur,Tada:2023rgp,Kawaguchi:2023mgk,Tomberg:2023kli,Ragavendra:2023ret,Zhai:2023azx}. 

The outline of this work is as follows: In Sec.\ref{s2}, we begin with a brief layout of the Galileon theory with the important Covariantized Galileon action for the underlying setup to work. In Sec.\ref{s3}, we incorporate inflation within Galileon by presenting the action and general scalar field solution in a background de Sitter spacetime. In Sec.\ref{s4}, we focus in great detail on the implementation of the sharp transition feature with Galileon in an EFT framework. We discuss the features in the slow-roll parameters and analyze the EFT coefficients necessary for executing the SR/USR/SR setup. In Sec.\ref{s5}, we focus on building the one-loop corrected scalar power spectrum and the importance of the non-renormalization theorem in understanding the relevant one-loop corrections. In Sec.\ref{s6}, we introduce the Equation of State parameter $w$ and its impact on the PBH formation mechanism. In Sec.\ref{s7}, we discuss the impact of $w$ on the scalar-induced gravity waves production by presenting first a motivation followed by a concise derivation of the energy density of GWs but for a general cosmological background having constant $w$ and speed of propagation $c_{s}$. In Sec.\ref{s8}, we first outline the overproduction issue and discuss the possible resolutions for this problem. In Sec.\ref{s9}, we present the numerical outcomes concerning the overproduction issue and the $w$-SIGW spectra. The discussions and overall summary of this work is provided in Sec.\ref{s10}. Lastly, we mention some of the key features concerning the various integrals encountered during the general treatment of the kernel for the GW energy density spectrum in the appendix \ref{app:A} and \ref{app:B} with their special limits in \ref{app:C}.

\section{Galileon EFT Set-up}
\label{s2}

In this section, we briefly outline inflation within Galileon theory and provide the necessary information to construct the one-loop corrected scalar power spectrum. Galileon theory has the property that the underlying action conceals terms multi-linear in first and second derivatives, but the resulting non-linear equations of motion are still second-order thereby eliminating the presence of ghost-like instabilities in the Hamiltonian and also preserving unitarity. We will primarily focus on a curved de-sitter space to study inflation. 

The action of the Galileon theory is embedded with a Galilean Shift symmetry which allows the scalar field $\phi$ transformation as:
\bea
\phi \longrightarrow \phi + b_{\mu}x^{\mu} + d,
\eea
where $b_{\mu}$ and $d$ are vector and scalar constants and $x^{\mu}$ represents the $3+1$ space-time coordinates. To conduct inflation requires a mild breaking of this Galilean shift symmetry such that the effects due to gravity become highly suppressed by powers of $1/M_{p}$. In ref. \cite{Deffayet:2009wt}, the authors show how to construct an action in a de-Sitter background, which preserves unitarity and eliminates ghost instabilities by introducing a non-minimal coupling to gravity, giving rise to the following Covariantized Galileon Theory (CGT) action \cite{Nicolis:2008in, Deffayet:2009wt,deRham:2010eu,deRham:2011by}:  
\bea \label{CovGal}
S = \int d^{4}x\sqrt{-g}\left[\frac{M^{2}_{pl}}{2}R - V_{0} +\sum^{5}_{i=1}c_{i}{\cal L}_{i}\right]
\eea
where the Lagrangians ${\cal L}_{i},\;\forall i=1,\cdots,5$, are given by the expressions:
\bea
{\cal L}_1 & = & \phi, \nonumber\\
 {\cal L}_2 &=&-\frac{1}{2} (\grad \phi)^2, \nonumber\\
	{\cal L}_3 &=&\frac{1}{\Lambda^3} (\grad \phi)^2 \Box \phi ,\nonumber\\
	{\cal L}_4 &=& -\frac{1}{\Lambda^6} (\grad \phi)^2 \Big\{
					(\Box \phi)^2 - (\grad_\mu \grad_\nu \phi)
					(\grad^\mu \grad^\nu \phi)
					- \frac{1}{4} R (\grad \phi)^2
				\Big\},\nonumber\\
	{\cal L}_5&=& \frac{1}{\Lambda^9} (\grad \phi)^2 \Big\{
					(\Box \phi)^3 - 3 (\Box \phi)( \grad_\mu \grad_\nu \phi)
					(\grad^\mu	 \grad^\nu \phi)
     + 2 ( \grad_\mu  \grad_\nu \phi)
					(\grad^\nu	 \grad^\alpha \phi)
					(\grad_\alpha \grad^\mu \phi)
					- 6 G_{\mu \nu} \grad^\mu \grad^\alpha \phi
					\grad^\nu \phi \grad_\alpha \phi
				\Big\}.
	\eea
In the above, $G_{\mu\nu}$ and $R$ are the Einstein tensor and Ricci scalar for the curved background, while $V_{0}$ is a constant term that is also involved in the soft Galilean symmetry-breaking. See refs. \cite{Kobayashi:2010wa,Jain:2010ka,Gannouji:2010au,Ali:2010gr,deRham:2011by,Tsujikawa:2010sc,DeFelice:2010gb,DeFelice:2010pv,DeFelice:2010nf,Kobayashi:2010cm,Deffayet:2010qz,Burrage:2010cu,Mizuno:2010ag,Nesseris:2010pc,Khoury:2010xi,DeFelice:2010as,Kimura:2010di,Hirano:2010yf,Kamada:2010qe,Hirano:2011wj,Li:2011sd,Kobayashi:2011pc,DeFelice:2011zh,Burrage:2011bt,Liu:2011ns,Kobayashi:2011nu,PerreaultLevasseur:2011wto,deRham:2011by,Clifton:2011jh,Gao:2011mz,DeFelice:2011uc,Gao:2011qe,DeFelice:2011hq,Qiu:2011cy,Renaux-Petel:2011rmu,DeFelice:2011bh,Wang:2011dt,Kimura:2011dc,DeFelice:2011th,Appleby:2011aa,DeFelice:2011aa,Zhou:2011ix,Shirai:2012iw,deRham:2012az,Ali:2012cv,Liu:2012ww,Choudhury:2012yh,Choudhury:2012whm,Barreira:2012kk,Gubitosi:2012hu,Arroja:2013dya,Sami:2013ssa,Khoury:2013tda,Burrage:2015lla,Koyama:2015vza,Saltas:2016nkg,Herrera:2022tad} for more details. The coefficients $c_{i},\;\forall i=1,\cdots,5$ present along with the Lagrangian have a crucial role when demanding an SR/USR/SR-like setup having sharp transitions between each phase. We will primarily work with a sharp transition scenario when going from SRI to USR and USR to SRII phases, and we will discuss this construction in detail soon. After the covariantization, the action in eqn.(\ref{CovGal}) maintains the quadratic nature of the equations of motion. This formulation can be extended to higher dimensions also but here we are concerned with the above version in the $3+1$ dimensions. Galileon has observed much attention as models for dark energy. It allows for an infrared modification of gravity and comes as a subclass of the Horndeski theory which gives the most general theory of a scalar field interacting with gravity having second-order equations of motion. In our further analysis, we will apply the Galileon theory in a cosmological setting by studying the formation of primordial black holes in the framework of single-field inflation. The non-renormalization theorem present for the Galileon will prove as the most important feature to account for the quantum loop effects and provide interesting results related to PBH formation. 

We would also like to mention the underlying connection with the action for fluctuations in Galileon inflation and the most general action for fluctuations in a quasi de Sitter background based on unbroken spatial diffeomorphisms and non-linear realization of Lorentz invariance and first proposed by \cite{Cheung:2007st}. Before going into the details, we present the general EFT action under consideration:
\bea \label{EFTaction}
  S &=& \int d^4 x \sqrt{-g} \Bigg(\frac{M_p^2}{2}R+M_{p}^2 \dot H g^{00}-M_p ^2 (3H^2 +\dot H)+ \frac{M_{2}^{4}(t)}{2!}(g^{00}+1)^{2} + \frac{M_{3}^{4}(t)}{3!}(g^{00}+1)^{3} \nonumber\\ 
&&\quad\quad\quad\quad\quad\quad\quad\quad\quad\quad\quad\quad\quad\quad\quad\quad\quad\quad -\frac{\overline{M}_{1}^3 (t)}{2} (g^{00}+1) \delta K_{\mu}^{\mu}-\frac{\overline{M}_{2}^2 (t)}{2} (\delta K_{\mu}^{\mu})^2 - \frac{\overline{M}_{3}^2 (t)}{2} \delta K_{\nu}^{\mu}\delta K_{\mu}^{\nu}+\cdots\Bigg).  \eea
The above-truncated version of the EFT action is constructed for a single scalar field $\phi$ model with the gauge chosen such that the constant time slices coincide with the uniform $\phi$ slices. This gauge makes it easier for one to study the metric perturbations only. Notice that the above expansion in the action consists of terms with powers of $g^{00}+1$, which show the fluctuations around an unperturbed FLRW background with quasi de Sitter solution. The remaining higher-order terms in the action are omitted by the use of ellipsis. 

The conditions for the action in eqn.(\ref{EFTaction}) require the use of the unit normal vector $n^{\mu}$ on the constant time slice, and the fluctuations coming from the extrinsic curvature tensor $\delta K_{\mu}^{\nu}$. The terms $M_{2}(t),\;M_{3}(t),\;\overline{M}_{1}(t),\;\overline{M}_{2}(t),$ and $\overline{M}_{3}(t)$ represent the time-dependent Wilson's EFT coefficients. To restore the gauge symmetry due to broken time-diffeomorphisms, a new scalar field, the Goldstone mode $\pi(t,\mbf{x})$, is introduced via the St\"{u}ckelberg mechanism, which non-linearly transforms under time diffeomorphism. Between this Goldstone mode and the comoving curvature perturbation variable $\zeta(t,\mbf{x})$, there exists a one-to-one correspondence in the super-Horizon regime. This can be realized since using a combination of the above Wilson coefficients one can construct the Galileon EFT coefficients, $c_{i},\;\forall\;i=1,\cdots,5$, with the underlying Galilean shift symmetry as slightly broken to consider an inflationary scenario. Similarly, one can go around and, from the Galileon EFT coefficients, construct the above Wilson's EFT coefficients. In the general EFT, the amount of symmetry breaking gets described by the Goldstone $\pi(t,\mbf{x})$ which can also be identified with the comoving curvature perturbation using the relation $\zeta(t,\mbf{x}) = -H\pi(t,\mbf{x})$, at the linear order in perturbations, where $H$ is the Hubble expansion rate.

This establishes a direct correspondence between the Goldstone EFT and the CGEFT, which requires the presence of the decoupling limit approximation where the gravity sector does not interfere with the nonlinear Galileon self-interactions. Thus, at the level of the perturbations, the calculations of the correlation functions can be performed either with the Goldstone $\pi$ or the curvature perturbation $\zeta$. We need the background Galileon setup to implement this in the Galileon EFT paradigm, which we describe in the next section. One can also compute another critical parameter known as the effective sound speed $c_{s}$ for both the EFT sectors, Galileon and the general one with the Goldstone, and one can show a correspondence of a similar nature, which we also shed light on in the next section.

\section{Quasi de Sitter solution from Galileon EFT}
\label{s3}

We start our discussion with the action for a background time-dependent and homogeneous Galileon field $\bar{\phi(t)}$ in de Sitter spacetime:
\bea \label{bcgaction}
S^{(0)}=\int d^4x\,a^3 \,\Bigg\{\dot{\bar{\phi}}^2\Bigg(\frac{c_2}{2}+2c_3Z+\frac{9c_4}{2}Z^2+6c_5Z^3\Bigg)+\lambda^3\bar{\phi}\Bigg\}
\eea
where the de Sitter scale factor $a(t)=e^{Ht}$ is used, $H$ is the Hubble parameter, $t$ is the cosmic time, and the coefficient $c_{1}=\lambda^{3}$ for Galileon which brings the mild symmetry breaking due to the linear term in eqn.(\ref{CovGal}). The parameter $Z$ represents the coupling constant for the Galileon theory and is given by:
\bea \label{Z}
Z \equiv \frac{H\dot{\bar{\phi}}}{\Lambda^{3}},
\eea
with $\Lambda$ representing the physical cut-off scale of the theory only below which their effective description remains valid. The evolution of such a scalar field in de Sitter background is defined properly when the decoupling limit $M_{p} \rightarrow \infty$ is considered keeping $H$ fixed. This limit works provided the variation in inflationary potential satisfies the constraint $|\Delta V/V| \ll 1$. Depending on the value of the Galileon coupling parameter, the theory can be studied into two separate regimes, the strongly $(Z \gg 1)$ and the weakly $(Z \ll 1)$ coupled. Interestingly, a solution exists between the two mentioned regimes. We mention the following smooth solution from the above action:
\bea \label{bcgsoln}
\dot{\bar{\phi}} = \frac{\Lambda^3}{12H}\frac{c_2}{c_3}\Bigg[-1+\sqrt{1+\frac{8c_3}{c^2_2}\frac{\lambda^3}{\Lambda^3}}\Bigg]=
\left\{
	\begin{array}{ll}
		\displaystyle \sqrt{\frac{\Lambda^3}{18c_3}\frac{\lambda^3}{H^2}}\quad\quad\quad & \mbox{when}\quad  Z\gg 1\;\text{(strong limit)}    \\ 
			\displaystyle 
			\displaystyle \frac{\lambda^3}{3c_2H}\quad\quad\quad & \mbox{when }\quad  Z\ll 1\;\text{(weak limit)} 
	\end{array}
\right. \eea
where we mention the solutions in the strong and weak coupling limits. In the regime $Z \ll 1$, the theory approaches the canonical slow-roll inflation, while in $Z \gg 1$, the theory approaches the DGP-like model. For the case with $ \simeq 1$, we have a theory interpolating between the strong and weak regimes. This solution is important in the sense that the non-linear interaction in Galileon becomes significant while any mixing with the gravity sector becomes irrelevant. Since we are not concerned with studying the effects of gravity mixing terms and the significant changes in the canonical slow-roll picture, we will stick with the intermediate regime provided by studying $Z \simeq 1.$ 

\textcolor{black}{Focusing on the decoupling limit again, due to the presence of an exact shift symmetry in this scenario, the equation of motion for the Galileon scalar field in background de Sitter space follows from the use of the current conservation equation as follows:}
\bea \label{currentconserve}
\nabla_{\mu}J^{\mu} = \dot{J}^{t} + 3H\left(J^{t} - \frac{\lambda^{3}}{3H}\right) = 0,
\eea
\textcolor{black}{where $J^{t}$ is the temporal part of the Noether current. Upon comparing the above equation with the continuity equation found for the case of a perfect fluid we can make the following identification:}
\bea
\rho=J^{t},\quad\quad P=\frac{-\lambda^{3}}{3H}
\eea
\textcolor{black}{where $\rho$ and $P$ are the energy density and pressure, respectively. These will further become useful in realising the equation of state (EoS) parameter $w$ which is defined as $P=w\rho$. To involve the remaining Galileon EFT coefficients into determining $w$, we make use of the equation of motion as coming from the background action in eqn. \ref{bcgaction}. From variation of the action, $\delta S^{(0)}/\delta\bar{\phi}= 0$, we can obtain the following equation \cite{Burrage:2010cu}:}
\bea
\dot{\bar{\phi}}(c_{2}+6c_{3}Z+18c_{4}Z^{2}+30c_{5}Z^{3}) = \frac{\lambda^{3}}{3H},
\eea
\textcolor{black}{where if we identify $J^{t}=\lambda^{3}/3H$ as a solution of eqn. \ref{currentconserve}, we include other coefficients into the definition of the current $J^{t}$ (or energy density $\rho$). From the above, we finally write the equation of state in the framework of Galileon EFT as follows:}
\bea \label{eosGal}
w = \frac{P}{\rho} = \frac{-\lambda^{3}/3H}{\dot{\bar{\phi}}(c_{2}+6c_{3}Z+18c_{4}Z^{2}+30c_{5}Z^{3}) },
\eea
\textcolor{black}{and this provides us one way to analyse the range of values that the Galileon EFT coefficients can take to specify a certain the EoS. In the subsequent discussions we will attempt to identify possible ranges the EFT coefficients $c_{i},\;\forall\;i=1,\cdots,5$ can take in the context of the sharp transition set up for the present work.} In the next section we will show in detail how to successfully implement an SR/USR/SR-like construction within Galileon.

\section{Implementation of sharp transition using Galileon EFT} \label{s4}

In this section, we analyze the construction behind the sharp transition feature in the underlying Galileon theory. Our setup involves a sharp transition from the first slow-roll (SRI) to the USR phase and again from the USR to the second slow-roll (SRII) phase. This setup will provide a means to accurately study the generation of PBHs due to the large enhancements in the primordial fluctuations brought by the sharp transitions and the USR phase; therefore, we discuss its construction explicitly. The deviations from exact de sitter during inflation are signaled by the  slow-roll parameters defined as:
\bea \label{srparams}
\epsilon = -\frac{\dot{H}}{H^{2}} = -\frac{d\ln{H}}{d{\cal N}}, \quad\quad \eta \equiv -\frac{\ddot{\phi}}{\dot{\phi}H} = \epsilon - \frac{1}{2}\frac{d\ln{\epsilon}}{d{\cal N}}
\eea
where $\epsilon$ and $\eta$ denote the first and second slow-roll parameters and ${\cal N}$ denotes the number of e-folding. 
Note that the Hubble rate $H$ contains the information about the non-zero constant term $V_{0}$ and the correction due to the linear terms in the scalar field $\phi$. The mild symmetry breaking implies the potential of the form: $V = V_{0} - \lambda^{3}\phi$. Now, from the Friedman equations we have:
\bea
H = \frac{1}{M_{p}}\sqrt{\frac{V}{3}} &=& \frac{1}{M_{p}}\sqrt{\frac{V_{0}-\lambda^{3}\phi}{3}} \nonumber\\
&=& \frac{1}{M_{p}}\bigg[\sqrt{\frac{V_{0}}{3}}\bigg(1-\frac{\lambda^{3}}{V_{0}}\phi\bigg)^{1/2}\bigg] \nonumber\\
&=& \frac{1}{M_{p}}\sqrt{\frac{V_{0}}{3}}\bigg(1-\frac{\lambda^{3}}{2V_{0}}\phi - \frac{\lambda^{6}}{8V_{0}^{2}}\phi^{2} + \cdots\bigg)
\eea
where $H$ is the Hubble parameter for the quasi de Sitter spacetime and is not a constant but is written using the Hubble parameter in exact de Sitter, $H_{\rm dS} = \sqrt{V_{0}/3M_{p}^{2}}$, as seen above. The constant term satisfies $V_{0} > 0$ and $\lambda^{3}/V_{0} \ll 1$ is the condition which justifies the above expansion in the last line. The terms after the leading de Sitter contribution $H_{\rm dS}$ represent the subsequent corrections and this is reflected in the slow-roll paradigm where one calculates the SR parameters considering the above expansion.

The behavior of these slow-roll parameters changes during each phase, and they are connected to the smooth solution of the background scalar field in eqn.(\ref{bcgsoln}) involving the coefficients $c_{1},c_{2},c_{3}$. We will suggest possible values for these coefficients in this section. To constrain the other two remaining coefficients, $c_{4},c_{5}$, requires further knowledge of the second-order action for the scalar perturbations, which we will elaborate on later. Lastly, we introduce the effective sound speed $c_{s}$ as another essential parameter to consider. The definition for this requires the following time-dependent coefficients:
\bea \label{coeffA}
{\cal A}&\equiv& \frac{\dot{\bar{\phi}}^2}{2}\Bigg(c_2+12c_3Z+54c_4Z^2+120c_5Z^3\Bigg),\\
\label{coeffB}  {\cal B}&\equiv& 
   \frac{\dot{\bar{\phi}}^2}{2}\Bigg\{c_2+4c_{3}Z\Big(2-\eta\Big)+2c_{4}Z^{2}\Big(13-6\big(\epsilon+2\eta\big)\Big)-24c_5Z^{3}\big(2\epsilon+1\big)\Bigg\}.
\eea
In terms of these the effective sound speed is defined as \cite{Choudhury:2023hvf,Choudhury:2023kdb}:
\bea \label{soundspeed}
c_{s}=\sqrt{\frac{\cal B}{\cal A}}.
\eea
where $\dot{\bar{\phi}}$ is defined previously in eqn.(\ref{bcgsoln}) and the constant $Z$ is defined in eqn.(\ref{Z}). As promised earlier at the end of Sec.\ref{s2}, we now mention the effective sound speed obtained from the general Goldstone EFT framework \cite{Cheung:2007st,Choudhury:2017glj,Choudhury:2023hvf,Choudhury:2023jlt,Choudhury:2023rks}:
\bea \label{speedEFT}
c_{s} = \frac{1}{\sqrt{1-\displaystyle{\frac{2M_{2}^{4}}{\dot{H}M_{p}^{2}}}}},
\eea
Upon comparison with eqn.(\ref{soundspeed}) one notices the direct correspondence between the $c_{i}'$s used for ${\cal A}$ and ${\cal B}$ and the EFT coefficient $M_{2}$:
\bea
\frac{M_{2}^{2}}{M_{p}} = \sqrt{\frac{\dot{H}}{2}\bigg(1-\frac{{\cal A}}{{\cal B}}\bigg)}
\eea
The sound speed $c_{s}$ will explicitly arise later and its effect propagates into the computations of the tree and one-loop corrected scalar power spectrum. The Galileon EFT  coefficients $c_{i},\;\forall i=1,\cdots,5,$ appear explicitly in $c_{s}$. Hence, once we establish constraints on $c_{s}$, we indirectly include the contributions of the coefficients $c_{i}$. However, we must remember that putting constraints has to be performed for each of the three phases involved. By demanding the respective slow-roll conditions for each phase, we can constrain values of $c_{1},c_{2},c_{3}$ for each phase. Thus, the nature of slow-roll and ultra-slow roll conditions in the setup of SRI, USR, and SRII together provide us with three sets of parameter values. The other two of the Galileon EFT coefficients $c_{4},c_{5}$ values can be fixed by demanding the unitarity and causality conditions on $c_{s}$ and using information about the sufficient tree-level scalar power spectrum amplitude in the three phases. Current bounds on the parameter $c_{s}$ under the mentioned conditions appear to fall within $0.024 \leq c_{s} < 1$ \cite{Planck:2015sxf}. The tree-level power spectrum, denoted hereon by $\Delta^{2}_{\zeta,{\bf Tree}}(k)$, have the values $\big[\Delta^{2}_{\zeta,{\bf Tree}}\big (k)]_{\rm SRI} \sim {\cal O}(10^{-9})$ in SRI where the CMB sensitive scales appear, $\big[\Delta^{2}_{\zeta,{\bf Tree}}\big (k)]_{\rm USR} \sim {\cal O}(10^{-2})$ in the USR, and $\big[\Delta^{2}_{\zeta,{\bf Tree}}\big (k)]_{\rm SRII} \sim {\cal O}(10^{-5})$ in the SRII.  By imposing such conditions, we can come up with ranges of constraint on the parameter space of $c_{4},c_{5}$ values. In the end, these two values do not directly affect  $\epsilon$ and $\eta$ but by further working with $c_{s}$ and $\Delta^{2}_{\zeta,{\bf Tree}}(k)$ we effectively cover the whole space of coefficients $c_{i}$.

\subsection{First Slow Roll (SRI) phase}

Inflation is initiated with the help of the SRI phase where a scalar field slowly rolls down the inflationary potential with the slow-roll behaviour characterized by the two parameters $\epsilon$ and $\eta$ defined as before in eqn(\ref{srparams}). The CMB-scale fluctuations correspond to those early e-folds probed near the pivot scale $k_{*}=0.02{\rm Mpc}^{-1}$ where we realize the SRI conditions. During SRI, we assume $\epsilon$ to be small while $\eta \simeq 0$ is a small, almost constant value. From the eqn.(\ref{srparams}) one can write for the behaviour of $\epsilon$ the following:
\bea
\int\frac{d\epsilon}{\epsilon(\epsilon - \eta)} = \int 2d{\cal N}\implies
\ln{\bigg(1 - \frac{\eta}{\epsilon}\bigg)}\Biggr|_{\epsilon_i}^{\epsilon_f} = 2\eta({\cal N}_{f}-{\cal N}_{i}). 
\eea
where for the SRI phase the initial condition on $\epsilon_{i}$ is fixed with its value at the CMB-scale entering the Horizon with $k_{*}$ and this also corresponds to the instant of time denoted by e-folds ${\cal N}_{i} = {\cal N}_{*}$. The expression for $\epsilon$ results as follows:
\bea \label{epsN1}
\epsilon_{\rm SRI}({\cal N}) = \eta_{\rm SRI}\bigg(1-\bigg(1-\frac{\eta_{\rm SRI}}{\epsilon_{i,\rm SRI}}\bigg)e^{2\eta_{\rm SRI}\Delta{\cal N}}\bigg)^{-1}, \eea
with $\Delta{\cal N} = {\cal N}_{f}-{\cal N}_{*}$. We take $\eta_{\rm SRI} \sim -0.001$ and $\epsilon_{i,\rm SRI} \sim {\cal O}(10^{-3})$ for the purpose of numerical simplification and with these get their respective behaviour in SRI. Their behaviour with the number of e-foldings is extremely slowly varying which is expected and this will be coupled with the eqs.(\ref{bcgsoln},\ref{srparams}) so as to constrain the values of $c_{1},c_{2},c_{3}$ in SRI. For this purpose, we start by writing eqn.(\ref{bcgsoln}) as:
\bea \label{phiftn}
\dot{\bar{\phi}} = \frac{f(c_{1},c_{2},c_{3},\Lambda)}{H},\quad\quad\quad {\rm where}\quad\quad f(c_{1},c_{2},c_{3},\Lambda) \equiv \frac{\Lambda^3}{12}\frac{c_2}{c_3}\Bigg[-1+\sqrt{1+\frac{8c_3}{c^2_2}\frac{c_{1}}{\Lambda^3}}\Bigg]. 
\eea
This includes the underlying mass scale $\Lambda$ for a valid EFT description. For future calculation purposes we choose $\Lambda \sim {\cal O}(10^{-3}-10^{-1})M_{p}$. Now, after combining the above with the eqn(\ref{srparams}), we can obtain the following relations:
\bea
\eta = -\frac{\ddot{\phi}}{\dot{\phi}H} = -\frac{1}{\dot{\phi}}\frac{d}{d{\cal N}}\bigg(\frac{f}{H}\bigg) = -\frac{1}{f}\bigg(\frac{df}{d{\cal N}} - \frac{f}{H}\frac{dH}{d{\cal N}}\bigg),
\eea
where we have used the fact $d{\cal N}=Hdt$ and
chain rule in the second equality to convert from time $t$ to e-folds ${\cal N}$. Since $\eta$ is almost constant, this provides a differential equation for the function $f$ solving which will give us the combined behaviour of the coefficients within a particular interval of e-folds. Solving the above equation for SRI with appropriate initial conditions leads us to the following result:
\bea \label{ftnN1}
f_{\rm SRI}({\cal N}) = \frac{H_{\rm SRI}({\cal N})}{H({\cal N}_{*})}\exp{(-\eta_{\rm SRI}\Delta{\cal N})}. \eea
We notice that to completely identify $f_{\rm SRI}({\cal N})$ from the above expression requires dependence on the Hubble parameter with e-folds, $H({\cal N})$. This can be achieved via plugging the solution from eqn.(\ref{epsN1}) into the second definition of $\epsilon$ from eqn.(\ref{srparams}) in the following manner to give:
\bea
-\frac{dH}{H} = \eta_{\rm SRI}\bigg(1-\bigg(1-\frac{\eta_{\rm SRI}}{\epsilon_{i, \rm SRI}}\bigg)e^{2\eta_{\rm SRI}\Delta{\cal N}}\bigg)^{-1}d{\cal N}, \eea
Upon integrating both sides of the above-mentioned equation we arrive at the following result:
\bea
\ln{(H)}\Biggr|_{H_{*}}^{H} = -\frac{1}{2}\ln{\bigg(\frac{\exp{(2\eta_{\rm SRI}{\cal N})}}{-\epsilon_{i,\rm SRI}+\exp{(2\eta_{\rm SRI}{\cal N})}(\epsilon_{i,\rm SRI}-\eta_{\rm SRI})}\bigg)}\Biggr|_{{\cal N}_{*}}^{{\cal N}},
\eea
and after taking the suitable integration limits this reduces to give us the following expression for the Hubble parameter in terms of the number of e-foldings ${\cal N}$ and other necessary important parameters as:
\bea \label{HubbleN1}
\frac{H_{\rm SRI}({\cal N})}{H_{*}} = \bigg(\frac{\eta_{\rm SRI}\exp{(2\eta_{\rm SRI}{\cal N})}}{\epsilon_{i, \rm SRI}-\exp{(2\eta_{\rm SRI}{\cal N})}(\epsilon_{i, \rm SRI}-\eta_{\rm SRI})}\bigg)^{-\frac{1}{2}},
\eea
where the term $\eta_{\rm SRI}$ in the inverse square root numerator comes after imposing the initial condition at ${\cal N}={\cal N_{*}}$, and we choose to keep ${\cal N_{*}}=0$ for convenience. Information concerning the Galileon EFT coefficients gets contained within the solution from eqn.(\ref{ftnN1}), based on which we can identify the possible set of values for the three coefficients $c_{1},c_{2},c_{3}$.

\begin{figure*}[htb!]
    	\centering
    \subfigure[]{
      	\includegraphics[width=8.5cm,height=7.5cm]{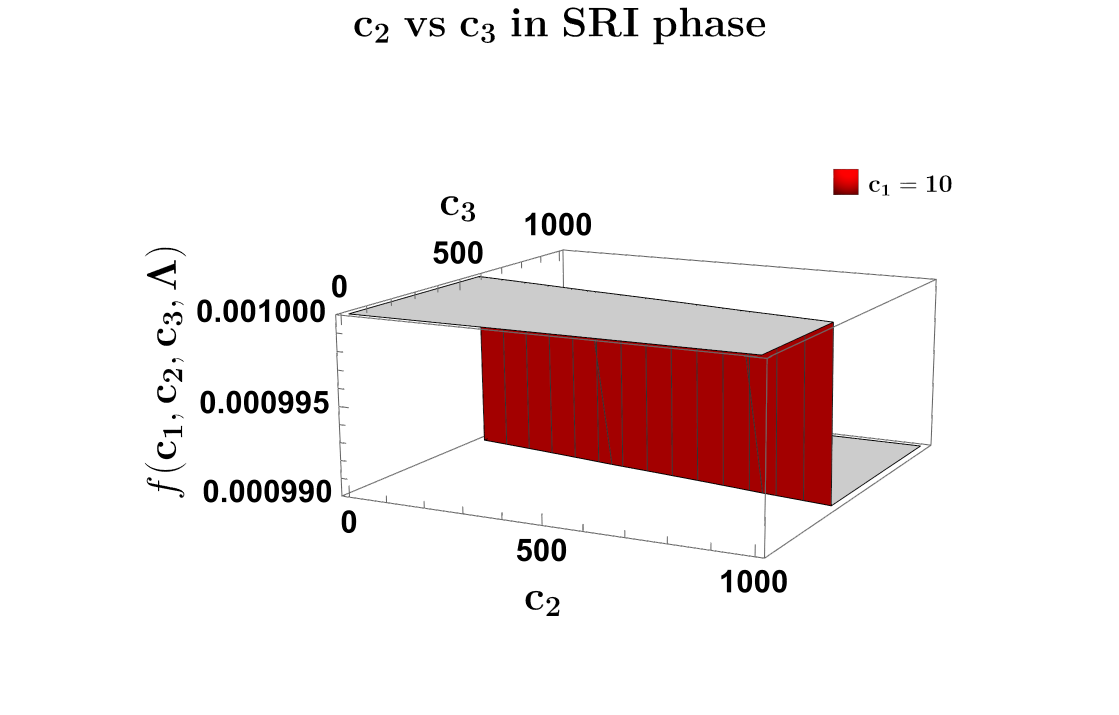}
        \label{sr1c1}
    }
    \subfigure[]{
       \includegraphics[width=8.5cm,height=7.5cm]{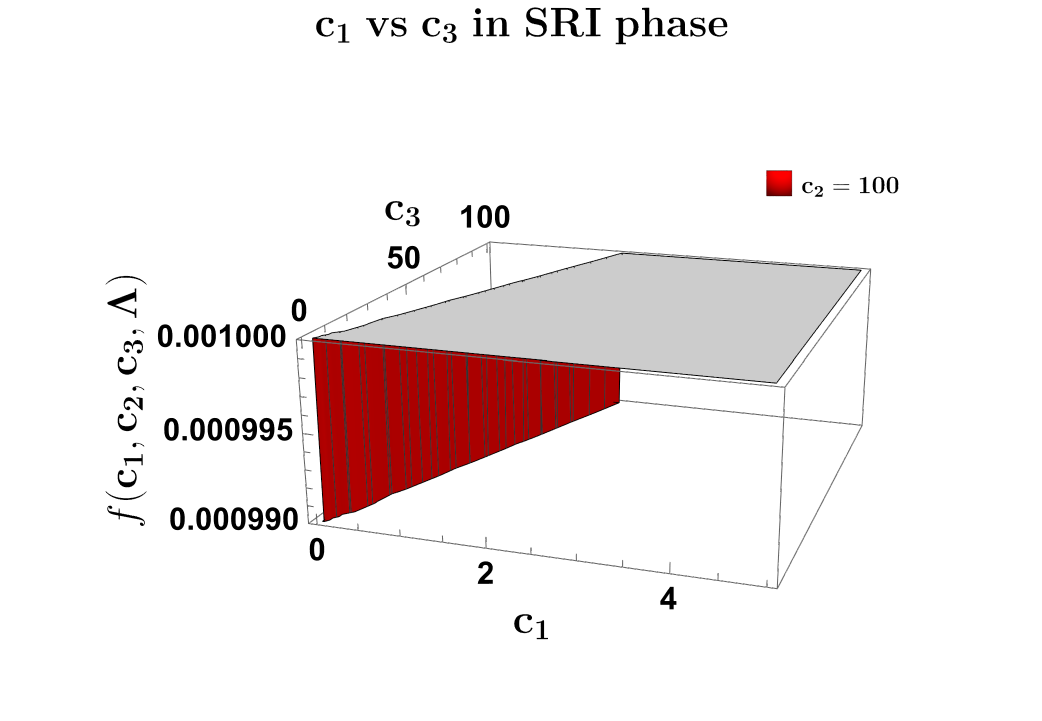}
        \label{sr1c2}
       }\\
   \subfigure[]{
       \includegraphics[width=8.5cm,height=7.5cm]{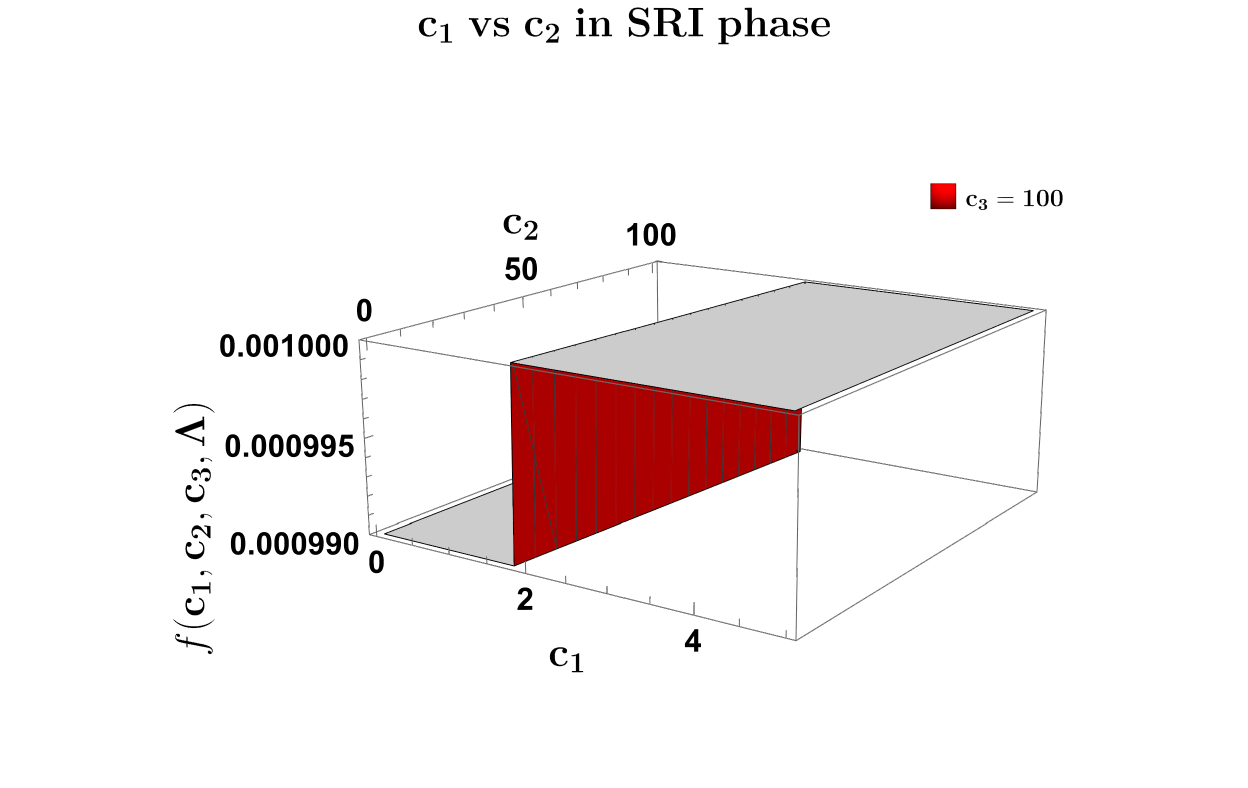}
        \label{sr1c3}
       }\hfill
    
    	\caption[Optional caption for list of figures]{Behaviour of the three Galileon EFT coefficients $c_{1},c_{2},c_{3}$ to satisfy $f\equiv f(c_{1},c_{2},c_{3},\Lambda)$ in the SRI phase. In (a), $c_{1}=10$ is fixed which results in $c_{3}\sim {\cal O}(10^{3})$ while change in $c_{2}$ less sensitive to obtain $f$. In (b), $c_{2}=100$ is fixed which confirms $c_{3}\sim {\cal O}(10^{3})$ in order to have $c_{1}\sim {\cal O}(10)$. In (c), $c_{3}=100$ is fixed and shows less sensitivity to $c_{2}$ but $c_{1}\sim {\cal O}(10)$ is sufficient to satisfy $f$. } 
    	\label{sr1}
    \end{figure*}
We now begin with the analysis of $c_{1},c_{2},c_{3}$. Recall that the regime where non-linearities of the Galileon remain relevant while the mixing with gravity can be neglected occurs for $Z \gtrsim 1$. From eqs.(\ref{Z},\ref{phiftn}) one can see that the function must satisfy $f(c_{1},c_{2},c_{3},\Lambda) \sim \Lambda^{3} \sim {\cal O}(10^{-9}-10^{-3})M_{p}^{3}$. The result from eqn.(\ref{ftnN1}) also predicts this similar behaviour for  $f(c_{1},c_{2},c_{3},\Lambda)$ till the SRI phase functions.
We start with fixing $c_{2}=1$ implying canonical normalization for the kinetic term in eqn.(\ref{CovGal}). With $c_{2}$ fixed we get a large range of $c_{3}$ values for a given $c_{1}$ but the reciprocal of this is not true with $c_{1}$ if we consider a particular $c_{3}$. Using the allowed range for the cut-off scale $\Lambda$ as mentioned before we can obtain the possible ranges for the coefficients. For the case with $\Lambda \sim {\cal O}(10^{-1})M_{p}$ and keeping $c_{1}=\lambda^{3}$, the term mildly breaking Galilean symmetry, within $c_{1} \sim {\cal O}(10^{-2}-10^{2})$ gives us set of points for $c_{3}$ within $c_{3} \sim {\cal O}(10^{-1}-10^{4})$. As we go below for $\Lambda < {\cal O}(10^{-1})M_{p}$ then compared to the ranges mentioned earlier, the allowed values of $c_{3}$ increases at a larger rate in magnitude for the mentioned ranges of $c_{1}$ values. The effects of change in $c_{2}$ is found to be always minimal when fixing either $c_{1}$ or $c_{2}$ and varying the other.  

The fig.(\ref{sr1}) depicts how changes in the coefficients $c_{1},c_{2},c_{3}$ allow for us to satisfy the behaviour of $f(c_{1},c_{2},c_{3},\Lambda)$ coming from eqn.(\ref{ftnN1}). The red surfaces show the allowed values for the coefficients in the range depicted in the plots. From the overall analysis of the three plots present within the figure, we conclude that values of $c_{1} \sim (10^{-2}-10^{2})$ suffice the need to achieve the desired $f$. This also includes having $c_{3} \sim {\cal O}(10^{-1}-10^{4})$ where decreasing $c_{3}$ also demands even lower $c_{1}$ values. Lastly, changes in $c_{2}$ is least sensitive and do not improve constraints on the values of $c_{1},c_{3}$. 

To determine the allowed values of the other two coefficients $c_{4}$ and $c_{5}$, we use the effective sound speed $c_{s}$ and the amplitude of the scalar power spectrum $\big[\Delta^{2}_{\zeta,{\bf Tree}} (k)\big]_{\rm SRI} \sim {\cal O}(10^{-9})$. The analytic expression for the scalar power spectrum contains the effective sound speed and this will be made clear through explicit computations in later sections.  
The scalar power spectrum and eqn.(\ref{soundspeed}) both contain $c_{4},c_{5}$ information, and with the known values of the two quantities we can establish constraints on the allowed values of the coefficients. We use the observational constraint $0.024 \leq c_{s} < 1$ to obtain a range of values. 

After performing the above-mentioned procedure to generate the values we find both coefficients lie within the interval of magnitude ${\cal O}(10^{-2}) < \{c_{4},c_{5} \}< {\cal O}(10^{2})$ and can carry negative and positive signatures. For the specific case of $c_{s}(\tau_{*})=c_{s,*}=0.05$ which will be chosen by us during the calculation of the scalar power spectrum and further estimation of the mass of PBH and the GW spectrum, we determine the values $(c_{4},c_{5})\simeq(-2.8,0.27)$ where $c_{2}=1$, and $c_{3}=10$.

\textcolor{black}{Based on the above discussions about the range of values for the various Galileon EFT coefficients in the SRI, below we examine some possible values for each of the $c_{i}$'s that can allow to realise a given EoS $w$ for which we gave an expression using these coefficients in eqn.(\ref{eosGal}).  }

\begin{table}[H]

\centering
\begin{tabular}{|c|c|c|c|c|c|c|}

\hline\hline
\multicolumn{7}{|c|}{\normalsize \textbf{Galileon EFT coefficients for a given equation of state $w$ in SRI}} \\

\hline

EoS $(w)$ & $f$ &\hspace {0.5cm}  $c_{1}$ \hspace {0.5cm} &\hspace {0.5cm}  $c_{2}$ \hspace {0.5cm} &\hspace {0.5cm}  $c_{3}$ \hspace {0.5cm} &\hspace {0.5cm}  $c_{4}$ \hspace {0.5cm} &\hspace {0.5cm}  $c_{5}$ \hspace {0.5cm}  \\
\hline
$1/3$ & & $0.1$ & $10$ & $50$ & $15$ & $-22.5$ \\ 
$0.25$ & ${\cal O}(10^{-3})$  & $0.5$ & $20$ & $50$ & $23$ & $-46.6$ \\
$0.16$ & & $0.6$ & $25$ & $50$ & $24$ & $-67$ \\ \hline 
\hline

\end{tabular}

\caption{ Table describes possible set of values for the EFT coefficients, $c_{i}\;\forall\;i=1,\cdots,5$, satisfying a given EoS $w$. }

\label{tab1eos}

\end{table}

\textcolor{black}{In the above table \ref{tab1eos}, we highlight the different values the EFT coefficients may take in order to realise a specific value of the EoS $w$ and also lie within the ranges necessary to realise the USR phase. }

\subsection{Ultra Slow Roll (USR) phase}

In this section we analyze the parameter space of the coefficients $c_{i},\;\forall\;i=1,\cdots,5$ using the similar procedure as done for the SRI phase substituted with the features of the USR. During the USR regime, the scalar field encounters an extremely flat nature of the inflationary potential which leads to a rapid enhancement in the amplitude of the primordial fluctuations generated during and after the sharp transition from the early SR phase. In our case, since we are not working with any model case for the inflationary potential, the Galileon EFT coefficients do the job of realizing the USR region.

The slow-roll approximation breaks during USR with the SR parameters having the behaviour $\epsilon_{\rm USR} \propto a^{-6}$ which is extremely small, whereas $\eta_{\rm USR} \simeq -6$ becomes very large. As a result of this, we find using eqn.(\ref{srparams}) that the corresponding $\epsilon_{\rm USR}$ falls sharply in magnitude from the previous value of ${\cal O}(10^{-3})$. The initial conditions for $\epsilon$ will now become $\epsilon_{i,{\rm USR}} = \epsilon_{\rm SRI}({\cal N}_{s})$ where ${\cal N}_{s}$ denotes the e-foldings marking the beginning of the USR and end of the SRI phase. The USR persists in between ${\cal N}_{s}$ to ${\cal N}_{e}$ till it transitions sharply into another SRII phase. For the purpose of our calculations, we choose ${\cal N}_{s}\sim {\cal O}(18)$ significance of which will become clear later during the analysis of PBH mass as this wavenumber position corresponding to this instant of time can help generate solar mass PBHs. 

The equation of the parameter $\epsilon$ modifies here to:
\bea \label{epsN2}
\epsilon_{\rm USR}({\cal N}) = \eta_{\rm USR}\bigg(1-\bigg(1-\frac{\eta_{\rm USR}}{\epsilon_{i,\rm USR}}\bigg)e^{2\eta_{\rm USR}({\cal N}-18)}\bigg)^{-1}, \eea
this is able to generate the sharp decreasing behaviour for $\epsilon$ in the USR phase where $\eta \simeq -6$. Using the above into the eqn.(\ref{srparams}) provides us with a new version of eqn.(\ref{HubbleN1}):
\bea \label{HubbleN2}
\frac{H_{\rm USR}({\cal N})}{H_{\rm SRI}({\cal N}_{s})} = \bigg(\frac{\eta_{\rm USR}\exp{(2\eta_{\rm USR}\Delta{\cal N}_{\rm USR})}}{\epsilon_{i,\rm USR}-\exp{(2\eta_{\rm USR}\Delta{\cal N}_{\rm USR})}(\epsilon_{i,\rm USR}-\eta_{\rm USR})}\bigg)^{-\frac{1}{2}},
\eea
the term $\eta_{\rm USR}$ in the inverse square root numerator comes after imposing the initial condition at ${\cal N}={\cal N}_{s}$. Here $\Delta{\cal N}_{\rm USR}={\cal N}-{\cal N}_{s}$ with the initial condition of $\epsilon_{i,\rm USR}$ and this ultimately gives us $f_{\rm USR}({\cal N})$ similar to the expression in eqn.(\ref{ftnN1}):
\bea \label{ftnN2}
f_{\rm USR}({\cal N}) = \frac{H_{\rm USR}({\cal N})}{H_{\rm SRI}({\cal N}_{s})}\exp{(-\eta_{\rm USR}({\cal N}-{\cal N}_{s}))}. \eea
We will analyze the behaviour of the above equation for the USR phase to constrain $c_{1},c_{2},c_{3}$ by working with $\Lambda \sim {\cal O}(10^{-1})M_{p}$. The USR phase persists upto from ${\cal N}_{s} = 18$ to ${\cal N}_{e} = 20.4$ from our numerical analysis which is equivalent to the value from $\Delta{\cal N}_{\rm USR}=\ln{(k_{e}/k_{s})} \sim {\cal O}(2)$ required from perturbativity constraints. To study the allowed values more clearly, we focus on two cases within the USR: Firstly, for $f_{\rm USR}$ close to the instant of transition at ${\cal N}_{s}$ and, secondly, for the remaining amount of e-folds left to completer USR till ${\cal N}_{e}$ is reached. For the first case, we find that the value $f_{\rm USR}$ changes much more quickly than an order of magnitude before it was in the SRI. This allows us to consider a large region of values for $c_{1},c_{2},c_{3}$ for an extremely short interval of e-folds and find the common acceptable values upon considering variation with each coefficient. For the second case, the change occurs much slower compared to the first. 

\begin{figure*}[htb!]
    	\centering
    \subfigure[]{
      	\includegraphics[width=8.5cm,height=7.5cm]{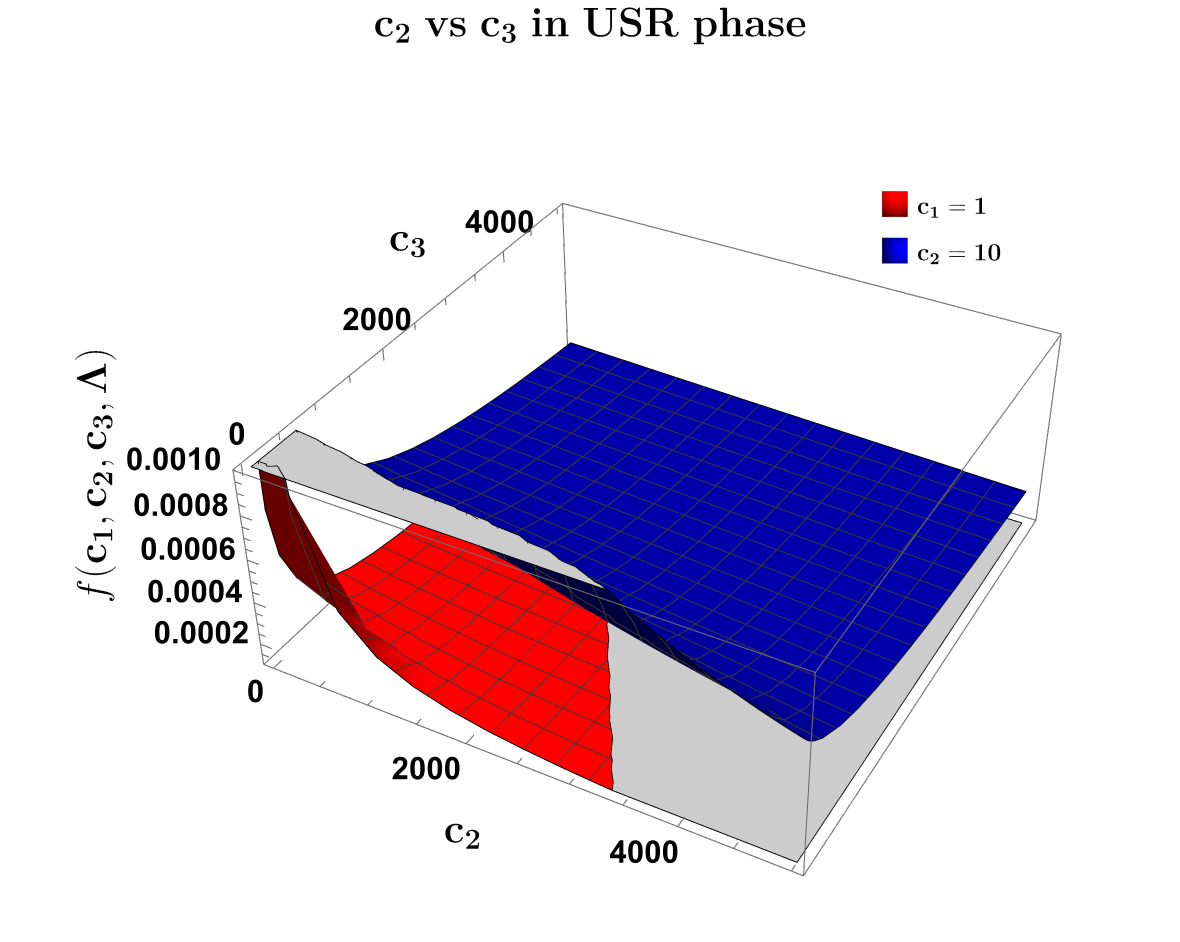}
        \label{usrc1}
    }
    \subfigure[]{
       \includegraphics[width=8.5cm,height=7.5cm]{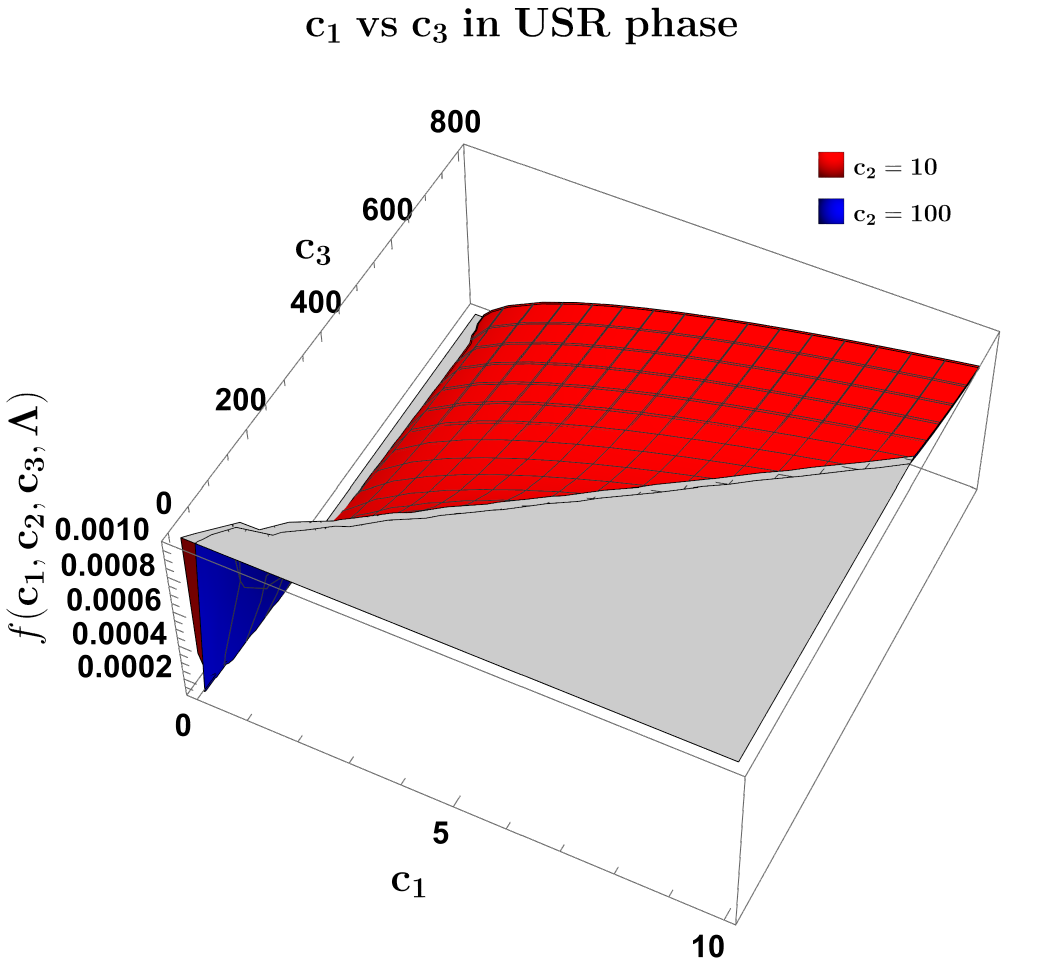}
        \label{usrc2}
       }\\
   \subfigure[]{
       \includegraphics[width=9cm,height=7.5cm]{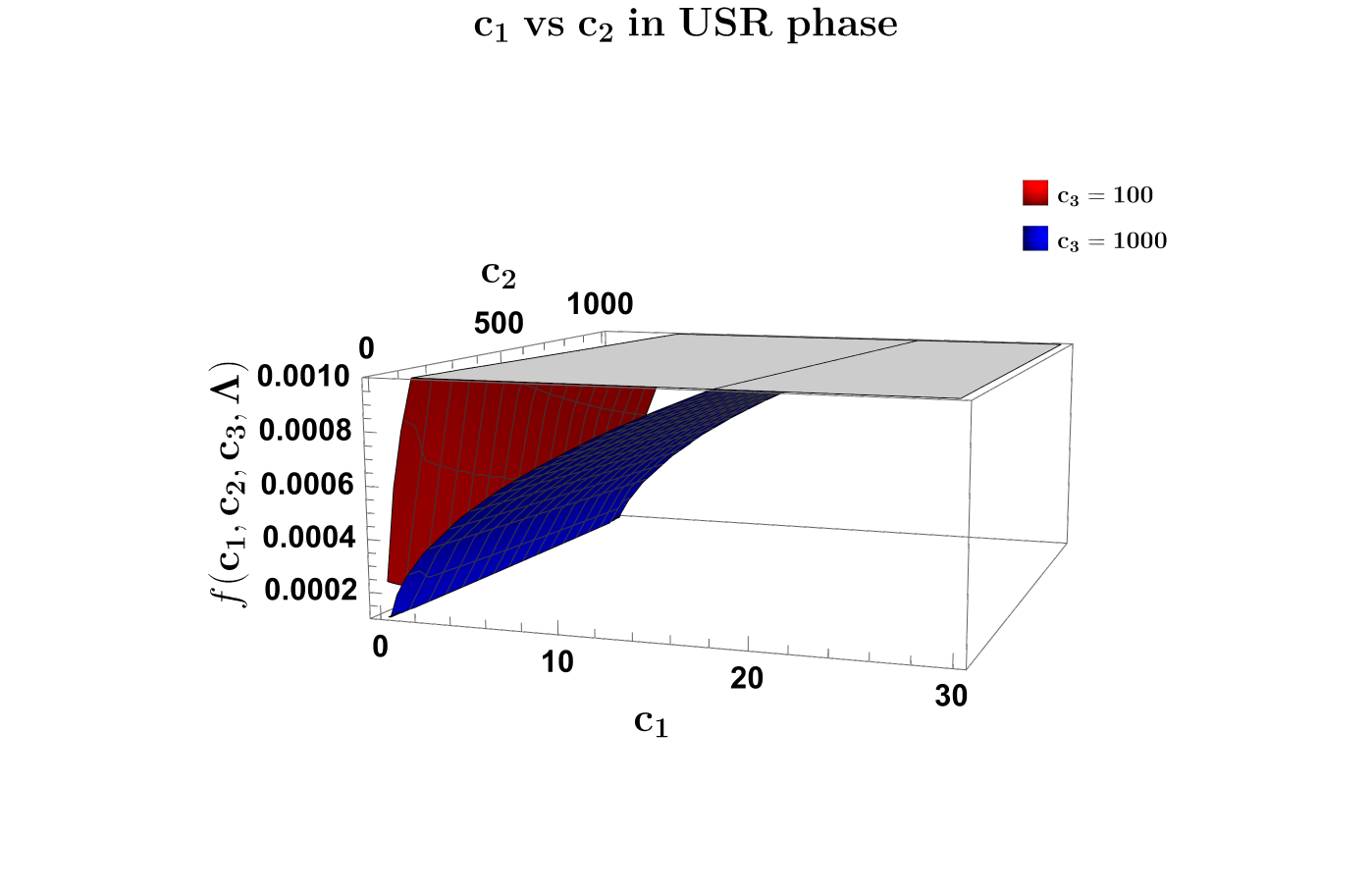}
        \label{usrc3}
       }\hfill
    
    	\caption[Optional caption for list of figures]{Behaviour of the three Galileon EFT coefficients $c_{1},c_{2},c_{3}$ to satisfy $f\equiv f(c_{1},c_{2},c_{3},\Lambda)$ right after transition into the USR phase. In (a), $c_{1}\sim {\cal O}(10)$ is able to generate the desired values of $f$ by keeping $c_{2},c_{3} \sim {\cal O}(10^{-3}).$ In (b), changing from $c_{2}=10$ (red) to $c_{2}=100$ (blue) does not affect the interval allowed for $c_{1},c_{3}.$ In (c), we consider $c_{3}\sim {\cal O}(10^{2}-10^{3})$ where large $c_{3}$ corresponds to larger values of $c_{1}$. } 
    	\label{usr1}
    \end{figure*}

In fig.(\ref{usr1}), we highlight the behaviour of the EFT coefficients $c_{1},c_{2},c_{3}$ when considering the e-folding instant near the sharp transition and the corresponding value of eqn.(\ref{ftnN2}). The surfaces in red and blue show the allowed space of values for the coefficients to satisfy $f$. We see that lower values of $c_{1} < {\cal O}(10)$ are much preferred for not having to consider large $c_{2},c_{3}$. We also find that both $c_{2},c_{3}$ remain similar in magnitude which is now increased relative to their previous values in the SRI case to satisfy the $f$ near the transition. Both coefficients remain to satisfy $c_{2},c_{3} \sim {\cal O}(10^{2})$ for $c_{1} < {\cal O}(10)$ and for $c_{1} \gtrsim {\cal O}(10)$ the plots show that for both we must have $c_{2},c_{3} \gtrsim {\cal O}(10^{3})$.  Larger values of coefficients like $c_{3}$ can mean the significance of higher derivative interactions on the dynamics of the scalar field increases during that particular interval. In the process of the sharp transition at the beginning of USR, these effects coming from the small scales encoded in the higher derivative operators tend to increase. 

\begin{figure*}[htb!]
    	\centering
    \subfigure[]{
      	\includegraphics[width=8.5cm,height=7.5cm]{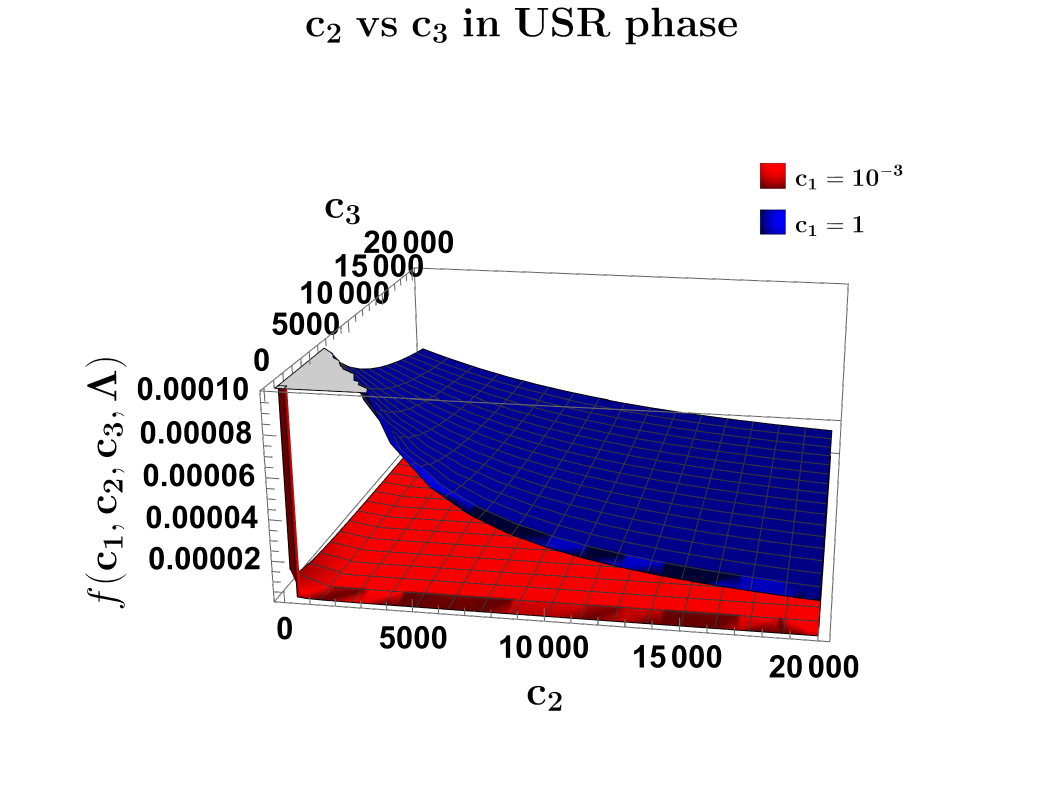}
        \label{usrc11}
    }
    \subfigure[]{
       \includegraphics[width=8.5cm,height=7.5cm]{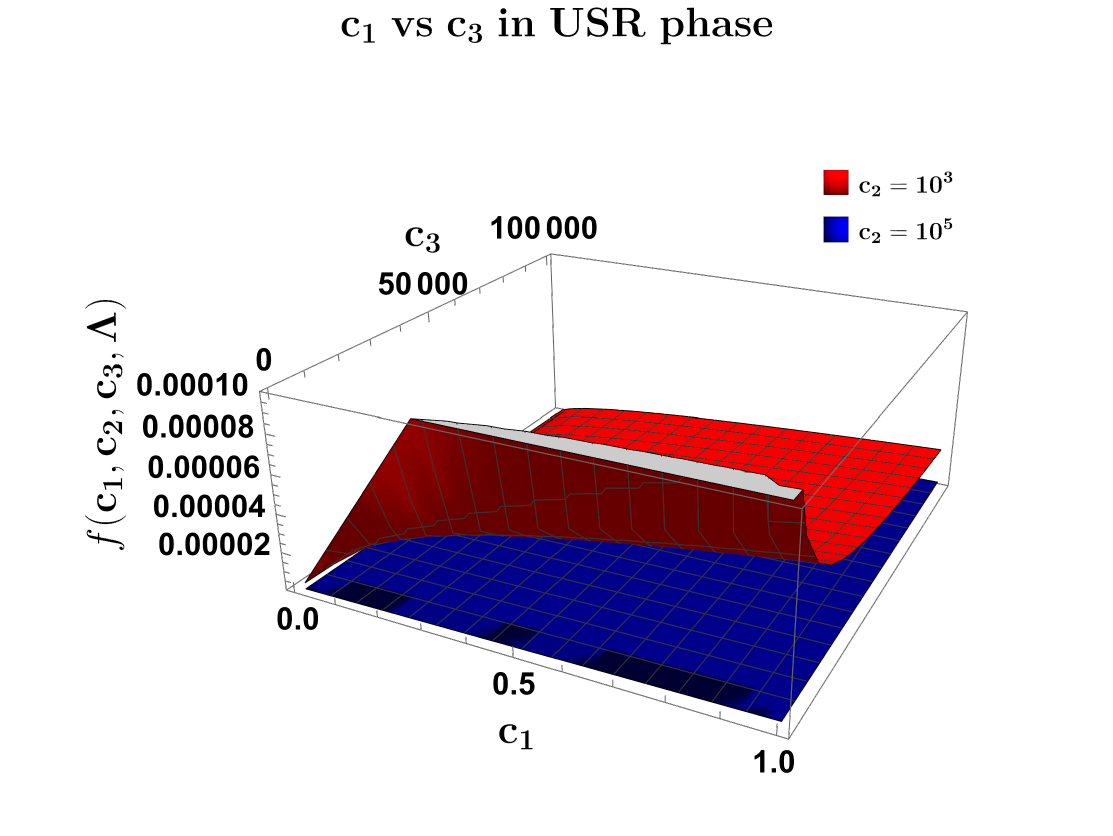}
        \label{usrc21}
       }\\
   \subfigure[]{
       \includegraphics[width=9cm,height=7.6cm]{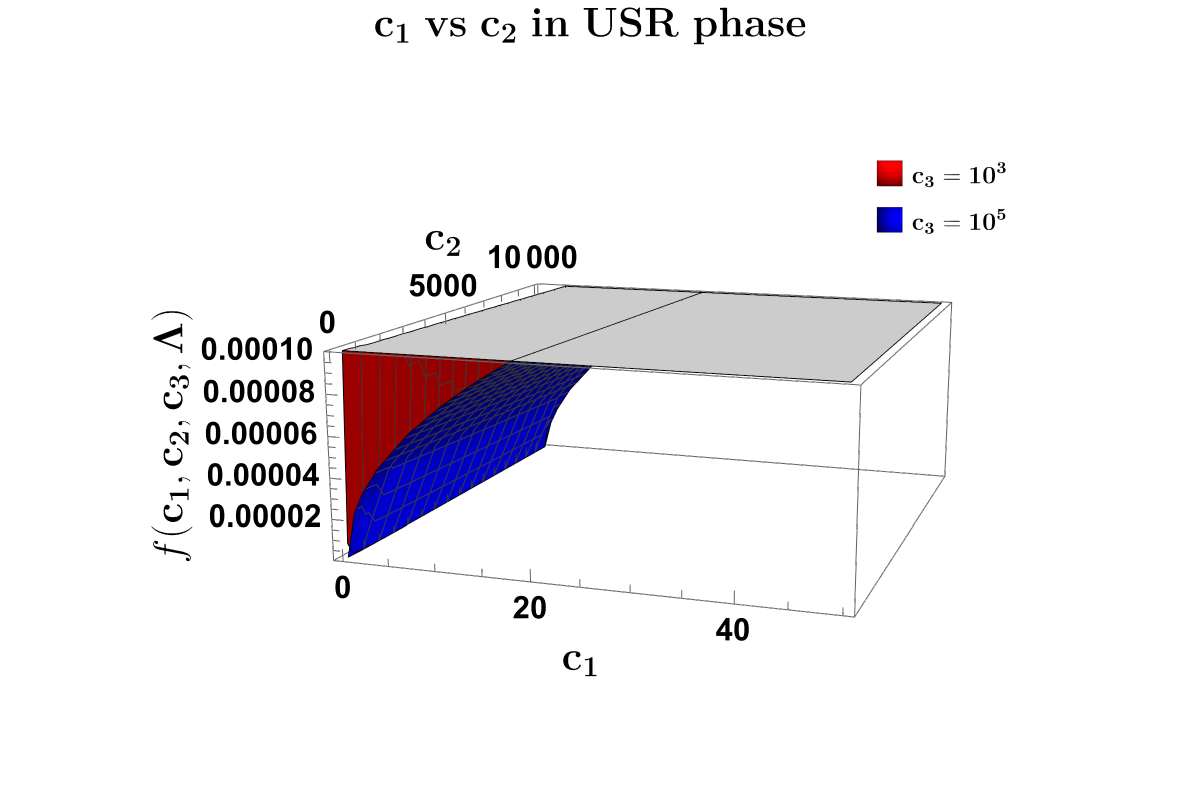}
        \label{usrc31}
       }\hfill
    
    	\caption[Optional caption for list of figures]{Behaviour of the three Galileon EFT coefficients $c_{1},c_{2},c_{3}$ to satisfy $f\equiv f(c_{1},c_{2},c_{3},\Lambda)$ during the USR phase. In (a), $c_{1}=1$ (blue) and $c_{1}=10^{-3}$ (red) are fixed to show the allowed values of $c_{2}$ and $c_{3}$. In (b), $c_{2}=10^{3}$ (red) and $c_{2}=10^{5}$ (blue) are kept fixed. In (c), $c_{3}=10^{3}$ (red) and $c_{1}=10^{5}$ (blue) are kept fixed. } 
    	\label{usr11}
    \end{figure*}

Now we visualize the coefficients in the rest of the USR phase. In fig.(\ref{usr11}), we present the behaviour of the Galileon EFT coefficients during the USR phase after the sharp transition has taken place. From keeping $c_{1}$ fixed we conclude that higher values of $c_{1} \gtrsim {\cal O}(1)$ demand even larger $c_{2},c_{3} \sim {\cal O}(10^{5})$ to achieve the values of $f$ at the lower end. This fact is seen more closely from the other plots where $c_{2}$ and $c_{3}$ are individually fixed. If we decrease $c_{2}$, then $c_{1}$ must be greatly reduced; otherwise, lower values of $f$ will always remain inaccessible. Hence, by further increasing the magnitude of the coefficients, it is implied that the non-linear interactions within the Galileon sector dominate even further. In contrast, the linear order and quadratic interactions remain suppressed. The decrease in $c_{1}$ signifies that the shift symmetry breaking is milder than in the previous SRI phase. After analyzing the outcomes for various ranges of the coefficients $c_{1},c_{2},c_{3}$, we now move towards discussing the behaviour of the other two $c_{4},c_{5}$ coefficients in the manner similar to the above.

We use the fact that the effective sound speed $c_{s}$ changes sharply at the transition scale with $\tilde{c_{s}}=c_{s}(\tau_{s})  = c_{s}(\tau_{e}) = 1 \pm \delta$ where $\delta \ll 1$, while it remains at the value $c_{s}(\tau)=c_{s,*}$ during the conformal time interval $\tau_{s} \leq \tau \leq \tau_{s}$. This parameterization of the sound speed coupled with the constraint on the amplitude of the scalar power spectrum in the USR as $\big[\Delta^{2}_{\zeta,{\bf Tree}} (k)\big]_{\rm USR} \sim {\cal O}(10^{-2})$ will provide us with a range of values for $c_{4},c_{5}$ during the instant of the sharp transition and in the remaining interval of the USR. From the fig.(\ref{usr1}), we choose to keep within $c_{2},c_{3} \sim {\cal O}(10-10^{2})$ and impose the constraints from $\tilde{c_{s}}=1\pm \delta$ and the power spectrum in the USR. As a result, we obtain the allowed interval of values as ${\cal O}(10^{-1}) \leq \{c_{4},c_{5}\} \leq {\cal O}(10^{2})$ where the magnitudes increase requiring also large values of $c_{2},c_{3}$ as we transition into the USR. While this range just tells about the magnitude, both the coefficients together carry positive and negative signatures.

For the case of the remaining duration of the USR, we have the following analysis of $c_{4},c_{5}$. From the plots in fig.(\ref{usr11}) it can be seen that for both $c_{2},c_{3} \lesssim {\cal O}(10^{3})$ remains a good range to produce the values of $f$ as shown. These values when used for the effective sound speed and the scalar power spectrum later help us to find the interval for $c_{4},c_{5}$. Since now we have entered the USR and $c_{s}(\tau)=c_{s,*}$ remains throughout with its value constrained in the interval $0.024 \leq c_{s} < 1$, we find that for both $\{ c_{4},c_{5}\} \sim {\cal O}(10^{2}-10^{3})$. Their values increase quickly relative to the change in $c_{2},c_{3}$ and achieve larger magnitudes for higher $c_{s}$ where both can have positive and negative signatures. As we progress into USR, the $\epsilon_{\rm USR}$ decreases extremely fast for $\eta \sim -6$ and this reflects significantly in the allowed values for $c_{4},c_{5}$ as they change within ${\cal O}(10^{5}) \leq c_{4},c_{5} \leq {\cal O}(10^{16})$ with $c_{5}$ always a few order of magnitudes higher than $c_{4}$. Although such large values will equally remain highly suppressed by powers of the cut-off scale $\Lambda$, as seen from eqn.(\ref{CovGal}), such behaviour can still signal the increase of higher order non-linear interactions in the Galileon sector.

\begin{table}[H]

\centering
\begin{tabular}{|c|c|c|c|c|c|c|}

\hline\hline
\multicolumn{7}{|c|}{\normalsize \textbf{Galileon EFT coefficients for a given equation of state $w$ in USR }} \\

\hline

EoS $(w)$ & $f$ &\hspace {0.5cm}  $c_{1}$ \hspace {0.5cm} &\hspace {0.5cm}  $c_{2}$ \hspace {0.5cm} &\hspace {0.5cm}  $c_{3}$ \hspace {0.5cm} &\hspace {0.5cm}  $c_{4}$ \hspace {0.5cm} &\hspace {0.5cm}  $c_{5}$ \hspace {0.5cm}  \\
\hline
$1/3$ & & $0.01$ & $100$ & $500$ & $-10^3$ & $-10^5$ \\ 
$0.25$ & $\leq{\cal O}(10^{-4})$  & $0.03$ & $100$ & $500$ & $10^3$ & $-2.9\times 10^5$ \\
$0.16$ & & $0.03$ & $250$ & $700$ & $-10^3$ & $-4.4\times 10^5$ \\ \hline 
\hline

\end{tabular}

\caption{ Table describes possible set of values for the EFT coefficients $c_{i}\;\forall\;i=1,\cdots,5$ satisfying a given EoS $w$. }

\label{tab2eos}

\end{table}

\begin{figure*}[htb!]
    	\centering
    \subfigure[]{
      	\includegraphics[width=8.5cm,height=7.5cm]{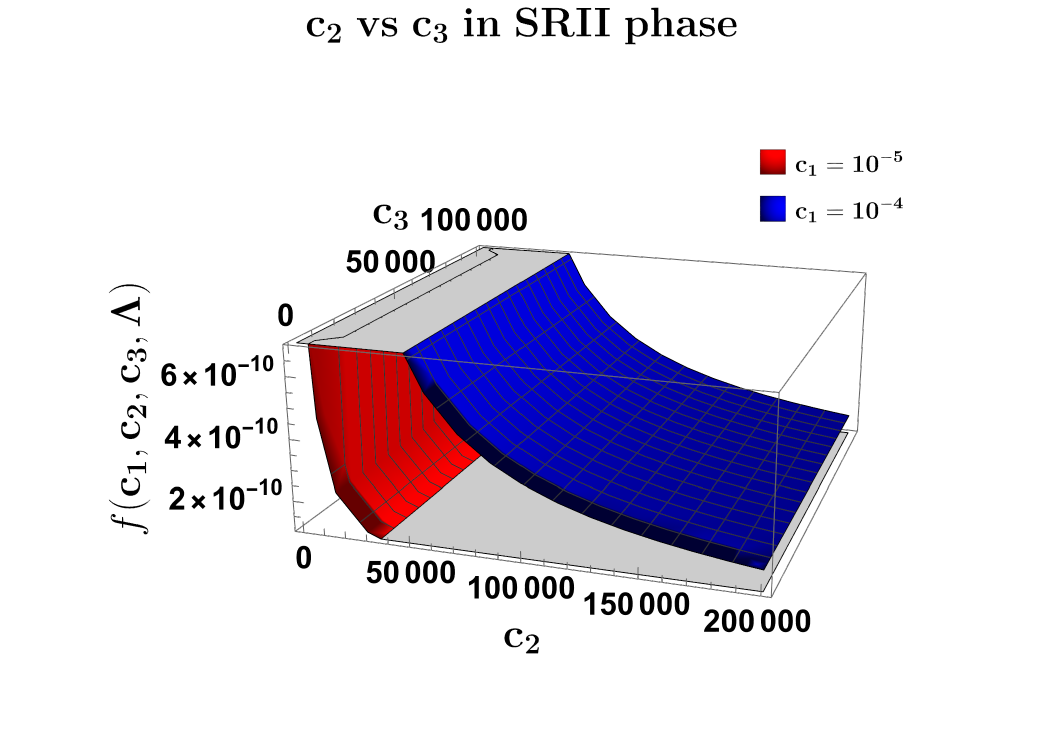}
        \label{sr2c1}
    }
    \subfigure[]{
       \includegraphics[width=8.5cm,height=7.5cm]{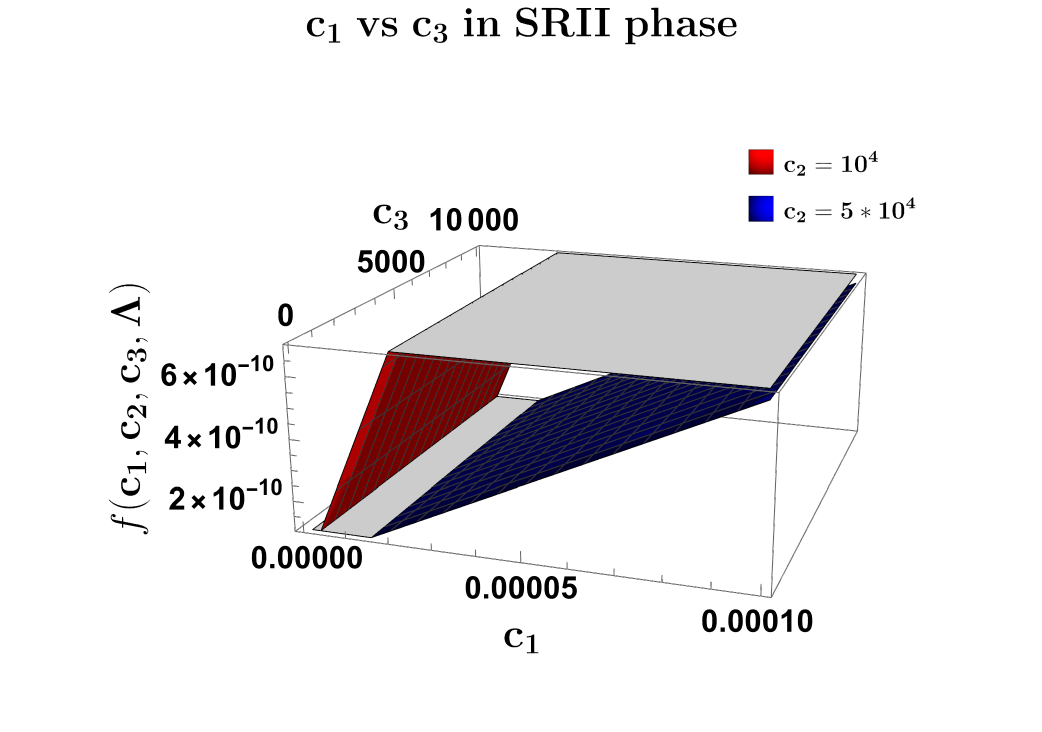}
        \label{sr2c2}
       }\\
   \subfigure[]{
       \includegraphics[width=9cm,height=8cm]{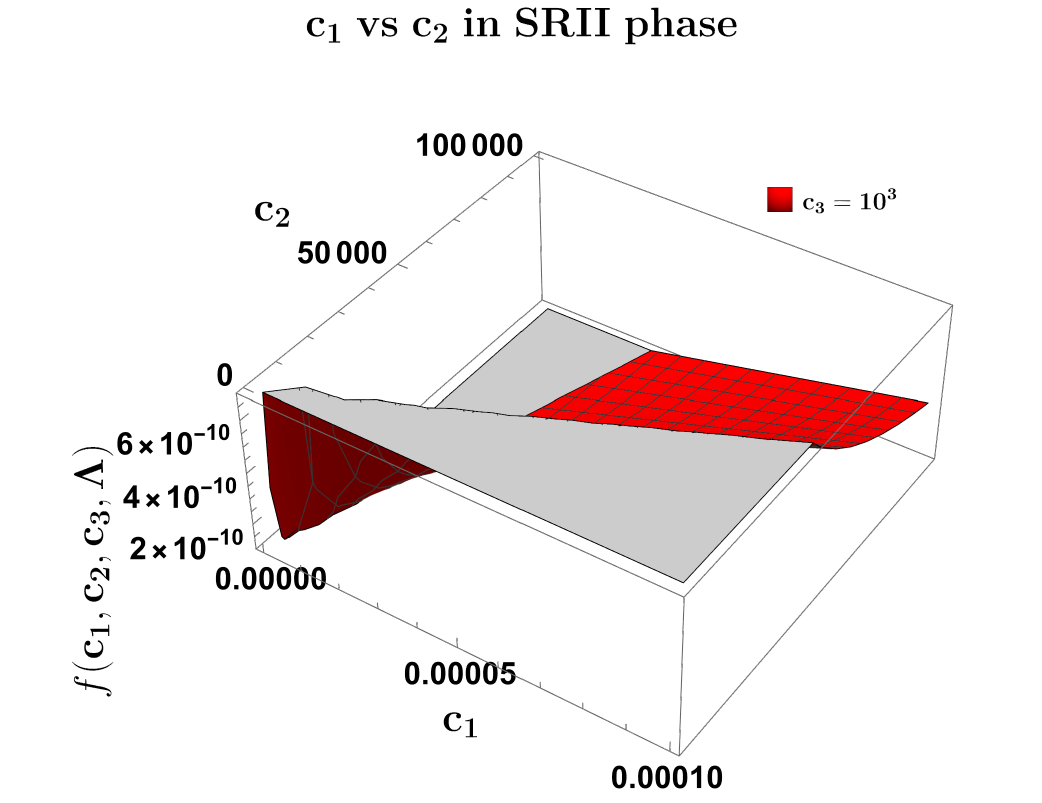}
        \label{sr2c3}
       }\hfill
    
    	\caption[Optional caption for list of figures]{Behaviour of the three Galileon EFT coefficients $c_{1},c_{2},c_{3}$ to satisfy $f\equiv f(c_{1},c_{2},c_{3},\Lambda)$ near the sharp transition into the SRII phase. In (a), $c_{1}=10^{-5}$ (red) and $c_{1}=10^{-4}$ (blue) are fixed to show the allowed values of $c_{2}$ and $c_{3}$. In (b), $c_{2}=10^{4}$ (red) and $c_{2}=5\times 10^{4}$ (blue) are kept fixed. In (c), $c_{3}=10^{3}$ (red) is kept fixed.   } 
    	\label{sr2}
    \end{figure*}

\textcolor{black}{The table \ref{tab2eos} shows the possible values for the Galileon EFT coefficients in the USR phase that can realise a specific EoS $w$. We can appreciate the fact that the the coefficients still lie within the interval important for realising the USR phase. The choice of coefficients is much sensitive to lower values of $c_{1}$ as compared to similar changes in either $c_{2}$ or $c_{3}$ and as $f$ decreases inside USR, higher values of $c_{4},c_{5}$ can be also be achieved.  }

\subsection{Second Slow Roll (SRII) phase}

We now analyze the parameter space of the coefficients $c_{i},\;\forall\;i=1,\cdots,5$ in a similar fashion as done previously by focusing on the two regimes, near the sharp transition and during the remaining e-folds of SRII to understand how these Galileon EFT coefficients can change. The SRII phase operates until inflation comes to an end, at $\tau=\tau_{\rm end}$ after the scalar field exits the USR phase at $\tau=\tau_{s}$. The exit from the USR is met by another sharp transition into the SRII but now the features of the potential or the mentioned coefficients change such that the slow-roll parameters start to increase in value and reach the required value of unity to successfully end inflation. 

The first slow-roll parameter $\epsilon_{\rm SRII}$ in the SRII remains directly proportional to the $\epsilon_{\rm SRI}$ and behaves as a non-constant number till the end of SRII. The second slow-roll parameter $\eta_{\rm SRII}$ also begins to climb from its previous value in the USR and reaches the value $\eta_{\rm SRII} \sim -1$ marking the end of inflation. The sudden change in the magnitude of the $\eta$ parameter just after the transition is also essential for analyzing the significant one-loop corrections to the tree-level scalar power spectrum. The discussions regarding the one-loop computations can be found in the future when talking about the scalar power spectrum. 

For this phase, we again analyze the form of $\epsilon_{\rm SRII}({\cal N})$ as follows:
\bea \label{epsN3}
\epsilon_{\rm SRII}({\cal N}) = \eta_{\rm SRII}\bigg(1-\bigg(1-\frac{\eta_{\rm SRII}}{\epsilon_{i,\rm SRII}}\bigg)e^{2\eta_{\rm SRII}({\cal N}-{\cal N}_{e})}\bigg)^{-1}, \eea
with the initial condition now chosen as $\epsilon_{i,\rm SRII} = \epsilon_{\rm USR}({\cal N}_{e})$ which includes the instant in e-folds ${\cal N}_{e}=20.4$ that signals the end of USR and also the moment of sharp transition into the SRII. The value for ${\cal N}_{e}$ was discussed before in USR and came out due to maintaining perturbativity and the numerical analysis. The above equation further gets used in the eqn.(\ref{srparams}) which then leads to the following form of the Hubble parameter:
\bea \label{HubbleN3}
\frac{H_{\rm SRII}({\cal N})}{H_{\rm USR}({\cal N}_{e})} = \bigg(\frac{\eta_{\rm SRII}\exp{(2\eta_{\rm SRII}\Delta{\cal N}_{\rm SRII})}}{\epsilon_{i,\rm SRII}-\exp{(2\eta_{\rm SRII}\Delta{\cal N}_{\rm SRII})}(\epsilon_{i,\rm SRII}-\eta_{\rm SRII})}\bigg)^{-\frac{1}{2}},
\eea
the term $\eta_{\rm SRII}$ in the inverse square root numerator comes after imposing the initial condition at ${\cal N}={\cal N}_{e}$. Here $\Delta{\cal N}_{\rm SRII}={\cal N}-{\cal N}_{e}$ and this gets used to finally give us the following expression for $f_{\rm SRII}({\cal N})$ similar to the one in eqn.(\ref{ftnN1}) as:
\bea \label{ftnN3}
f_{\rm SRII}({\cal N}) = \frac{H_{\rm SRII}({\cal N})}{H_{\rm USR}({\cal N}_{e})}\exp{(-\eta_{\rm SRII}({\cal N}-{\cal N}_{e}))}. \eea

Based on the above expression, we can now analyze the behaviour of the eqn.(\ref{ftnN3}) for the SRII phase to constrain $c_{1},c_{2},c_{3}$. From fig.(\ref{sr2}) one can see that, for $c_{1}$, ranges decrease to much lower magnitudes where $c_{1}\sim {\cal O}(10^{-5})$ such that the desired $f$ is achieved. The steepness of the surface also tells about the sensitivity of $f$ to $c_{1}$. We observe that a decrease of an order of magnitude in $c_{1}$ can greatly affect the values for $c_{2}$ keeping $c_{3}$ barely affected. In comparison to the previous values, from plots in fig.(\ref{usr11}) before the transition, $c_{1}$ decreases greatly while conversely $c_{2}$ increases with $c_{3}$ not showing much difference in magnitude. The subplot having $c_{3}=10^{3}$ fixed confirms the behaviour for the other two coefficients mentioned before, and other values of $c_{3}$, within two orders of magnitude, are not used as their result overlaps with the one present and hence not illuminating new information than already presented in this discussion. 

\begin{figure*}[htb!]
    	\centering
    \subfigure[]{
      	\includegraphics[width=8.5cm,height=7.5cm]{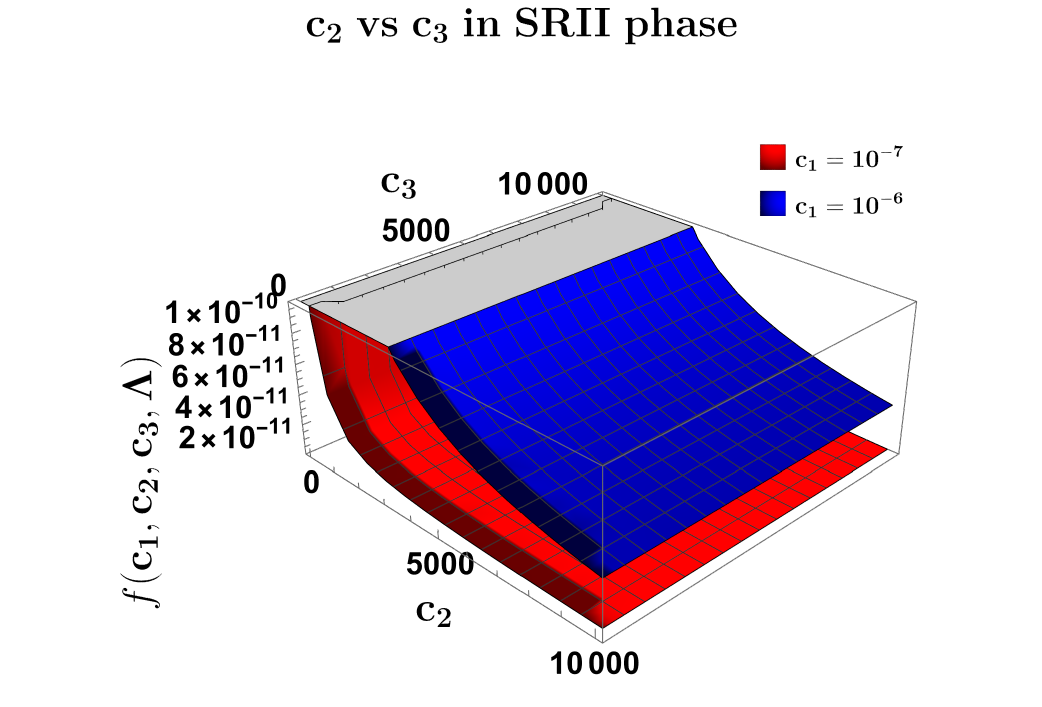}
        \label{sr2c11}
    }
    \subfigure[]{
       \includegraphics[width=8.5cm,height=7.5cm]{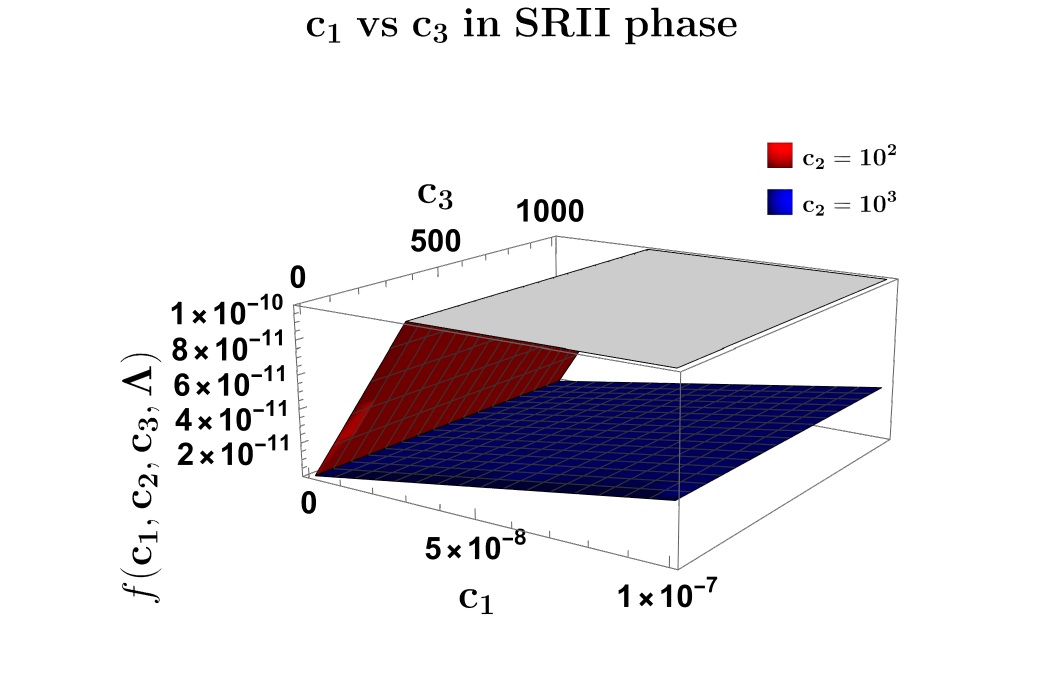}
        \label{sr2c21}
       }\\
   \subfigure[]{
       \includegraphics[width=9cm,height=8cm]{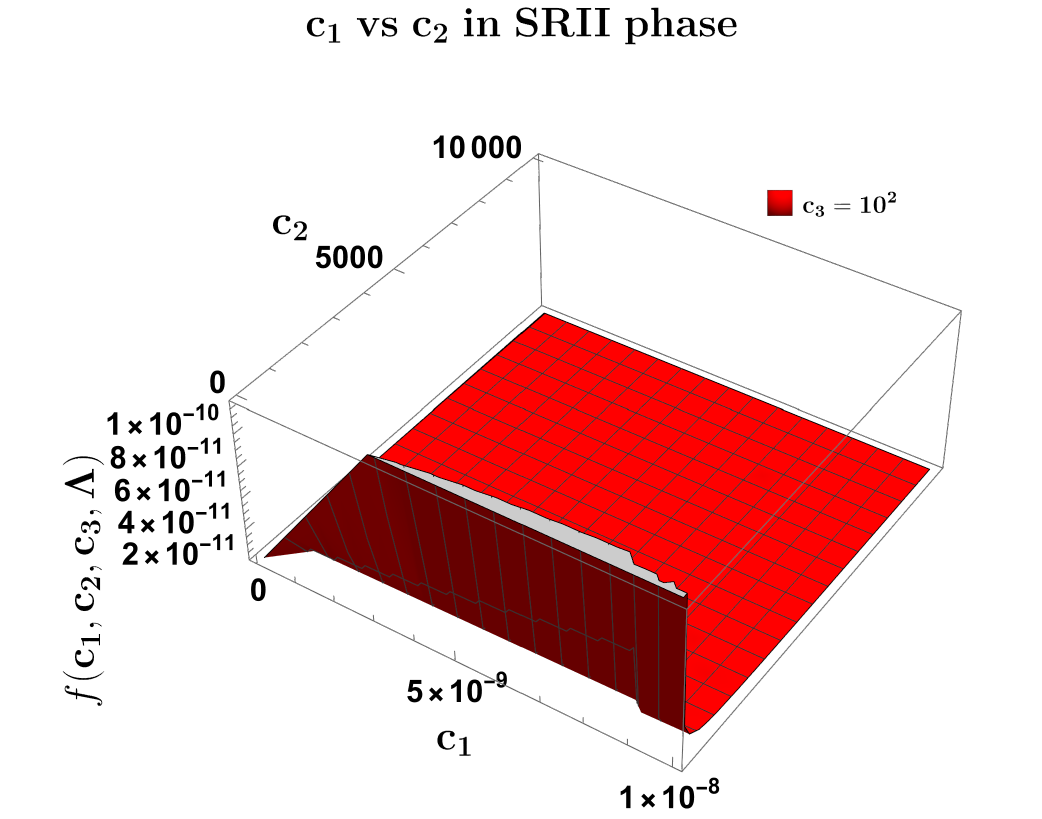}
        \label{sr2c31}
       }\hfill
    
    	\caption[Optional caption for list of figures]{Behaviour of the three Galileon EFT coefficients $c_{1},c_{2},c_{3}$ to satisfy $f\equiv f(c_{1},c_{2},c_{3},\Lambda)$ after the sharp transition into the SRII phase. In (a), $c_{1} = 10^{-6}$ (blue) and $c_{1} = 10^{-7}$ (red) are fixed to achieve the values of $f$. In (b), $c_{2}=10^{2}$ (red) and $c_{2}=10^{3}$ (blue) is kept fixed. In (c), $c_{3}=10^{2}$ is kept fixed. The magnitudes of the parameters provide the respective interval to satisfy the behaviour of $f$ during SRII till the end of inflation. } 
    	\label{sr21}
    \end{figure*}

More constraint on $c_{3}$ can be obtained from analyzing the function $f$ during the remaining of the SRII and the effective sound speed $c_{s}$. We first analyze the coefficients for $f$ in the remaining SRII phase. The fig.(\ref{sr21}) helps to understand the coefficients $c_{1},c_{2},c_{3}$ during SRII. The coefficient $c_{1}$ suffers a further decrease in magnitude to allow for the increasingly small values needed for $f$ during SRII, signaling an even milder shift symmetry breaking relative to its case during the USR. The range where $c_{1} \lesssim {\cal O}(10^{-7})$ lies can give rise to the lower values of $f$ if $c_{2}$ is also kept within $c_{2} \gtrsim {\cal O}(10^{2})$ and increased values of $c_{2} \sim {\cal O}(10^{3})$  corresponds to $c_{1} \gtrsim {\cal O}(10^{-6})$. The magnitude of $c_{3}$ in particular does not receive much changes even throughout the SRII with the values $c_{3} \sim {\cal O}(10^{3})$ still remaining acceptable. Further constraints follow from the analysis of the effective sound speed.

The effective sound speed $c_{s}$ takes on the value $c_{s}(\tau_{e}) = \tilde{c_{s}} = 1\pm \delta$, where $\delta \ll 1$, as per our parameterization during the sharp transition at the e-fold instant ${\cal N}_{e}$ which marks the exit from USR and entry into the SRII. During the remaining SRII phase, we have $c_{s}(\tau)=c_{s,*}$ for $\tau_{e} \leq \tau \leq \tau_{\rm end}$ which again respects the same causality and unitarity constraints mentioned before for $c_{s}$ and can in turn set bounds on $c_{4},c_{5}$. Last but not least, we also use the scalar power spectrum amplitude as another constraint on $c_{4},c_{5}$ where the amplitude satisfies $\big[\Delta^{2}_{\zeta,{\bf Tree}} (k)\big]_{\rm SRII} \sim {\cal O}(10^{-5})$.

When near the sharp transition at the end of USR, we previously saw huge magnitudes for $c_{4},c_{5}$. Since after crossing SRII, function $f$ again decreases quickly within a few e-folds, we observe a further rise in magnitude for $c_{4},c_{5}$. Such behaviour only gets enhanced as the value of $f$ keeps decreasing inside SRII. Initially, both of these start with their values in the interval ${\cal O}(10^{13}) \leq \{ c_{4},c_{5} \} \leq {\cal O}(10^{19})$ which is slightly increased relative to their values during the end of USR phase. As the SRII progresses, one observes a steep rise in values such that one must have at most $c_{4} \sim {\cal O}(10^{30})$ while $c_{5} \sim {\cal O}(10^{42})$. Such magnitudes of coefficients are extremely large and are also used to satisfy the constraint on the amplitude of the power spectrum in the SRII phase. In relation with the eqn.(\ref{CovGal}), the interval for $c_{4},c_{5}$ presented here shows that the strength of the higher derivative, non-linear Galileon self-interactions increases drastically, though the overall coefficient is suppressed by increasingly large powers of the cut-off $\Lambda$, and compared to the previous case of USR the large change in $\eta_{\rm USR}$ and corresponding change in $\epsilon_{\rm USR}$ leads to the requirement of higher values of Galileon EFT coefficients. 

\textcolor{black}{The table \ref{tab3eos} shows the possible values for the Galileon EFT coefficients in the SRII phase that can realise a specific EoS $w$. Here both $c_{1},c_{2}$ have similar impact on the other coefficients as they are either increased or decreased while changes in $c_{3}$ are least sensitive for any case. Also, similar to the previous phases, further decreasing $f$ allows for greater values of both $c_{4},c_{5}$ as seen from the entries in the table.  }

\begin{table}[H]

\centering
\begin{tabular}{|c|c|c|c|c|c|c|}

\hline\hline
\multicolumn{7}{|c|}{\normalsize \textbf{Galileon EFT coefficients for a given equation of state $w$ in SRII }} \\

\hline

EoS $(w)$ & $f$ &\hspace {0.5cm}  $c_{1}$ \hspace {0.5cm} &\hspace {0.5cm}  $c_{2}$ \hspace {0.5cm} &\hspace {0.5cm}  $c_{3}$ \hspace {0.5cm} &\hspace {0.5cm}  $c_{4}$ \hspace {0.5cm} &\hspace {0.5cm}  $c_{5}$ \hspace {0.5cm}  \\
\hline
$1/3$ &  & $10^{-7}$ & $10^{3}$ & $2\times 10^{3}$ & $-10^{16}$ & $-2.5\times 10^{16}$ \\ 
$0.25$ & $\leq {\cal O}(10^{-10})$  & $10^{-7}$ & $10^{3}$ & $5\times 10^{3}$ & ${\cal O}(10^{29})$ & ${\cal O}(10^{30})$ \\
$0.16$ & & $10^{-7}$ & $10^{3}$ & $7\times 10^{3}$ & ${\cal O}(10^{31})$ & ${\cal O}(10^{33})$ \\ \hline 
\hline

\end{tabular}

\caption{ Table describes possible set of values for the EFT coefficients $c_{i}\;\forall\;i=1,\cdots,5$ satisfying a given EoS $w$. }

\label{tab3eos}

\end{table}

After gaining an improved picture of the behaviour of the EFT coefficients that can facilitate the construction of our setup, we finally visualize the slow-roll parameters as functions of the number of e-folds elapsed during each phase and the Hubble parameter behaviour along with that of the effective sound speed as a function of the conformal time which is crucial from the purpose of parameterization of the required setup.

\begin{figure*}[htb!]
    	\centering
    {
      	\includegraphics[width=17cm,height=12cm]{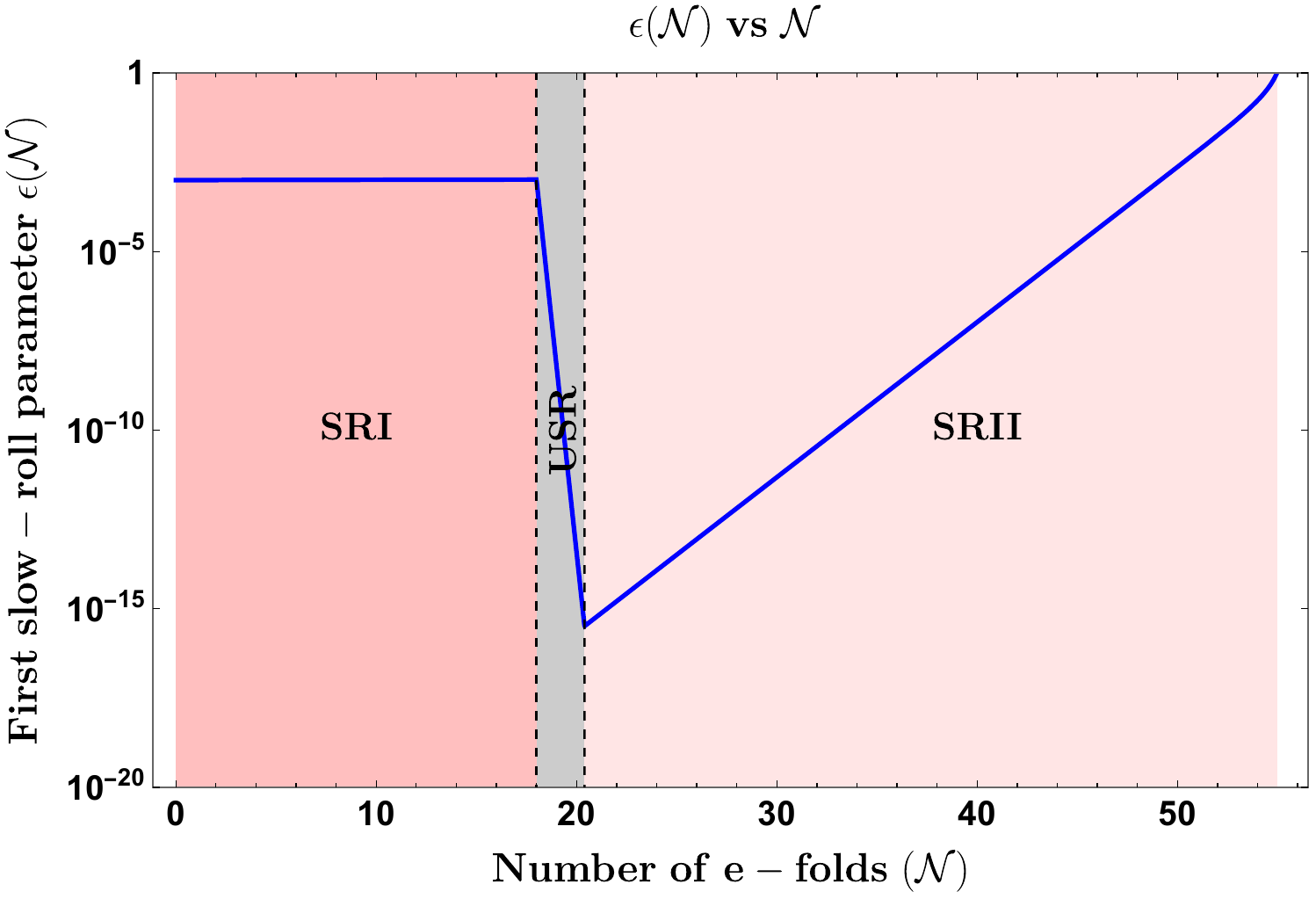}
        \label{epsilon}
    }
    	\caption[Optional caption for list of figures]{First slow-roll parameter $\epsilon$ plotted against the number of e-folds in the presence of sharp transitions. The variation includes its behaviour across the three phases of interest in our setup, SRI, USR, and SRII, with a drastic change when entering into and exiting from the USR phase. } 
    	\label{s4d1}
    \end{figure*}

\subsection{Numerical Outcomes: Behaviour of slow-roll parameters in the three consecutive phases}

In this section, we present numerical outcomes for the slow-roll parameters $\epsilon({\cal N})$ and $\eta({\cal N})$ across the three phases during inflation where we distinguish their behaviour for each phase as a function of the e-folds parameter ${\cal N}$. The corresponding nature of the Hubble parameter $H({\cal N})$ results from the use of eqs.(\ref{HubbleN1},\ref{HubbleN2},\ref{HubbleN3}) for the SRI, USR, and SRII phases and there we also incorporate our specific parameterization for the effective sound speed $c_{s}$ with the conformal time which we also mention in this section.

\begin{figure*}[htb!]
    	\centering
    {
       \includegraphics[width=17cm,height=12cm]{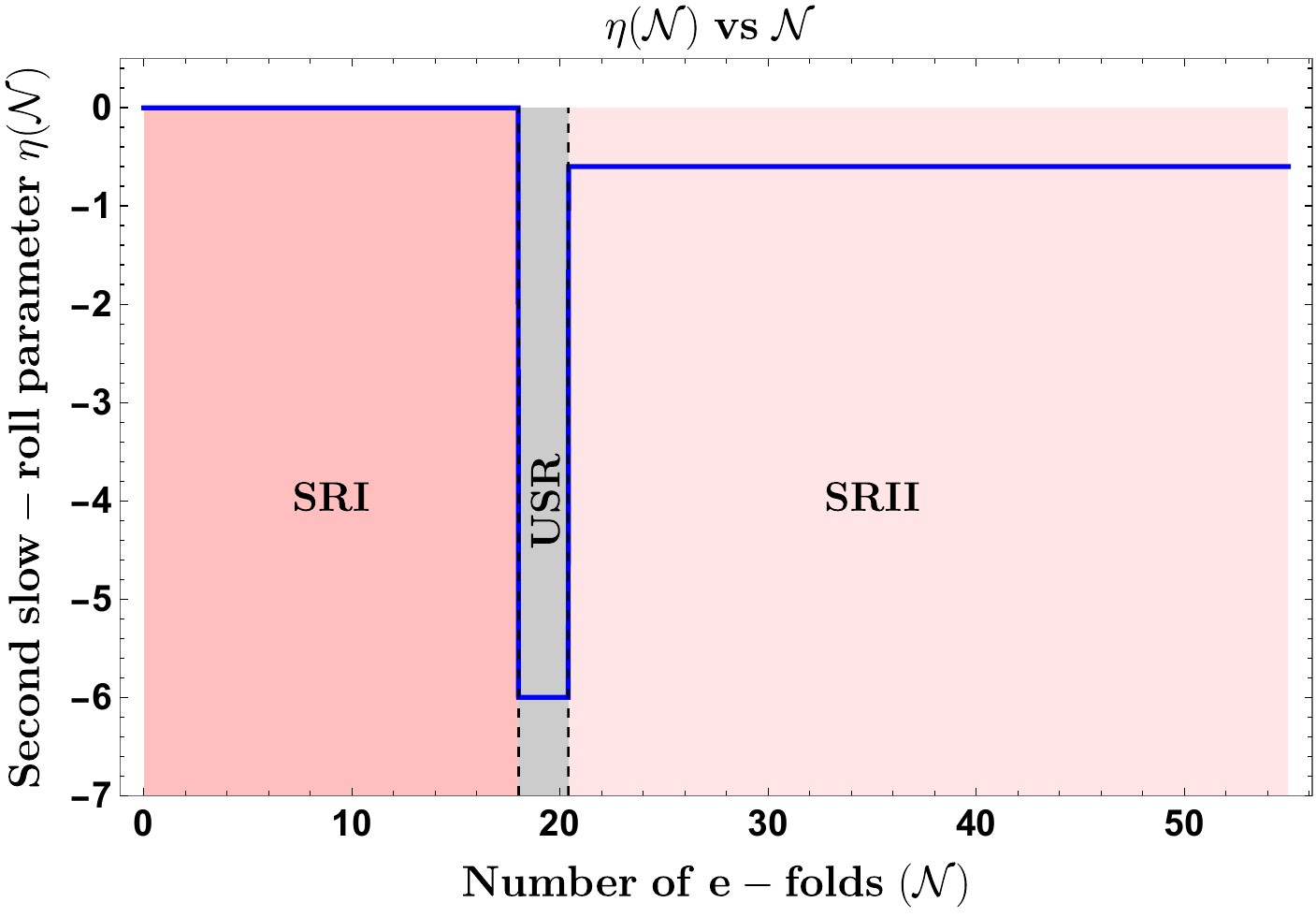}
        \label{eta}
       }
       \caption[Optional caption for list of figures]{Second slow-roll parameter $\eta$ plotted against the number of e-folds in the presence of sharp transitions. The variation includes its behaviour across the three phases of interest in our setup, SRI, USR, and SRII, with a sudden jump in its value for the USR phase. }
       \label{s4d2}
    \end{figure*}

The fig.(\ref{s4d1}) shows the evolution of the first slow-roll parameter during inflation in our SRI/USR/SRII setup.  We start with ${\cal N}_{*}=0$ as our reference during which the modes with $k_{*}=0.02 {\rm Mpc^{-1}}$ exits the Horizon and where we set the initial conditions with $\epsilon_{\rm SRI}({\cal N}_{*}) \sim {\cal O}(10^{-3})$. As we progress into the SRI, $\epsilon_{\rm SRI}$ is almost a constant and extremely slowly varying in the order of its initial value. Then, as we encounter the first sharp transition into the USR, $\epsilon_{\rm SRI}$ observes a sharp change in its magnitude and quickly decreases from $\epsilon_{\rm USR} \sim {\cal O}(10^{-3})$ to $\epsilon_{\rm USR} \sim {\cal O}(10^{-15})$ within a span of few e-folds, $\Delta{\cal N}_{\rm USR} \sim {\cal O}(2)$. This drastic change is a feature of the USR where $\epsilon_{\rm USR} \propto a^{-6}$ and the sudden change in the $\eta_{\rm USR}$ parameter which is visible from fig.(\ref{s4d2}). After another sharp transition during exit from the USR, the $\epsilon_{\rm SRII}$ value rises and reaches to $\epsilon_{\rm SRII} \sim {\cal O}(1)$ right at the end of inflation. The cumulative nature is a result of using the eqs.(\ref{epsN1},\ref{epsN2},\ref{epsN3}). 

In the next fig.(\ref{s4d2}), we depict the behaviour of the $\eta$ parameter across the three phases of interest. Initially in SRI, where the slow roll approximations are completely valid, $\eta_{\rm SRI}$ has a negative signature and starts off with a very small value which we have chosen here as $\eta_{\rm SRI}({\cal N}_{*}) \sim -0.001$. The behaviour of $\eta_{\rm SRI}$ continues as such soon as we encounter the sharp transition into the USR, where it jumps at the same instant to acquire $\eta_{\rm USR} \sim -6$. This almost sudden change in the value between $\eta_{\rm SRI}$ to $\eta_{\rm USR}$ signals the sharp transition nature for this parameter, which gets implemented via the use of the Heaviside Theta function. After exiting from the USR, $\eta_{\rm USR}$ drops back to a negative value of magnitude $|\eta_{\rm SRII}| \sim{\cal O}(1)$ and continues as such till inflation ends.    

\begin{figure*}[htb!]
    	\centering
    {
      	\includegraphics[width=17cm,height=12cm]{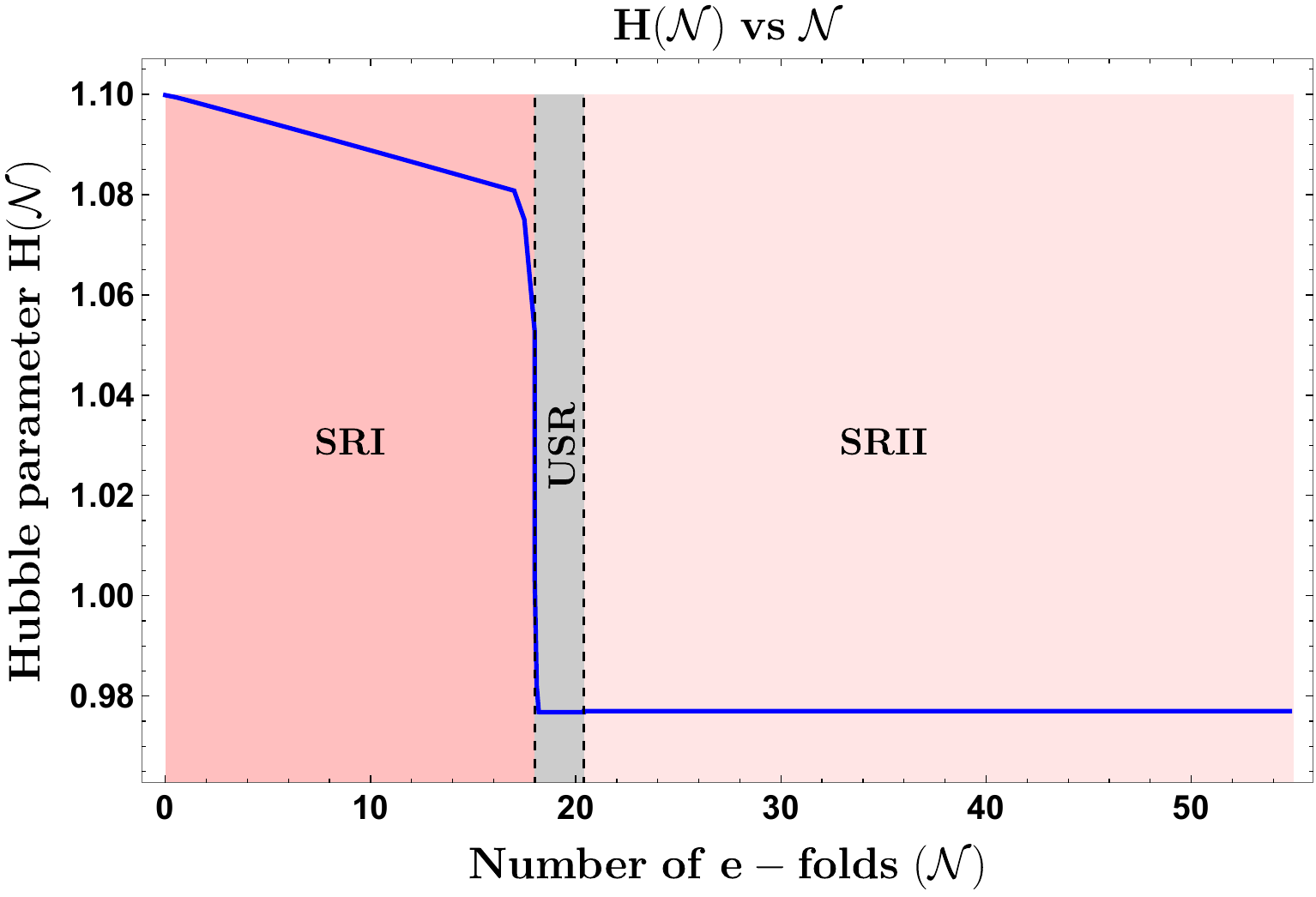}
        \label{hubble}
    }
    \caption[Optional caption for list of figures]{Variation of the Hubble rate $H$ with the number of e-folds. }
    \label{s4d3}
    \end{figure*}

From fig.(\ref{s4d3}), we can visualize the variation of the Hubble parameter $H({\cal N})$ with the number of e-foldings across the three phases of interest. We find that after imposing the initial conditions on the slow-roll parameters and the condition of starting with ${\cal N}_{*}=0$, the Hubble rate starts with a value of unity which stays the same until it falls down as we enter into the USR. Notice that the order of change in magnitude occurs at the second decimal place which is very small and this is the consequence of the initial conditions and the extremely large value of the $\eta_{\rm USR}$ parameter in the USR. Further into the USR and even after its exit and entry into the SRII phase, the Hubble rate stays constant throughout till inflation comes to its end.

\begin{figure*}[htb!]
    	\centering
    {
       \includegraphics[width=17cm,height=12cm]{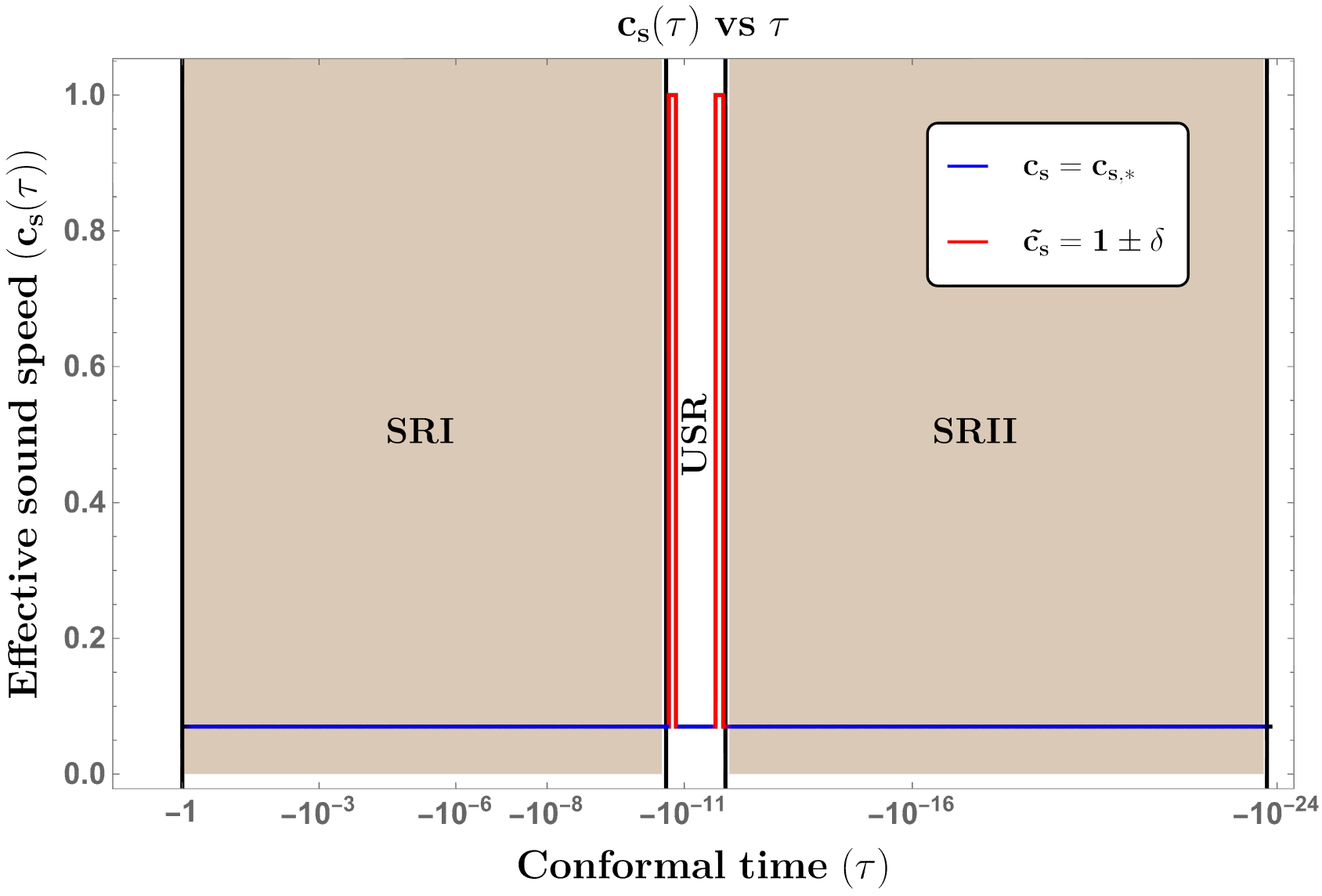}
        \label{sound}
    } 
    \caption[Optional caption for list of figures]{Schematic of the parameterization used for the effective sound speed $c_{s}$ with the conformal time $\tau$ where the red color highlights the sudden jumps in its value corresponding to the sharp transitions. }
\label{s4d4}
    \end{figure*}

In the last fig.(\ref{s4d4}), we provide a schematic representation of the chosen parameterization through which we implement the sharp transition feature in our setup. The sudden change in $c_{s}$ to its value $\tilde{c_{s}}=1\pm \delta$ at both $\tau_{s}$ and $\tau_{e}$, highlighted in red, facilitates the construction of USR phase conditions in the present Galileon inflation framework. The existing observational constraints on the value for $c_{s}$ is already previously mentioned with $0.024 \leq c_{s} < 1$ and we have chosen this range to constrain the parameter space of the EFT coefficients in the Lagrangian eqn.(\ref{CovGal}). The value $c_{s,*}$ remains constant in between the duration of the phases of interest and for the purpose of our future calculations we have chosen $c_{s,*}=0.05$. 

We highlight the features and implications of including a sharp transition in our setup with the USR. The sharp transition is implemented throughout with the help of Heaviside Theta functions, $\Theta(\tau-\tau_{s})$ and $\Theta(\tau-\tau_{e})$, right at the moment of transition $\tau_{s}$ and $\tau_{e}$ (or ${\cal N}_{s}$ and ${\cal N}_{e}$). This function is used to depict the nature of the $\eta$ parameter which later gets used to generate the features visible for the $\epsilon$ parameter. Now, we choose a specific parameterization for the $\eta$ which looks like:  \bea
\eta(\tau) &=&  \eta_{\rm SRI}\Theta(\tau-(\tau_{*}-\tau_{s})) +  \eta_{\rm USR}\Theta(\tau-(\tau_{s}-\tau_{e})) + - \Delta\eta(\tau)\{\Theta(\tau-\tau_{s}) - \Theta(\tau-\tau_{e})\} + \eta_{\rm SRII}\Theta(\tau-(\tau_{e}-\tau_{\rm end})),   \nonumber\\ 
&=& \eta_{\rm SRI}(\tau_{*} < \tau < \tau_{s}) + \eta_{\rm USR}(\tau_{s} < \tau < \tau_{e}) - \Delta\eta(\tau)\{\Theta(\tau-\tau_{s}) - \Theta(\tau-\tau_{e})\} + \eta_{\rm SRII}(\tau_{e} < \tau < \tau_{\rm end}).
\eea
If we further take the conformal time derivative of this parameterization of $\eta$ at the transition moments, we obtain the following:
\bea
\bigg(\frac{\eta(\tau)}{c^{2}_{s}}\bigg)' &=& \underbrace{\bigg(\frac{\eta_{\rm SRI}(\tau_{*} < \tau < \tau_{s})}{c^{2}_{s,*}}\bigg)'}_{=0} + \underbrace{\bigg(\frac{\eta_{\rm USR}(\tau_{s} < \tau < \tau_{e})}{c^{2}_{s,*}}\bigg)'}_{=0} \nonumber\\
&& \quad\quad\quad\quad\quad\quad\quad\quad\quad\quad - \frac{\Delta\eta(\tau)}{\tilde{c_{s}}^{2}}\{\delta(\tau-\tau_{s}) - \delta(\tau-\tau_{e})\} + \underbrace{\bigg(\frac{\eta_{\rm SRII}(\tau_{e} < \tau < \tau_{\rm end})}{c^{2}_{s,*}}\bigg)'}_{=0},
\eea
where the prime notation denotes a conformal time derivative. 
It will come to notice during discussions on the non-renormalization theorem that the dominant interaction term, when calculating one-loop corrections, is of the form $\zeta'\zeta^{2}$ and has the coefficient $(\eta/c_{s}^{2})'$. From the above equation we see that this introduces Dirac delta like enhancements at the transitions. Fortunately, the procedure of softly breaking the Galilean shift symmetry prevents such terms in the final calculations of the one-loop contributions. Still, the above construction makes it clear the way $\eta$ parameter appears in our analysis, with sharp transitions at the two transition moments $\tau=\tau_{s}\;( {\cal N}={\cal N}_{s})$ and $\tau=\tau_{e}\;( {\cal N}={\cal N}_{e})$ visible in the fig.(\ref{s4d2}). There have been attempts to integrate the SR/USR and USR/SR scenarios using a smooth transition \cite{Riotto:2023hoz,Riotto:2023gpm,Firouzjahi:2023ahg,Firouzjahi:2023aum}, also the impact of a bump/dip-like feature in the inflationary potential \cite{Mishra:2019pzq}, to study the impact of quantum corrections on the scalar power spectrum and the production of PBHs.

Another essential quantity considered to constrain the coefficients of the higher-derivative interaction terms was the scalar power spectrum amplitude for each of the three phases. In the next section, we provide the discussions for constructing the total power spectrum for the scalar modes after adding the one-loop corrections for each phase mentioned in our setup.

\section{Computation of scalar power spectrum from Galileon EFT} \label{s5}

The power spectrum associated with the scalar modes requires using the eqn.(\ref{CovGal}) to perform perturbation theory up to second order in the comoving curvature perturbation. This procedure provides the evolution equation for the scalar modes in the Fourier space, whose solutions later help us to build the scalar power spectrum for all three phases, namely SRI, USR, and SRII. Correctly determining the solutions across the three phases involves using boundary conditions at the junctions of the sharp transition, further referred to as the Israel junction conditions. We look into this construction in this section and provide the expressions for the total scalar power spectrum using the individual contributions coming from each phase.  

\subsection{Second order perturbed action}

The second order action for the comoving curvature modes from eqn.(\ref{CovGal}) in our quasi de Sitter background, where we neglect any effects coming from mixing with the gravity sector, turns out to be:
\bea
S^{(2)}_{\zeta} = \int d\tau\;\frac{d^{3}\mbf{k}}{(2\pi)^{3}}a(\tau)^2\frac{\cal A}{H^{2}}\left(|\zeta_{\mbf{k}}^{'}(\tau)|^{2} - c_{s}^{2}k^{2}|\zeta_{\mbf{k}}(\tau)|^{2}\right) = \int d\tau\;\frac{d^{3}\mbf{k}}{(2\pi)^{3}}a(\tau)^2\frac{\cal B}{c_{s}^{2}H^{2}}\left(|\zeta_{\mbf{k}}^{'}(\tau)|^{2} - c_{s}^{2}k^{2}|\zeta_{\mbf{k}}(\tau)|^{2}\right),
\eea
where the time-dependent coefficients ${\cal A},\;{\cal B},$ are defined previously in eqs.(\ref{coeffA},\ref{coeffB}), the effective sound speed $c_{s}$ back in eqn.(\ref{soundspeed}), and $H$ is the Hubble parameter for our background spacetime which is not exactly a constant. Using the above action one can easily construct the evolution equation for the comoving curvature perturbation modes $\zeta_{\mbf{k}}$ which is commonly referred to as the Mukhanov-Sasaki equation and it has the form:
\bea \label{MS}
\zeta^{''}_{\bf k}(\tau)+2\frac{z^{'}(\tau)}{z(\tau)}\zeta^{'}_{\bf k}(\tau) +c^2_sk^2\zeta_{\mbf{k}}(\tau)=0.
\eea
which uses the variable $z(\tau) = a\sqrt{2{\cal A}}/H^{2}$. The solution of the above equation leads to the curvature perturbation modes in the three phases, and we will solve for these solutions in the next section. Here, we emphasize that the variable $c_{s}$ contains the specific parameterization for implementing each of the three phases. This variable will appear in each mode solution during the phases and its related coefficients, which we will also mention in the subsequent section.  

\subsection{Semi-Analytical behaviour of the perturbed scalar modes}

In this section, we take the eqn.(\ref{MS}) before and find from it the most general mode solutions in the three phases of interest in our setup and later reduce them by choosing suitable initial quantum vacuum state conditions to the form desirable for our calculations.

\begin{figure*}[htb!]
    	\centering
    \subfigure[]{
      	\includegraphics[width=8.5cm,height=7.5cm]{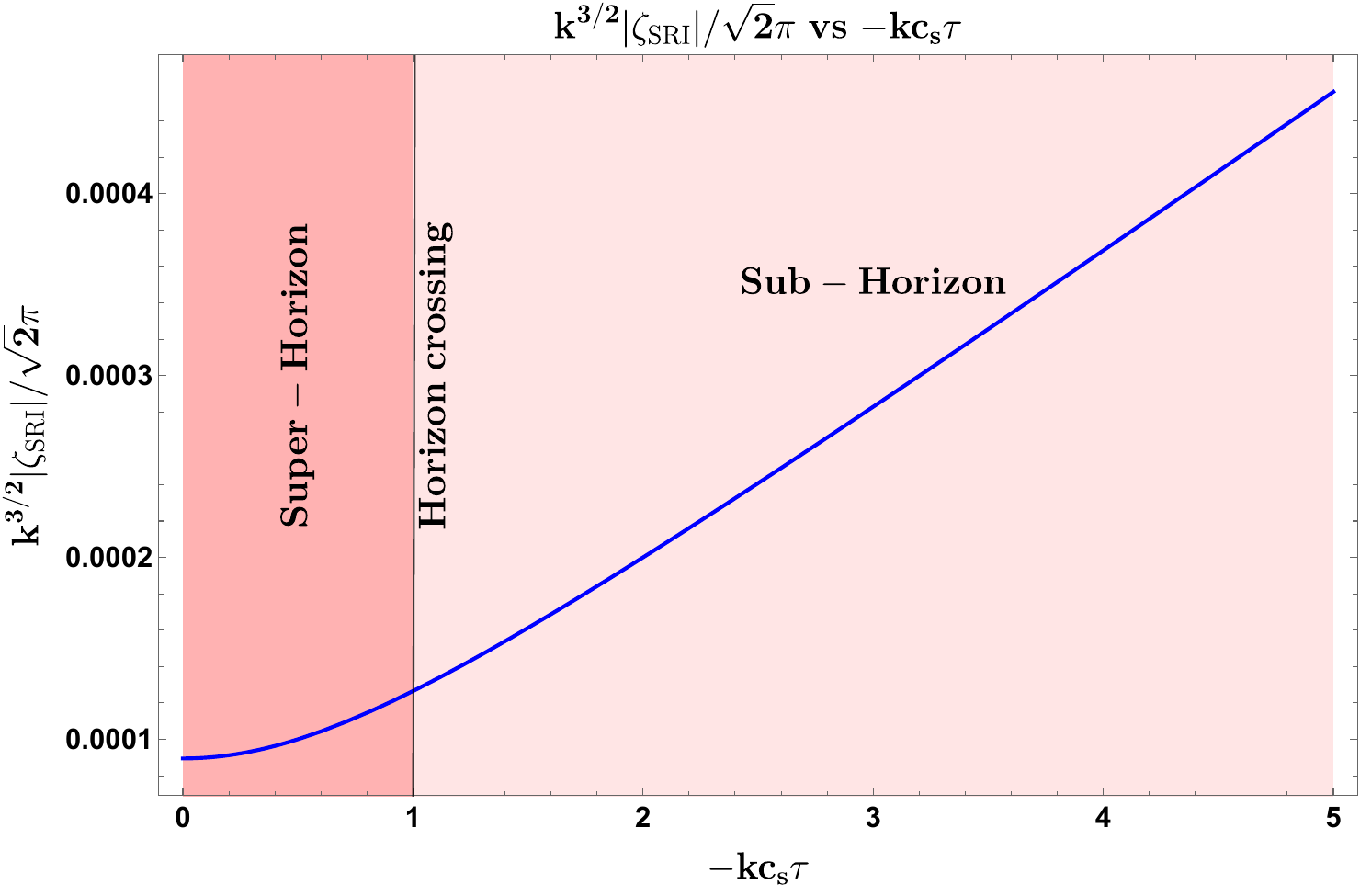}
        \label{zetasr1}
    }
    \subfigure[]{
       \includegraphics[width=8.5cm,height=7.5cm]{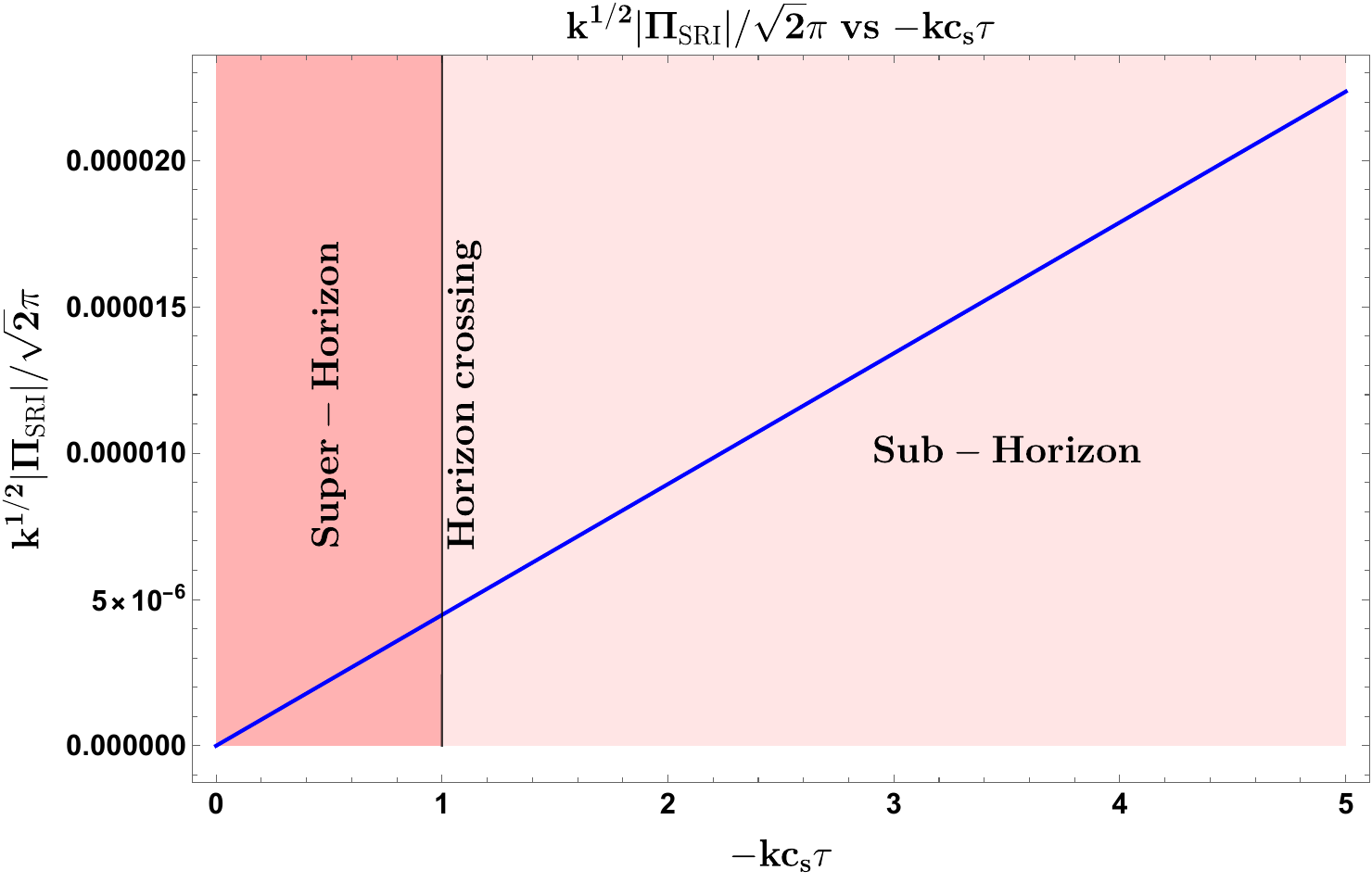}
        \label{pisr1}
    }
    	\caption[Optional caption for list of figures]{$k^{3/2}\abs{\zeta_{\mbf{k}}}/\sqrt{2}\pi$ (\textit{left panel}) and the conjugate momenta $k^{1/2}\abs{\Pi_{\mbf{k}}}/\sqrt{2}\pi$ (\textit{right panel}) plotted against $-kc_{s}\tau$ in the SRI phase. The expressions show monotonic behaviour since their start from the super-Horizon limit, ($-kc_{s}\tau \ll 1$), crossing the horizon ($-kc_{s}\tau = 1$), till their evolution in the sub-Horizon regime ($-kc_{s}\tau \gg 1$). } 
    	\label{modessr1}
    \end{figure*}

\subsubsection{In SRI phase}

The general mode solution for eqn.(\ref{MS}) and the corresponding canonically conjugate momentum during the SRI phase which operates within the conformal time window, $\tau_{*} \leq \tau < \tau_{s}$, or in e-foldings, ${\cal N}_{*} \leq {\cal N} < {\cal N}_{s}$, is given by the expression:
\bea \label{s5a}
    \zeta_{\mbf{k}}(\tau)&=& \left(\frac{iH^{2}}{2\sqrt{\cal A}}\right)\frac{1}{(c_{s}k)^{3/2}}\times\left\{\alpha^{(1)}_{\mbf{k}}\left(1+ikc_{s}\tau\right)\exp{\left(-ikc_{s}\tau\right)} - \beta^{(1)}_{\mbf{k}}\left(1-ikc_{s}\tau\right)\exp{\left(ikc_{s}\tau\right)}\right\},\\
\label{s5b}
\Pi_{\mbf{k}}(\tau) &=&\zeta^{'}_{\mbf{k}}(\tau)= \left(\frac{iH^{2}}{2\sqrt{\cal A}}\right)\frac{1}{(c_{s}k)^{3/2}}\times\frac{k^{2}c_{s}^{2}\tau^{2}}{\tau}\left\{\alpha^{(1)}_{\mbf{k}}\exp{\left(-ikc_{s}\tau\right)} - \beta^{(1)}_{\mbf{k}}\exp{\left(ikc_{s}\tau\right)} \right\},
\eea
where $\alpha^{(1)}_{\mbf{k}}$ and $\beta^{(1)}_{\mbf{k}}$ are the Bogoliubov coefficients which are also used to determine the initial vacuum state conditions for our mode solution. The above solutions consist of the curvature perturbation mode and its conjugate momenta and for these solutions $c_{s}(\tau)=c_{s,*}$ is satisfied throughout the SRI phase. We choose here the most commonly accepted Bunch-Davies initial vacuum condition for such coefficients which are written as:
\bea \label{s5a1}
    \alpha^{(1)}_{\bf k}=1,\\
    \label{s5a2}    \beta^{(1)}_{\bf k}=0. 
    \eea
The above choice of vacuum reduces the presently obtained mode solution to the version suited for our future calculations. The final form of the above solution is represented as:
\bea \label{s51} \zeta_{\bf{k}}(\tau) &=& \left(\frac{iH^{2}}{2\sqrt{\cal A}}\right)\frac{1}{(c_{s}k)^{3/2}}\left(1+ikc_{s}\tau\right)\exp{(-ikc_{s}\tau)}, \\
\label{s52} \Pi_{\mbf{k}}(\tau) &=&\zeta^{'}_{\mbf{k}}(\tau)= \left(\frac{iH^{2}}{2\sqrt{\cal A}}\right)\frac{1}{(c_{s}k)^{3/2}}\times\frac{k^{2}c_{s}^{2}\tau^{2}}{\tau}\exp{\left(-ikc_{s}\tau\right)}.
\eea
Throughout this phase, the parameter $\epsilon_{\rm SRI}$ remains as a slowly varying constant, and $\eta_{\rm SRI}$ is also of a fixed constant value with a negative signature. 

In fig.(\ref{modessr1}), the behaviour of the mode solution and related conjugate momenta are depicted with changing $-kc_{s}\tau$ using eqn.(\ref{s51},\ref{s52}). Starting with the solutions far in the super-Horizon, both evolve slowly but increasingly as they reach the horizon crossing where $-kc_{s}\tau=1$. After crossing, the solutions keep increasing as they get deep into the sub-Horizon regime. The conjugate momenta solution is relatively small from the curvature perturbation solution in the super-Horizon and remains so throughout its evolution in the sub-Horizon.

\subsubsection{In USR phase}

During this USR phase, the general solution for the curvature perturbation modes as well as the corresponding canonically conjugate momentum allowed within the conformal time interval, $\tau_{s} \leq \tau \leq \tau_{e}$, or in e-foldings, ${\cal N}_{s} \leq {\cal N} \leq {\cal N}_{e}$, is given by the expression:
\bea \label{s53}
    \zeta_{\mbf{k}}(\tau)&=&\left(\frac{iH^{2}}{2\sqrt{\cal A}}\right)\left(\frac{\tau_{s}}{\tau}\right)^{3}\frac{1}{(c_{s}k)^{3/2}}\times \left\{\alpha^{(2)}_{\bf k}\left(1+ikc_{s}\tau\right)\exp{\left(-ikc_{s}\tau\right)} - \beta^{(2)}_{\bf k}\left(1-ikc_{s}\tau\right)\exp{\left(ikc_{s}\tau\right)} \right\}, \\
\label{s54}
\Pi_{\mbf{k}}(\tau) &=&\zeta^{'}_{\mbf{k}}(\tau)= \left(\frac{iH^{2}}{2\sqrt{\cal A}}\right)\frac{1}{(c_{s}k)^{3/2}}\frac{\tau_{s}^{3}}{\tau^{4}}\left\{\alpha^{(2)}_{\mbf{k}}(k^{2}c_{s}^{2}\tau^{2}-3(1+ikc_{s}\tau))\exp{\left(-ikc_{s}\tau\right)}\right.\nonumber\\
&&\left.\quad\quad\quad\quad\quad\quad\quad\quad\quad\quad\quad\quad\quad\quad- \beta^{(2)}_{\mbf{k}}(k^{2}c_{s}^{2}\tau^{2}-3(1-ikc_{s}\tau))\exp{\left(ikc_{s}\tau\right)} \right\}.
\quad\quad \eea
The above solutions for both the modes and their conjugate momenta introduce two new Bogoliubov coefficients $\alpha^{(2)}_{\mbf{k}}$ and $\beta^{(2)}_{\mbf{k}}$ and the parameter $c_{s}=c_{s,*}$ remains satisfied during the phase but at the moment of the two sharp transitions it assumes the value of $c_{s}=\tilde{c_{s}}=1\pm \delta$.
We also require here using of the $\epsilon_{\rm USR}$ definition as:
\bea
\epsilon_{\rm USR}(\tau) = \epsilon_{\rm SRI}\left(\frac{\tau}{\tau_{s}}\right)^{6}.
\eea
The new Bogoliubov coefficients get determined after applying the continuity and differentiability boundary conditions, together known as the Israel junction conditions, for the modes at the conformal time of sharp transition $\tau=\tau_{s}$. The final form of these coefficients is expressed as follows:
\bea
    \label{s5b1}\alpha^{(2)}_{\bf k}&=&1+\frac{3k_{s}^{3}}{2 i k^{3}}\left(1+\left(\frac{k}{k_{s}}\right)^{2}\right),\\
    \label{s5b2}\beta^{(2)}_{\bf k}&=&\frac{3k_{s}^{3}}{2 i k^{3}}\left(1-i\left(\frac{k}{k_{s}}\right)^{2}\right)^{2}\; \exp{\left(2i\frac{k}{k_{s}}\right)}.
\eea
where $k_{s}$ refers to the wavenumber at the sharp transition scale at conformal time $\tau_{s}$. The new $\alpha^{(2)}_{\mbf{k}}$ and $\beta^{(2)}_{\mbf{k}}$ describes a shifted quantum vacuum state, from the previously chosen Bunch-Davies vacuum state, for the present USR phase and this will remain crucial in the analysis of the scalar power spectrum. The $\epsilon_{\rm USR}$ parameter takes on extremely small values while the $\eta_{\rm USR}$ parameter goes through a sharp jump to the value $\eta_{\rm USR} \sim -6$ as a result of our choice of transition. 

\begin{figure*}[htb!]
    	\centering
    \subfigure[]{
      	\includegraphics[width=8.5cm,height=7.5cm]{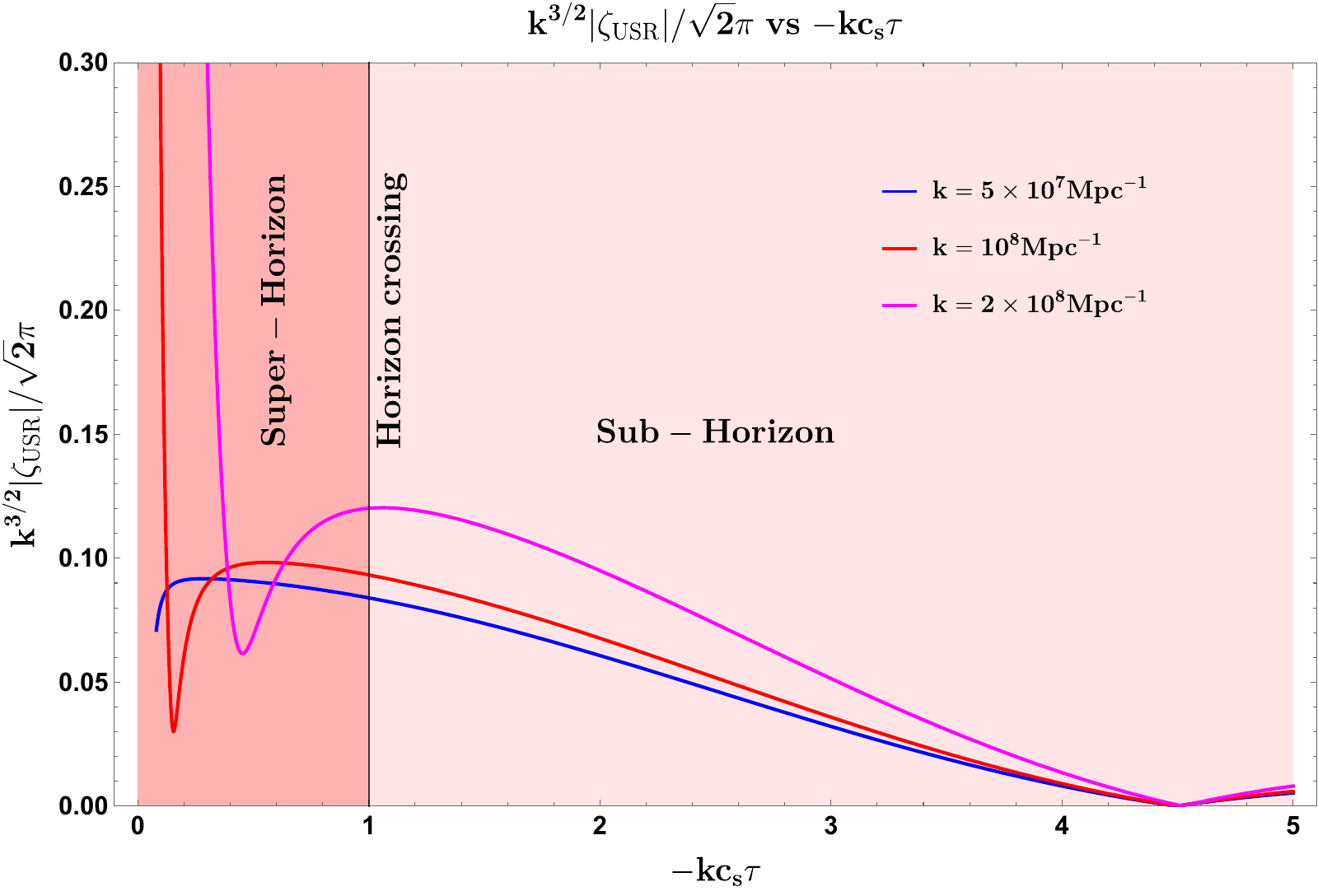}
        \label{zetausr}
    }
    \subfigure[]{
       \includegraphics[width=8.5cm,height=7.5cm]{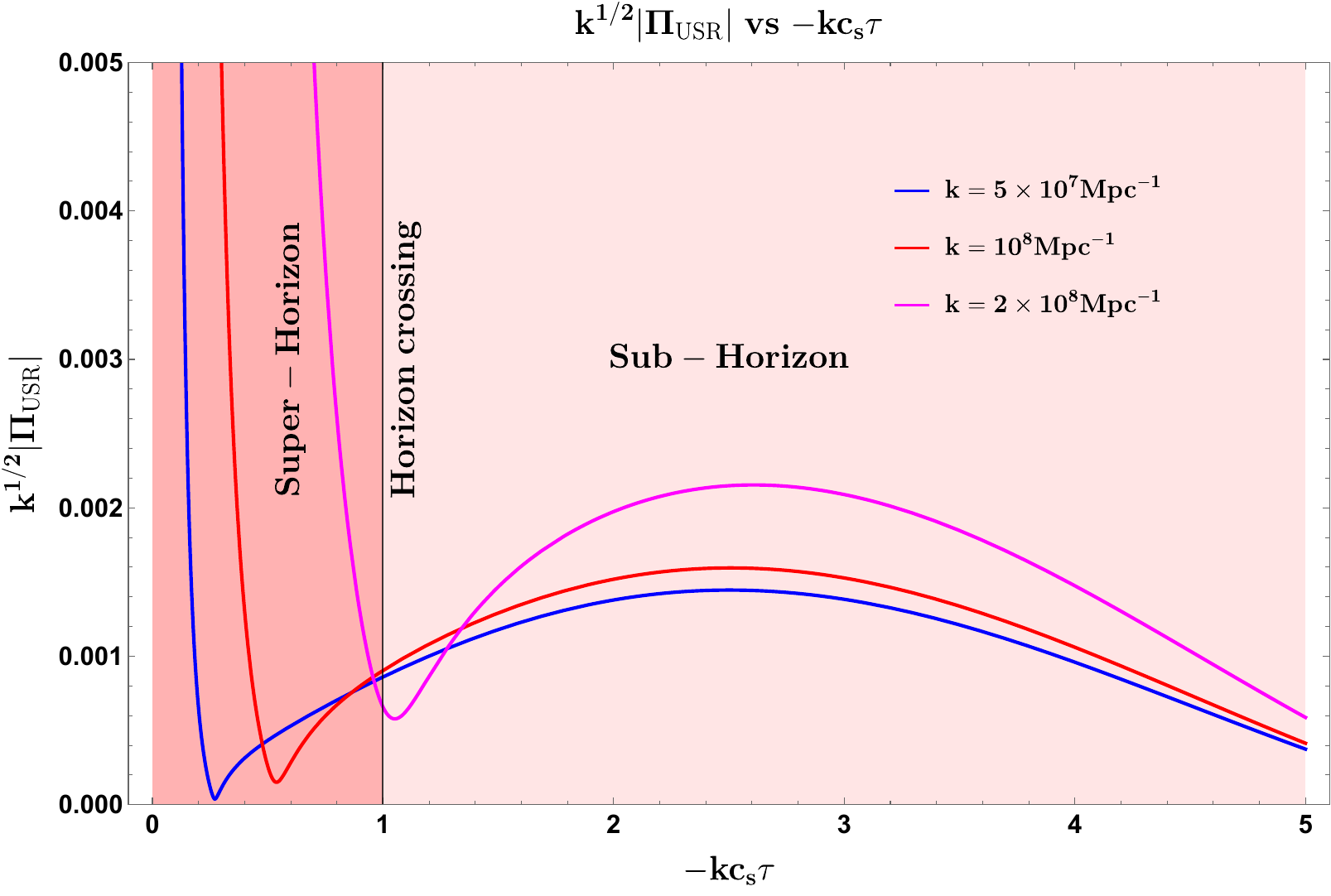}
        \label{piusr}
    }
    	\caption[Optional caption for list of figures]{$k^{3/2}\abs{\zeta_{\mbf{k}}}/\sqrt{2}\pi$ (\textit{left panel}) and $k^{1/2}\abs{\Pi_{\mbf{k}}}/\sqrt{2}\pi$ (\textit{right panel}) plotted against $-kc_{s}\tau$ in the USR phase. In the left, the mode solution behaviour is shown evolving through super-Horizon ($-kc_{s}\tau \ll 1$), crossing the horizon ($-kc_{s}\tau = 1$), and in the sub-Horizon ($-kc_{s}\tau \gg 1$) regimes. Similarly in the right, the evolution of the conjugate momenta of the mode solution is shown through the three regimes as mentioned before. Both panels show evolution when mode $k=5\times 10^{7}{\rm Mpc}^{-1}$ (blue), $k=10^{8}{\rm Mpc}^{-1}$ (red), and $k=2\times 10^{8}{\rm Mpc}^{-1}$ (magenta) is fixed. } 
    	\label{modesusr}
    \end{figure*}

In fig.(\ref{modesusr}), the behaviour for the mode solution and related conjugate momenta are depicted with changing $-kc_{s}\tau$ using eqn.(\ref{s53},\ref{s54}). This analysis requires setting $-k_{s}c_{s}\tau_{s} \sim {\cal O}(-0.01)$ for the nature shown and which comes out of necessity as our variable of interest is $-kc_{s}\tau$ which further brings a wavenumber dependence in the plots. For the left panel, if we focus on the super-Horizon, the curvature perturbation solution behaves asymptotically near $-kc_{s}\tau \ll 1$. We have plotted the behaviour for multiple wavenumbers as they evolve from the super-Horizon to the sub-Horizon regime. After getting closer to horizon crossing, the solution peaks at some value whereafter it tails down as it goes sub-Horizon. Larger wavenumber peaks close to the horizon crossing while smaller wavenumbers peak far in the super-Horizon. As we go deep inside the horizon, the modes start to show oscillations with highly suppressed amplitudes which is a result of the Bogoliubov coefficients. For the right panel, we notice a somewhat similar behaviour for the conjugate momenta where they asymptote sharply in the super-Horizon, drop quickly near horizon crossing, and the maximum value is achieved after going in the sub-Horizon. Here, for larger wavenumbers, greater amplitudes are encountered as we venture inside the sub-Horizon, while it is the opposite for smaller wavenumbers. Notice that the conjugate momenta has substantial amplitude in the super-Horizon till the moment of crossing is reached.  

\subsubsection{In SRII phase}

This phase remains the last in our setup which operates within the conformal time window, $\tau_{e} \leq \tau \leq \tau_{\rm end}$, or in e-foldings, ${\cal N}_{e} \leq {\cal N} \leq {\cal N}_{\rm end}$, and where the time $\tau_{\rm end}$ marks the end of inflation. The general solution for the evolution of curvature perturbation modes and the corresponding canonically conjugate momentum generated during this phase is given here as follows:
\bea \label{s55}
    \zeta_{\mbf{k}}(\tau)&=&\left(\frac{iH^{2}}{2\sqrt{\cal A}}\right)\left(\frac{\tau_{s}}{\tau_{e}}\right)^{3}\frac{1}{(c_{s}k)^{3/2}}\times \left\{\alpha^{(3)}_{\bf k}\left(1+ikc_{s}\tau\right)\exp{\left(-ikc_{s}\tau\right)} - \beta^{(3)}_{\bf k}\left(1-ikc_{s}\tau\right)\exp{\left(ikc_{s}\tau\right)} \right\}. \\
\label{s56}
\Pi_{\mbf{k}}(\tau) &=&\zeta^{'}_{\mbf{k}}(\tau)= \left(\frac{iH^{2}}{2\sqrt{\cal A}}\right)\frac{1}{(c_{s}k)^{3/2}}\left(\frac{\tau_{s}}{\tau_{e}}\right)^{3}\times \frac{k^{2}c_{s}^{2}\tau^{2}}{\tau}\left\{\alpha^{(3)}_{\mbf{k}}\exp{\left(-ikc_{s}\tau\right)} - \beta^{(3)}_{\mbf{k}}\exp{\left(ikc_{s}\tau\right)} \right\},
\eea
which introduces the new set of Bogoliubov coefficients $\alpha^{(3)}_{\mbf{k}}$ and $\beta^{(3)}_{\mbf{k}}$ and the value for the parameter $c_{s}=\tilde{c_{s}}$ is satisfied only at $\tau=\tau_{e}$, and for the subsequent interval after the value returns to give us $c_{s}=c_{s,*}$. The above solution also requires using of the following definition of $\epsilon$ as:
\bea
\epsilon_{\rm SRII}(\tau) = \epsilon_{\rm SRI}\left(\frac{\tau_{e}}{\tau_{s}}\right)^{6}.
\eea
Solving these Bogoliubov coefficients requires further use of the Israel junction conditions at the boundary with the conformal time $\tau=\tau_{e}$ for the modes that exit the USR and those that enter into SRII. At this instant, another sharp transition emerges, and the quantum vacuum state shifts further from the conditions during USR. The resulting expressions are as follows:  
\bea
    \label{s5c1} \alpha^{(3)}_{\bf k}&=&-\frac{k_{s}^{3}k_{e}^{3}}{4k^6}\Bigg[9 \left(-\frac{k}{k_e}+i\right)^2 \left(\frac{k}{k_s}+i\right)^2 \exp{\left(2 i k
   \left(\frac{1}{k_s}-\frac{1}{k_e}\right)\right)}\nonumber\\
    &&\quad\quad\quad\quad\quad\quad\quad\quad\quad\quad\quad\quad\quad\quad-
    \left\{\left(\frac{k}{k_{e}}\right)^2\left(-2\frac{k}{k_{e}}-3i\right)-3i\right\}\left\{\left(\frac{k}{k_{s}}\right)^2\left(-2\frac{k}{k_{s}}+3i\right)+3i\right\}\Bigg],\\
    \label{s5c2}
    \beta^{(3)}_{\bf k}&=&\frac{3k_{s}^{3}k_{e}^{3}}{4k^6}\Bigg[\left(\frac{k}{k_{s}}+i\right)^2\left\{\left(\frac{k}{k_{e}}\right)^{2}\left(2i\frac{k}{k_{e}}\right)+3\right\}\exp{\left(2i\frac{k}{k_{s}}\right)}\nonumber\\
    &&\quad\quad\quad\quad\quad\quad\quad\quad\quad\quad\quad\quad\quad\quad\quad\quad+i\left(\frac{k}{k_{e}}+i\right)^2\left\{3i+\left(\frac{k}{k_{s}}\right)^{2}\left(-2\frac{k}{k_{s}}+3i\right)\right\}\exp{\left(2i\frac{k}{k_{e}}\right)}\Bigg].
\eea
where $k_{s}$ and $k_{e}$ are the wavenumber associated with the sharp transition scales. The $\epsilon_{\rm SRII}$ parameter deviates from constant behaviour and rises from its previous values in the USR till it reaches ${\cal O}(1)$ at the end of inflation. The $\eta_{\rm SRII}$ parameter changes again by a sudden jump from $\eta_{\rm USR} \sim -6$ at $\tau=\tau_{e}$ in the USR to $\eta_{\rm SRII} \sim -1$, which remains till inflation ends.

\begin{figure*}[htb!]
    	\centering
    \subfigure[]{
      	\includegraphics[width=8.5cm,height=7.5cm]{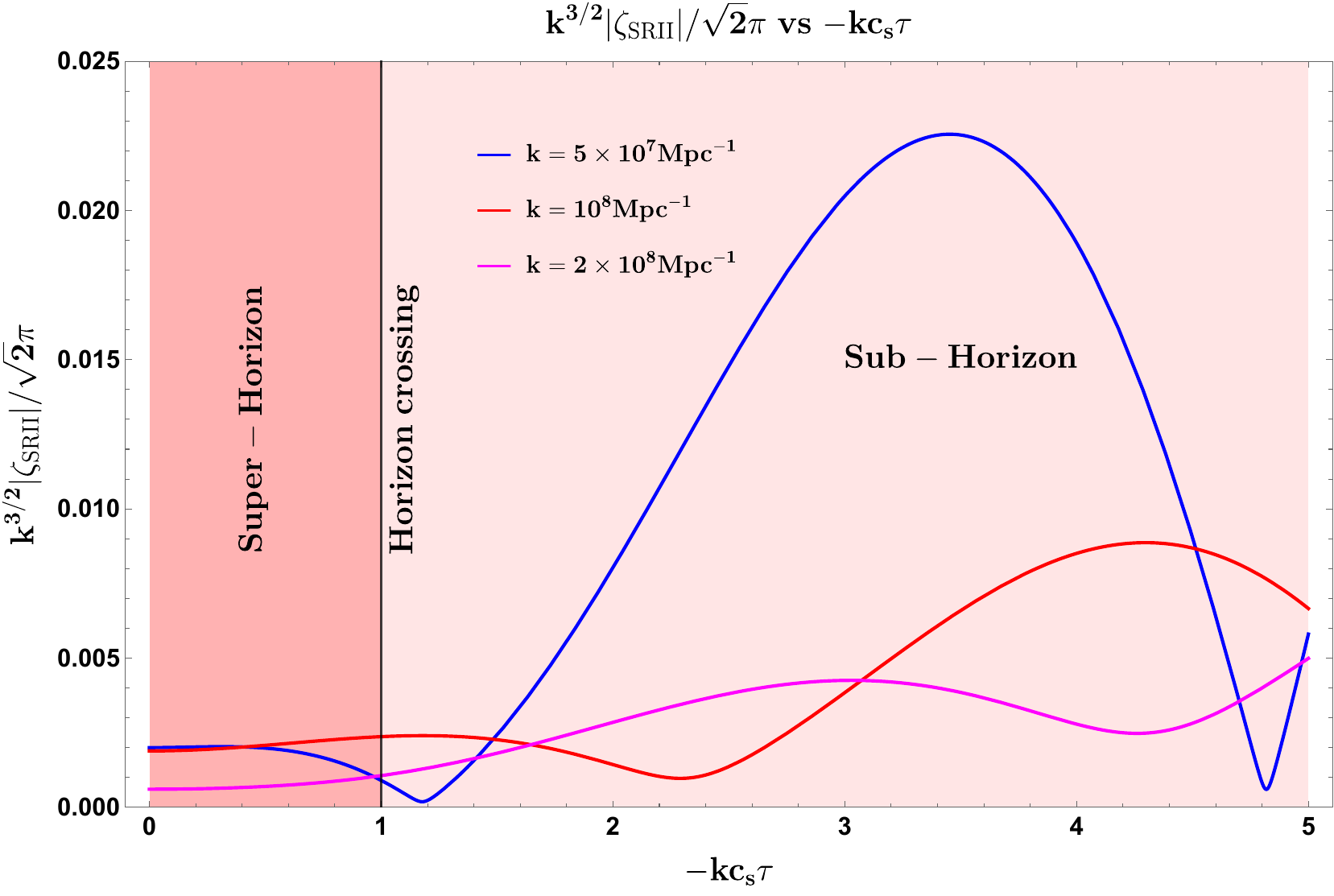}
        \label{zetasr2}
    }
    \subfigure[]{
       \includegraphics[width=8.5cm,height=7.5cm]{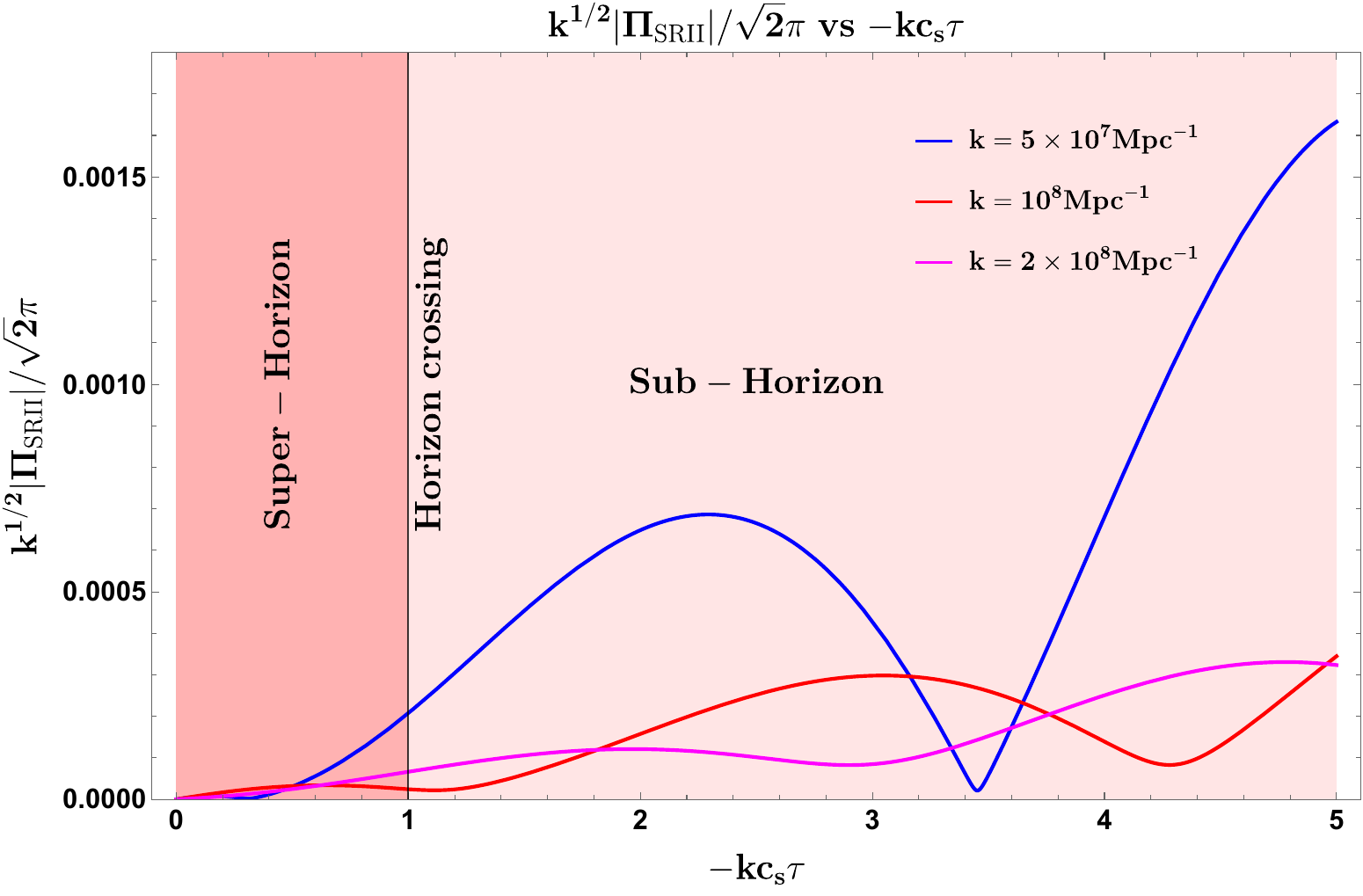}
        \label{pisr2}
    }
    	\caption[Optional caption for list of figures]{$k^{3/2}\abs{\zeta_{\mbf{k}}}/\sqrt{2}\pi$ (\textit{left panel}) and $k^{1/2}\abs{\Pi_{\mbf{k}}}/\sqrt{2}\pi$ (\textit{right panel}) plotted against $-kc_{s}\tau$ in the SRII phase. In the left, the mode solution behaviour is shown evolving through super-Horizon ($-kc_{s}\tau \ll 1$), crossing the horizon ($-kc_{s}\tau = 1$), and in the sub-Horizon ($-kc_{s}\tau \gg 1$) regimes. Similarly in the right, the evolution of the conjugate momenta of the mode solution is shown through the three regimes as mentioned before. Both panels show evolution when the mode $k=5\times 10^{7}{\rm Mpc}^{-1}$ (blue), $k=10^{8}{\rm Mpc}^{-1}$ (red), and $k=2\times 10^{8}{\rm Mpc}^{-1}$ (magenta) is fixed. } 
    	\label{modessr2}
    \end{figure*}

In fig.(\ref{modessr2}), the behaviour of the mode solution and related conjugate momenta are depicted with changing $-kc_{s}\tau$ using eqn.(\ref{s55},\ref{s56}). Similar to conditions in the USR, this analysis also requires setting both $-k_{s}c_{s}\tau_{s}$ and$-k_{e}c_{s}\tau_{e}$ as of $ \sim {\cal O}(-1)$ for the nature shown. For the curvature perturbation in the left panel, the overall behaviour remains almost constant till we encounter the horizon crossing. After the solutions enter the horizon, the oscillations become prominent with higher amplitudes for smaller wavenumbers, and the larger wavenumbers only start to increase deep inside the horizon. In the super-Horizon, the magnitude remains ${\cal O}(10^{-3})$, which then develops oscillations of increasing magnitudes further after horizon crossing. For the conjugate momentum modes in the right panel, the overall magnitude in the super-Horizon is relatively less than the curvature perturbation solutions, and here also we observe oscillatory features as we exit the horizon and progress in the sub-Horizon regime, with increased magnitudes for smaller wavenumbers.

\subsection{Quantifying the tree-level contribution to the scalar power spectrum}

In the previous section, we dealt with the semi-classical solutions for the comoving curvature perturbation modes for the three phases in our setup. From the information about the scalar modes and their associated Bogoliubov coefficients, we can now compute the tree-level scalar power spectrum. Constructing this power spectrum requires quantizing the curvature perturbation modes and evaluating the tree-level contribution to the two-point correlation function. This procedure introduces a set of creation and annihilation operators $\hat{a}_{\mbf{k}}$ and $\hat{a}^{\dagger}_{\mbf{k}}$ which when acted on the quantum vacuum state $\ket{0}$ of the Hilbert space either creates an excited state or annihilates it. The curvature perturbation modes are now promoted as operators and can be written as:
\bea
\hat{\zeta}_{\mbf{k}}(\tau) = \zeta_{\mbf{k}}\hat{a}_{\mbf{k}} + \zeta^{*}_{\mbf{k}}a^{\dagger}_{-\mbf{k}}\quad\quad\quad\quad \hat{\Pi}_{\mbf{k}}(\tau) = \Pi_{\mbf{k}}\hat{a}_{\mbf{k}} + \Pi^{*}_{\mbf{k}}a^{\dagger}_{-\mbf{k}}
\eea
with the conjugate momentum $\Pi_{\mbf{k}} = \partial_{\tau}\zeta_{\mbf{k}}$. 
The two-point correlation function can now be written in the following manner:
\bea \langle \hat{\zeta}_{\bf k}\hat{\zeta}_{{\bf k}^{'}}\rangle_{{\bf Tree}} =(2\pi)^{3}\;\delta^{3}\left({\bf k}+{\bf k}^{'}\right)\frac{2\pi^2}{k^3}\Delta^{2}_{\zeta,{\bf Tree}}(k),\quad\quad \text{where} \quad \Delta^{2}_{\zeta,{\bf Tree}}(k) =  \frac{k^{3}}{2\pi^{2}}|{\zeta}_{\bf k}(\tau)|^{2}_{\tau\rightarrow 0}.
\eea
which provides us the relevant tree-level contribution found after taking the late-time limit $\tau \rightarrow 0$. The term $\Delta^{2}_{\zeta,{\bf Tree}}(k)$ is the dimensionless tree-level scalar power spectrum. We are now in the position to mention the tree-level scalar power spectrum using the mode solutions mentioned before in eqs.(\ref{s51},\ref{s53},\ref{s55}) for the three phases of interest in our setup. The final form of the dimensionless scalar power spectrum comes out as follows:
\bea  \label{treespec}
\Delta^{2}_{\zeta,{\bf Tree}}(k)
&=& \displaystyle
\displaystyle\left\{
	\begin{array}{ll}
		\displaystyle\Delta_{\bf {Tree}}^{\text{SRI}}(k)=\left(\frac{H^{4}}{8\pi^{2}{\cal A} c^3_s}\right)_* \Bigg\{1+\Bigg(\frac{k}{k_s}\Bigg)^2\Bigg\}& \mbox{when}\quad  k\leq k_s  \;(\rm SRI)  \\  
			\displaystyle 
			\displaystyle\Delta_{\bf {Tree}}^{\text{USR}}(k)=\left(\frac{H^{4}}{8\pi^{2}{\cal A} c^3_s}\right)_* \left(\frac{k_e}{k_s }\right)^{6}\left|\alpha^{(2)}_{\bf k}-\beta^{(2)}_{\bf k}\right|^2 & \mbox{when }  k_s\leq k\leq k_e  \;(\rm USR)\\ 
   \displaystyle 
			\displaystyle\Delta_{\bf {Tree}}^{\text{SRII}}(k)=\left(\frac{H^{4}}{8\pi^{2}{\cal A} c^3_s}\right)_* \left(\frac{k_e}{k_s }\right)^{6}\left|\alpha^{(3)}_{\bf k}-\beta^{(3)}_{\bf k}\right|^2 & \mbox{when }  k_e\leq k\leq k_{\rm end}  \;(\rm SRII) 
	\end{array}
\right. \eea
where we use the conditions related to the super-Horizon limit for the modes that cross the Hubble horizon and obey $-kc_{s}\tau \ll 1$. Using the above, we mention below the total tree-level scalar power spectrum, which results from the fact that the different phases have their contributions connected via a Heaviside Theta function to signify the presence of sharp transitions in our setup: 
\bea \label{totpower} \left[\Delta^{2}_{\zeta,{\bf Tree}}(k)\right]_{\bf Total} &=&\Delta_{\bf {Tree}}^{\text{SRI}}(k)+\Delta_{\bf {Tree}}^{\text{USR}}(k)\Theta(k-k_s)+\Delta_{\bf {Tree}}^{\text{SRII}}(k)\Theta(k-k_e).
\eea
The total power spectrum will remain necessary for further analysis in the later sections. 
We must mention here again the fact that it is the amplitude of the above-mentioned scalar power spectrum in the three phases which also serves as another constraint for the Galileon EFT coefficients $c_{4},c_{5}$ and the same gets mentioned before in the analysis of section \ref{s4}.

\subsection{Non-renormalization theorem and suppression of the loop contributions in scalar power spectrum}

In this section we outline the importance of the non-renormalization theorem and how it enables us to accurately calculate the one-loop corrections to the scalar power spectrum. 

\subsubsection{Non-renormalization theorem}

Recall that successfully performing inflation in Galileon theory requires mildly breaking the Galilean shift symmetry. The linear term proportional to the scalar field and the constant potential term in the Lagrangian in eqn.(\ref{CovGal}) are responsible for this nature of symmetry breaking. However, it might happen that such terms, when introduced into the Lagrangian, must also bring significant quantum corrections, eventually spoiling the theory. The non-renormalization theorem in Galileon theory states that the loops of the fields do not renormalize the Galileon interactions at any order in the perturbation theory if the couplings to the heavy fields respect the underlying Galilean symmetry.  

This theorem profoundly impacts the results of our analysis related to Galileon inflation. Through this property, we can also observe the production of large non-gaussianities and control the duration of inflation. Let's look into the actual significance of this theorem through some expressions, starting with the way the comoving curvature perturbation transforms under the action of the Galileon symmetry:
\bea \label{s5d} \zeta \rightarrow \zeta - \frac{H}{\dot{\bar{\phi}}}b.\delta x, \quad\quad \partial_{i}\zeta \rightarrow \partial_{i}\zeta - \frac{H}{\dot{\bar{\phi}}}b_{i}, \quad\quad \zeta^{'} \rightarrow \zeta^{'} - \frac{H}{\dot{\bar{\phi}}}b_{0}, \quad\quad \partial^{2}\zeta \rightarrow \partial^{2}\zeta. \eea
where $\bar{\phi} \equiv \bar{\phi}(t)$ is the time-dependent background Galileon field previously mentioned in section \ref{s2}. The above transformations tell us that only the term $\partial^{2}\zeta$ remains invariant under Galilean symmetry, and hence, to achieve mild symmetry breaking, one requires combinations of terms that overall break this symmetry. Some combinations are removed from the multiple possible terms by either field re-definitions or vanishing at the boundary after performing integration by parts. An essential term from such possibilities to look out for is $\zeta'\zeta^{2}$. Due to the coefficient $\partial_{\tau}(\eta/c_{s}^{2})$ for the previous term, this quantity gives enormous one-loop contributions during the sharp transitions, which can be harmful to our perturbative analysis. However, this remains absent from the final third-order action even though the shift symmetry gets broken here. This particular absence is due to it ultimately vanishing at the boundary after breaking the Galilean symmetry softly; hence, we can evade the presence of significant loop corrections from the three-point correlations. 

\subsubsection{Galileon cubic action and one-loop contributions}

The final number of remaining terms allowed in the third order action are $\zeta^{'3}, \zeta^{'2}\partial^{2}\zeta, \zeta^{'}(\partial_{i}\zeta)^{2}, \partial^{2}\zeta(\partial_{i}\zeta)^{2}$ which when combined lead to the following action equation: 
\bea
S^{3}_{\zeta} = \int d\tau\; d^{3}x\frac{a(\tau)^{2}}{H^{3}}\bigg[\frac{{\cal G}_1}{a}\zeta^{'3}+\frac{{\cal G}_2}{a^2}\zeta^{'2}\left(\partial^2\zeta\right)+\frac{{\cal G}_3}{a}\zeta^{'}\left(\partial_i\zeta\right)^2+\frac{{\cal G}_4}{a^2}\left(\partial_i\zeta\right)^2\left(\partial^2\zeta\right)\bigg].
\eea
This contains the coupling constants defined as:
\bea
    {\cal G}_1&\equiv& \frac{2H                    \dot{\bar{\phi}}^3}{\Lambda^3}            \Bigg(c_3+9c_4Z+30c_5Z^2\Bigg),\\
    {\cal G}_2&\equiv& -\frac{2                    \dot{\bar{\phi}}^3}{\Lambda^3}\Bigg(c_3+6c_4Z+18c_5Z^2\Bigg),\\ 
    {\cal G}_3&\equiv& 
       -\frac{2H\dot{\bar{\phi}}^3}{\Lambda^3}\Bigg(c_3+7c_4Z+18c_5Z^2\Bigg)-\frac{2\dot{\bar{\phi}}^3H\eta}{\Lambda^3}\Bigg(c_3+6c_4Z+18c_5Z^2\Bigg),\\
    {\cal G}_4&\equiv& 
        \frac{\dot{\bar{\phi}}^3}{\Lambda^3}\bigg\{c_3+3c_4Z+6c_5\bigg[Z^2+\frac{\dot{H}\dot{\bar{\phi}}^2}{\Lambda^6}\bigg]\bigg\}-\frac{3\dot{\bar{\phi}}^4H\eta}{\Lambda^6}\bigg\{c_4+4c_5Z\bigg\},
\eea
where the parameter $Z$ is the same as defined previously in eqn.(\ref{Z}). The one-loop effects get calculated using the terms from the above cubic action and applying the Schwinger-Keldysh (In-In) formalism. This formalism is the commonly chosen method when calculating the cosmological correlation functions, and we incorporate this method to evaluate the one-corrections to the scalar power spectrum fully. After the necessary Wick contractions, we determine the correlations to evaluate using each interaction operator and performing the temporal or momentum integrals. We do not mention the complete calculations, details of which appear in the ref.\cite{Choudhury:2023hvf}, and condense the loop contributions to present the final one-loop corrected scalar power spectrum:
\bea \label{s4dtot}  
\Bigg[\Delta^{2}_{\zeta}(k)\Bigg]_{\bf Total}
&=&  \displaystyle \Bigg[\Delta^{2}_{\zeta,\bf {Tree}}(k)\Bigg]_{\rm \textbf{SRI}} +  \Bigg[\Delta^{2}_{\zeta,\bf {Tree}}(k)\Bigg]_{\rm \textbf{USR}}\Theta(k-k_{s})   + \Bigg[\Delta^{2}_{\zeta,\bf {Tree}}(k)\Bigg]_{\rm \textbf{SRII}}\Theta(k-k_{e}) + {\cal Q}_{c},\nonumber\\
&\approx& A\Bigg[\Delta^{2}_{\zeta,\bf {Tree}}(k)\Bigg]_{\rm \textbf{SRI}}\left\{\left(\frac{k_{s}}{k_{e}}\right)^{6}(1+{\cal Q}_{c}) + \left(\big|\alpha_{\bf k}^{(2)}-\beta_{\bf k}^{(2)}\big|^2 \;\Theta(k-k_s)+\big|\alpha_{\bf k}^{(3)}-\beta_{\bf k}^{(3)}\big|^2 \;\Theta(k-k_e)\right)\right\}\quad
\eea
where eqn.(\ref{treespec}) is used and the term ${\cal Q}_{c}$ acts as a label for the one-loop corrections which are of the form:
\bea \label{quantcorr}
{\cal Q}_{c} &=& \Bigg[\Delta^{2}_{\zeta,\bf {One-Loop}}(k)\Bigg]_{\rm \textbf{SRI}} + \Bigg[\Delta^{2}_{\zeta,\bf {One-Loop}}(k)\Bigg]_{\rm \textbf{USR}}\Theta(k-k_{s}) + \Bigg[\Delta^{2}_{\zeta,\bf {One-Loop}}(k)\Bigg]_{\rm \textbf{SRII}}\Theta(k-k_{e})\nonumber\\
&=& \Bigg[\Delta^{2}_{\zeta,\bf {Tree}}(k)\Bigg]_{\rm \textbf{SRI}}\times \frac{1}{8{\cal A}^{2}_{*}\pi^{4}}\bigg\{-\sum^{4}_{i=1}{\cal G}_{i,\mbf{SRI}}\mbf{F}_{i,\mbf{SRI}}(k_{s},k_{*}) + \sum^{4}_{i=1}{\cal G}_{i,\mbf{USR}}\mbf{F}_{i,\mbf{USR}}(k_{e},k_{s})\;\Theta(k-k_{s}) \nonumber\\
&+& \sum^{4}_{i=1}{\cal G}_{i,\mbf{SRII}}\mbf{F}_{i,\mbf{SRII}}(k_{\rm end},k_{e})\;\Theta(k-k_{e})\bigg\}.
\eea
We also define here the amplitude, $A=\displaystyle{\left(\frac{H^{4}}{8\pi^{2}{\cal A}c_{s}^{3}}\right)_{*}\left(\frac{k_{e}}{k_{s}}\right)^{6}}$, of the total power spectrum and the couplings  ${\cal G}_{i,\mbf{SRI}}, {\cal G}_{i,\mbf{USR}}, {\cal G}_{i,\mbf{SRII}}$ and terms $\mbf{F}_{i,\mbf{SRI}}, \mbf{F}_{i,\mbf{USR}}, \mbf{F}_{i,\mbf{SRII}}$ collectively describe the momentum dependent one-loop effects in the above equations and their explicit calculations can be found in \cite{Choudhury:2023hvf,Choudhury:2023kdb}.The figure (\ref{s5d}) shows the behavior of the one-loop corrected scalar power spectrum, which comes from eqn.(\ref{s4dtot}). We see that the power spectrum peaks during the USR to achieve the desired amplitude of ${\cal O}(10^{-2})$ and after another sharp transition into SRII, it reaches the amplitude of ${\cal O}(10^{-5})$ till the end of inflation. The one-loop corrections do not spoil the overall nature of the spectrum, especially in the USR, which is essential to studying PBH formation. 

\begin{figure*}[ht!]
    	\centering
    {
        \includegraphics[width=17cm,height=12cm]{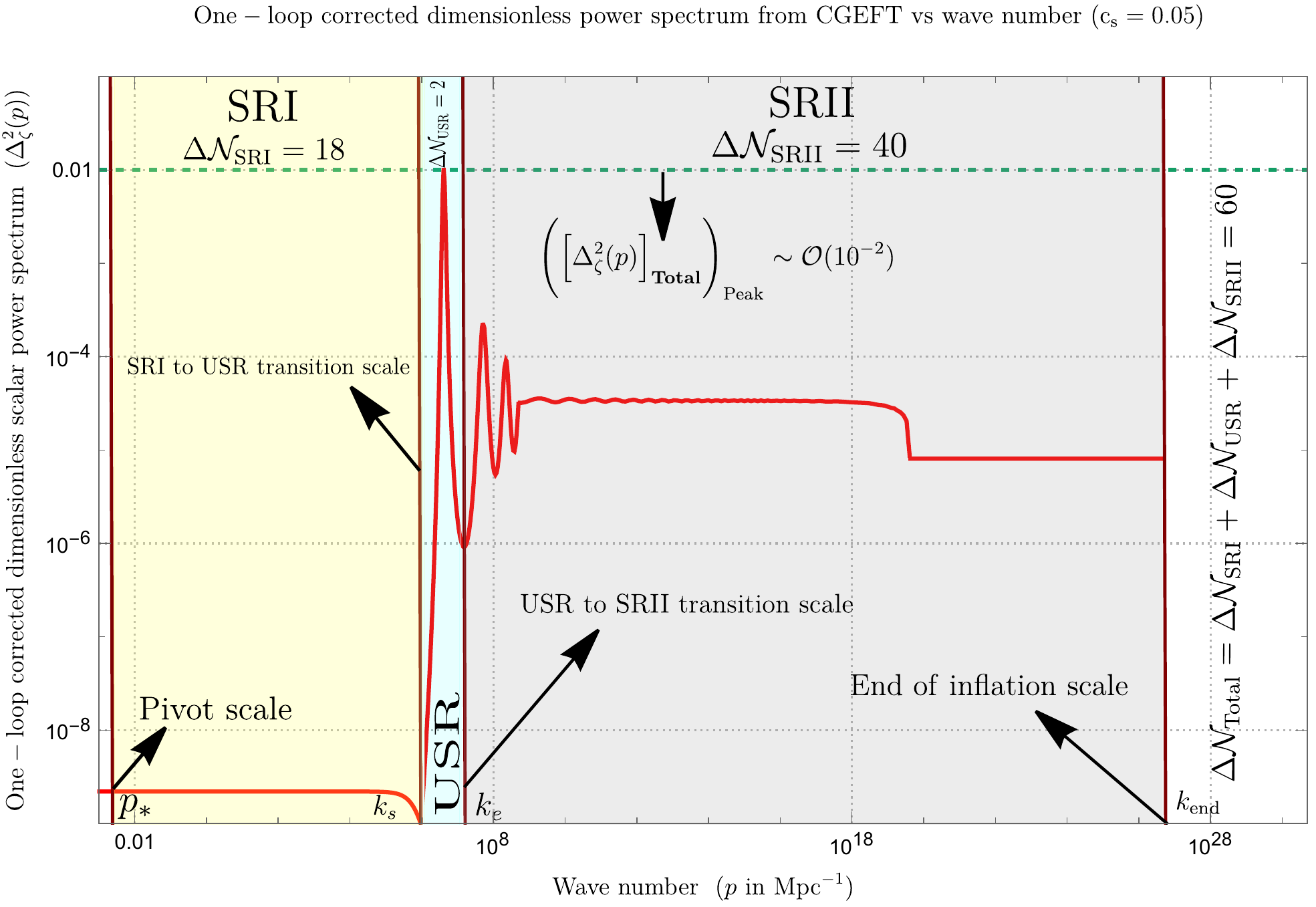}
        \label{scalarpspec}
    } 
    \caption[Optional caption for list of figures]{One-loop corrected dimensionless scalar power spectrum as a function of the wavenumber $p$. The effective sound speed $c_{s}=0.05$ is fixed. The transition wavenumbers $k_{s}=10^{6}{\rm Mpc^{-1}}$ and $k_{e}=10^{7}{\rm Mpc^{-1}}$ are fixed here. The power spectrum peaks at the scale $\sim 5\times 10^{6}{\rm Mpc^{-1}}$. A total of $60$ e-folds is achieved for successful inflation.  }
\label{s5d}
    \end{figure*}

Now that we have the expression for the total scalar power spectrum, we can analyze the PBH formation mechanism in the presence of a general cosmological background, which includes the total scalar power spectrum amplitude responsible for generating PBHs and subsequent generation of gravitational waves.

\section{Impact of Equation of State Parameter (EoS) in PBHs formation} \label{s6}

This section is concerned with the production of PBHs through the mechanism of the collapse of large density fluctuations when entered in a cosmological background with the equation of state (EoS) $w$. The significance of the EoS parameter $w$ has recently gained some attention \cite{Liu:2023pau,Liu:2023hpw,Balaji:2023ehk,Domenech:2021ztg,Domenech:2020ers,Domenech:2019quo,Altavista:2023zhw} as a way to extract information about the physics of the early Universe during the pre-BBN era by using the latest conclusive signature of a stochastic gravitational wave background (SGWB) reported by the PTA collaborations. Since the primordial content of the Universe is uncertain in theory at present, assuming a scenario of arbitrary EoS background where even the effective sound speed $c_{s}$ for the fluctuations is also unknown provides an exciting opportunity to explore the effects of such parameters in the present theory and corroborating their signatures with the observational data. Within the possible number of scenarios to model the theoretical outcomes regarding PBH production and GW generation, we will focus on the impact of the parameter $w$ on PBH production and further link it with the induced GWs from the underlying Galileon inflation framework.

\subsection{$w$-Press-Schechter Formalism}

We prefer to work with the standard mechanism of threshold statistics to understand PBH formation in an era of constant EoS $w$. This method involves the condition for the primordial density fluctuations to gravitationally collapse and form PBHs when a specific threshold condition on the perturbation overdensity is satisfied. The amplitude of the scalar power spectrum plays a crucial role here and will be elaborated on further in this section. We will focus on working with the Press-Schechter formalism modified here with the presence of the constant EoS $w$. 

The mass of the formed PBH remains proportional to the mass contained within the Horizon at the time of formation. However, to initiate the formation requires the threshold condition $\delta\rho/\rho \equiv \delta>\delta_{\rm th}$ on the perturbation overdensities. We choose to work with Carr's criteria of $c_{s}^{2}=1$ \cite{1975ApJ...201....1C}, which then gives the relation for the threshold as:
\bea
\label{deltath}
\delta_{\rm th} = \frac{3(1+w)}{5+3w}.
\eea
We also assume the linearity approximation between the comoving curvature perturbation and the density contrast in the super-Horizon regime:
\bea \label{deltalinear}
\delta(t,\mathbf{x}) \cong \frac{2(1+w)}{5+3w}\left(\frac{1}{aH}\right)^{2}\nabla^{2}\zeta(k).
\eea
The analysis of PBH production in the presence of non-linearities in the above relation can be found in refs. \cite{Ferrante:2022mui,Franciolini:2023pbf,Franciolini:2023wun}. The resulting mass of the formed PBH is modified with $w$ in the manner as shown  \cite{Alabidi:2013lya}:
\bea \label{mpbh}
M_{\rm PBH} = 1.13 \times 10^{15} \times \bigg(\frac{\gamma}{0.2}\bigg)\bigg(\frac{g_{*}}{106.75}\bigg)^{-1/6}\bigg(\frac{k_{*}}{k_{\rm s}}\bigg)^{\frac{3(1+w)}{1+3w}} M_{\odot}. \eea
where $\gamma \sim 0.2$ is the efficiency factor of collapse, $k_{*}=0.02{\rm Mpc^{-1}}$ labels the pivot scale value and $M_{\odot}$ refers to the solar mass. To obtain the PBH abundance requires the estimate of the variance in the distribution of the primordial overdensity. This variance can be calculated as follows: 
\bea
\sigma_{\rm M_{\rm PBH}} ^2 = \bigg(\frac{2(1+w)}{5+3w}\bigg)^2 \int \frac{dk}{k} \; (k
R)^4 \; W^2(kR) \;\left[\Delta^{2} _{\zeta }(k)\right]_{\bf Total}.\eea
where we see the significance of the amplitude $A$ of the total scalar power spectrum mentioned before in eqn.(\ref{s4dtot}). The change in the amplitude $A$ is quite sensitive to the variance estimates, which is reflected in our numerical outcomes for the abundance discussed in future sections. Here $W(kR)$ is the Gaussian smoothing function given by $\exp{(-k^2 R^2 /4)}$, over the scales of PBH formation, $R=1/(\Tilde{c_s} k_{s})$.
The assumption of working with the linear relation in eqn.(\ref{deltalinear}) comes with constraints on the allowed threshold regime where the collapse of perturbations, thereby generating a sizeable abundance of PBH, is achieved based on the initial shape of the power spectrum. This regime has been studied extensively using numerical studies and comes out as $2/5 \leq \delta_{\rm th} \leq 2/3$ \cite{Musco:2020jjb}. In terms of $w$, we investigate the range $-5/9 \leq w \leq 1/3$. We will use this estimate to find the acceptable regime that can help generate the desired abundance of PBHs and the signature of induced GWs compatible with the recent NANOGrav15 signal. The mass fraction of the PBHs \cite{Sasaki:2018dmp} now reads as follows:
\bea
\beta(M_{\rm PBH})= \gamma \frac{\sigma_{\rm M_{\rm PBH}}}{\sqrt{2\pi}\delta_{\rm th}}\exp{\bigg(-\frac{\delta_{\rm th}^2}{2\sigma_{\rm M_{\rm PBH}}^2}\bigg)},
\eea
The mass and $w$ dependence coming from the variance now gets included in the mass fraction also. The choice of ours where we neglect any non-linear contributions in the density contrast gets reflected in the mass fraction which is a result of Gaussian statistics for $\delta$. The present-day abundance of the PBHs is then written using the expression:
\bea
f_{\rm PBH} \equiv \frac{\Omega_{\rm PBH}}{\Omega_{\rm CDM}}= 1.68\times 10^{8} \bigg(\frac{\gamma}{0.2}\bigg)^{1/2} \bigg(\frac{g_{*}}{106.75}\bigg)^{-1/4} \left(M_{\rm PBH}\right)^{-\frac{6w}{3(1+w)}}\times \beta(M_{\rm PBH}),\eea
where $g_{*}=106.75$ represents the relativistic degrees of freedom. We note that the frequency and wavenumber are connected through $f \simeq 1.6\times 10^{-15}(k/{\rm Mpc}^{-1})$. We use the Galileon scalar power spectrum to find the abundance estimate and determine what is the allowed range for the EoS $w$ which can still provide sizeable abundance after keeping within the numerically allowed range of the threshold.

\begin{figure*}[htb!]
    	\centering
    {
        \includegraphics[width=17cm,height=12cm]{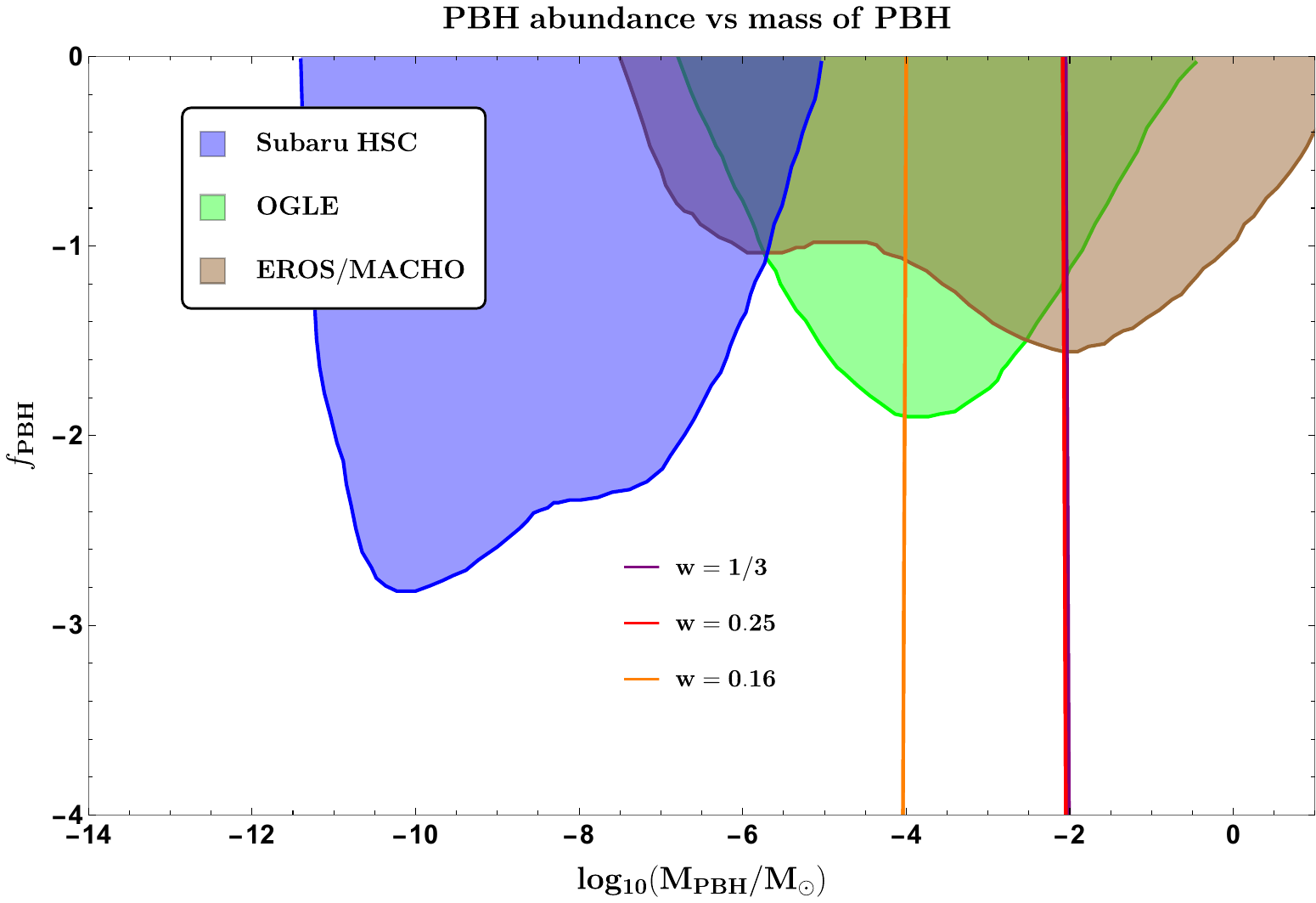}
        \label{fpbhvsmpbh}
    } 
    \caption[Optional caption for list of figures]{PBH abundance as a function of their mass in solar mass units. The Blue, green, and brown shaded background contours represent the current microlensing constraints from Subaru HSC (blue), OGLE (green), and EROS/MACHO (brown) respectively \cite{Niikura:2017zjd,Niikura:2019kqi,EROS-2:2006ryy}. The various cases of EoS $w$ are shown using purple ($w=1/3$), red ($w=0.25$), and orange ($w=0.16$) which can generate sizeable abundance of $f_{\rm PBH} \in (10^{-3},1)$. The transition scale for PBH formation is set with $k_{s}\sim {\cal O}(10^{7}){\rm Mpc^{-1}}.$   }
\label{s6d1}
    \end{figure*}
The fig.(\ref{s6d1}) depicts the abundance behavior for various masses of PBHs, each corresponding to a particular constant value of $w$. We see that for $w=1/3$, PBHs near solar mass get generated, which falls within the regime of having a sizeable abundance after applying constraints from the microlensing experiments. One reaches the same conclusion for a background with $w=0.25$ where we observe the generation of similar near solar mass PBH within our framework. Decreasing the $w$ value further, we found that for $w=0.16$ also, one can generate $M_{\rm PBH} \sim {\cal O}(10^{-4})$ with having enough abundance and be labeled as potential dark matter candidate. We do not analyze cases for $w$ less than $w=0.16$ since the resulting spectrum of the SIGWs does not comply well with the SGWB signal obtained by the PTA.

\subsection{Old vs New formalism: A comparative analysis}

In this section, we provide a comparative analysis between the methods used before considering any variation in the EoS parameter, primarily working only with the radiation-dominated (RD) era $w=1/3$ and the changes that occur after an arbitrary but constant $w$ background gets considered in our calculations for the PBH formation. 

When studying PBHs under the standard scenario of assuming an RD-era for the Universe, opting for the Press-Schechter formalism is not entirely correct to evaluate the mass fraction and obtain accurate predictions for the PBH abundance associated with the frequency of the NANOGrav-15 data. Primary reasons include the assumptions of a Gaussian distribution for the density contrast and the linearity approximations in the super-Horizon as stated before in eqn.(\ref{deltalinear}). Including non-linearities and non-gaussianities is essential to better estimate the PBH abundance without overproducing them, and this also requires a change in the usual threshold statistics, a prospective alternative of which is the compaction function formalism. In refs.\cite{Franciolini:2023pbf,Ferrante:2022mui,Choudhury:2023fwk}, the significance of non-Gaussianities from the perspective of PBH overproduction with the application of the compaction function can be found. For the present theme of this work, we investigate the possibility of avoiding overproduction by envisioning a scenario where a constant $w$ background dominates the very early Universe during PBH formation. We show from our outcomes that the $w$-Press-Schechter formalism works well to suffice our requirements of generating a sizeable abundance of PBHs, thereby avoiding overproduction. A more robust picture where the $w$ parameter gets incorporated within the use of non-linearities and the compaction function formalism is yet to be established, which can strengthen the results both from the theory and observational side. One can also consider calculating the higher point non-Gaussian correlations by which the respective higher-order non-Gaussianity parameters can further constrain the space of values to enable a refined calculation of the mass fraction. In this work, however, we play with the EoS $w$ to achieve the desired PBH abundance.  

\section{Impact of Equation of State Parameter (EoS) in Scalar Induced Gravitational Waves} \label{s7}

This section discusses the theory of scalar-induced gravitational waves (SIGW) generated in the presence of a general cosmological background characterized by a constant EoS $w$. We first examine the underlying theoretical setup while constructing the necessary mathematical framework, followed by applying the Galileon theory scalar power spectrum and analyzing the resulting GW spectrum. \textcolor{black}{Our aim in this section is to present the theoretical background in a self-consistent manner since we are exploring this specific theory for the first time and continue to incorporate its results towards the primary goal of this work. }

\subsection{The underlying theory of $w$-SIGWs: Motivation and details}
Gravitational Waves (GW) have attracted a lot of attention in recent literature due to their ability to explain phenomena in the primordial universe that cannot be probed by, for example, the CMB and BBN observations. For instance, the primordial fluctuations provide their imprints on the CMB anisotropies. However, that information exists on a larger scale, which gives insufficient information about the later stages of inflation. This is where GW physics comes into the picture. It enables exploration into the primordial Universe, even before the Big Bang Nucleosynthesis, and can provide details of the last stages of inflation. Now, these induced GWs can be studied from different theories like cosmic strings, domain walls, first-order phase transitions, and inflationary scenarios, to name a few  \cite{Choudhury:2023hfm,Bhattacharya:2023ysp,Franciolini:2023pbf,Inomata:2023zup,Wang:2023ost,Balaji:2023ehk,HosseiniMansoori:2023mqh,Gorji:2023sil,DeLuca:2023tun,Choudhury:2023kam,Yi:2023mbm,Cai:2023dls,Cai:2023uhc,Huang:2023chx,Vagnozzi:2023lwo,Frosina:2023nxu,Zhu:2023faa,Jiang:2023gfe,Cheung:2023ihl,Oikonomou:2023qfz,Liu:2023pau,Liu:2023ymk,Wang:2023len,Zu:2023olm, Abe:2023yrw, Gouttenoire:2023bqy,Salvio:2023ynn, Xue:2021gyq, Nakai:2020oit, Athron:2023mer,Ben-Dayan:2023lwd, Madge:2023cak,Kitajima:2023cek, Babichev:2023pbf, Zhang:2023nrs, Zeng:2023jut, Ferreira:2022zzo, An:2023idh, Li:2023tdx,Blanco-Pillado:2021ygr,Buchmuller:2021mbb,Ellis:2020ena,Buchmuller:2020lbh,Blasi:2020mfx, Madge:2023cak, Liu:2023pau, Yi:2023npi,Gangopadhyay:2023qjr,Vagnozzi:2020gtf,Benetti:2021uea,Inomata:2023drn,Lozanov:2023rcd,Basilakos:2023jvp,Basilakos:2023xof,Li:2023xtl,Domenech:2021ztg,Yuan:2021qgz,Chen:2019xse,Cang:2023ysz,Cang:2022jyc,Konoplya:2023fmh,Huang:2023chx}. We particularly focus our attention on GWs induced by the mode coupling between scalar perturbations, otherwise known as Scalar Induced Gravitational Waves. 
To obtain a sizeable abundance of the SIGWs at observationally relevant scales, the scalar perturbations have to be significantly enhanced at the smaller scales relative to the CMB scale. The majority of the studies have been performed where induced GWs are assumed to be formed in the RD epoch. However, there might exist possibilities for them to be generated in other epochs as well. Hence, it is interesting to examine the production of SIGWs for a general EoS parameter $w$, which we have performed in the scope of this analysis.

In the following discussions, we compute the formula for the tensor power spectrum amplitude for SIGWs generated at second order in cosmological perturbation theory. Let us first start with the spatially flat FLRW metric written in the transverse-traceless gauge:
\bea
ds^2 = a^2(\tau)[-(1+2\Phi)d\tau^2 + (\del_{ij}\left(1-2\Psi\right)+2h_{ij})dx^i dx^j],
\eea
where $a(\tau)$ represent the scale factor, $\Phi$ and $\Psi$ are the scalar potentials. $h_{ij}$ in the above equation represents the tensor modes in linear order, which get interpreted as the primordial gravitational waves. 
Assuming a perfect fluid to represent the matter content of the universe, we can write its energy-momentum tensor as :
\bea
T_{\mu \nu} =(\rho+P)u_{\mu}u_{\nu}+Pg_{\mu \nu},
\eea
where $u^{\nu}$ represents the four-velocity of the fluid and $g_{\mu \nu}$ is our spacetime metric tensor. This fluid has a characteristic equation of state given by $w=P/\rho$, which is the center of discussion throughout this paper. In the second order in perturbation theory, the scalar and tensor modes show mixing. We require solving the first-order equations of motion to solve for the induced GWs in the second order. The solutions then obtained act as a source for the GW; hence, SIGWs.

We briefly mention here the analysis of GWs and gravitational scalar potential when examined in the leading order. For both the quantities, in the absence of any anisotropies, are shown to satisfy the following equations:
\bea
\Phi'' + \frac{6(1+w)}{\tau(1+3w)}\Phi' + wk^{2}\Phi = 0, \quad\quad h''_{\lambda} + \frac{4}{\tau(1+3w)}h'_{\lambda} + k^{2}h_{\lambda} = 0.
\eea
for the two modes of polarization $\lambda=+,\times$. We now quote the solutions for the above equations \cite{Domenech:2021ztg,Domenech:2019quo}:
\bea \label{O1scalartensor}
\Phi(x) = (\sqrt{w}x)^{-s}\left[C_{1}(k)J_{s}(\sqrt{w}x) + C_{2}(k)Y_{s}(\sqrt{w}x)\right], \quad\quad 
h_{\lambda}(x) = x^{-t}\left[C_{1}(k)J_{t}(x) + C_{2}(k)Y_{t}(x)\right]
\eea
where $x\equiv k\tau$ and $\{J_{\alpha},Y_{\alpha}\}$ refer to the Bessel functions of the first and second kind respectively. The indexes $s,t$ further have the following $w$-dependent form:
\bea
s = \frac{5+3w}{2(1+3w)}, \quad\quad\quad t = \frac{3(1-w)}{2(1+3w)},
\eea
such that the solutions mentioned carry the effects of damped oscillations when modes become sub-Horizon. The equality $1+t=s$ will later become helpful, and $t$ should not be confused with any time variable. These solutions will later be essential when understanding the general solution for the tensor modes at the second order. Now, we discuss the induced GW scenario.  

We start with the Fourier space version of the equation of motion for the tensor modes obtained at second order in perturbation theory:
\bea \label{tensoreqn}
h''_{\lambda} + 2 {\cal H}h'_{\lambda}+k^2 h_{\lambda} = S_{\lambda}(\mbf{k}),
\eea
where $S_{\lambda}(\mbf{k})$ represents the source term given by:
\bea \label{source}
S_{\lambda} = 4 \int \frac{d^3q}{(2\pi)^3}e_{\lambda}^{ij}(k)q_{i}q_{j}\left\{\Phi_{\mbf{q}}\Phi_{\mbf{k-q}} + \frac{1+b}{2+b}\bigg[\Phi_{\mbf{q}}+\frac{\Phi'_{\mbf{q}}}{{\cal H}}\bigg]\;\bigg[\Phi_{\mbf{k-q}}+\frac{\Phi'_{\mbf{k-q}}}{{\cal H}}\bigg]\right\},
\eea
where $\Phi_{\mbf{q}}$ are the Fourier components of the scalar potential, and $b$ is a $w$-dependent combination given by:
\bea 
b= \frac{1-3w}{1+3w},
\eea
Also present are the polarization tensors of GWs given by:
\bea
e_{ij}^{+}(\mbf{k})= \frac{1}{\sqrt{2}}\bigg(e_{i}(\mbf{k})e_{j}(\mbf{k})-\bar{e}_{i}(\mbf{k})\bar{e}_{j}(\mbf{k})\bigg),\quad\quad
e_{ij}^{\times}(\mbf{k})= \frac{1}{\sqrt{2}}\bigg(e_{i}(\mbf{k})\bar{e}_{j}(\mbf{k})+\bar{e}_{i}(\mbf{k})e_{j}(\mbf{k})\bigg).
\eea
for the two modes of polarization.  Using the Green's function method one can obtain the solution for the tensor modes $h_{\lambda}(\tau)$ from eqn.(\ref{tensoreqn}) as:
\bea
h_{\lambda}(\tau) = \int_{\tau_i} ^{\tau} d\Tilde{\tau}\;{\cal G}(\tau,\Tilde{\tau})S_{\lambda}(\Tilde{\tau}).
\eea
The solution is presented with the initial conditions at time $\tau_i$ obeying $h_{\lambda}(\tau_i)= h_{\lambda}^{'}(\tau_i)=0$. 
The ultimate goal is to find the expression for the SIGW power spectrum for which we need the two-point correlation function of the tensor modes, which is presented as:
\bea
\big<h_{\mbf{k}}(\tau)h_{\mbf{k'}}(\tau)\big> = \int_{0}^{\tau}\;d\tau_1\;\int_{0}^{\tau}\;d\tau_2 \;{\cal G}(\tau,\tau_1)G(\tau,\tau_2)\big<S_{\lambda}(k,\tau_1)S_{\lambda}(k',\tau_{2})\big>.
\eea
This requires further knowledge of the two-point correlation function of the source term which can be computed by (excluding non-Gaussian features of the primordial power spectrum):
\bea
\langle S_{\lambda}(\mbf{k},\tau_{1})S_{\lambda}(\mbf{k'},\tau_{2})\rangle &=& 16 \int \frac{d^3 q}{(2\pi)^3} \int \frac{d^{3}q'}{(2\pi)^3} e_{\lambda}^{ij}(\mbf{k})q_{i}q_{j}e_{\lambda}^{ij}(\mbf{k'})q'_{i}q'_{j} \times f(\tau_{1},q,k)\; f(\tau_{2}, q', k')\langle\phi_{\mbf{q}}\phi_{\mbf{k-q}}\phi_{\mbf{q'}}\phi_{\mbf{k'-q'}}\rangle,
\eea
where we implement the split of the Fourier modes of scalar potential into the primordial fluctuations $\phi_{\mbf{q}}$ and the transfer function $\Phi(q\tau)$. Such a split enables the use of the primordial scalar power spectrum and information about the evolution of the scalar potential, respectively. The function $f(\tau,q,k)$ above refers to the source function in eqn.(\ref{source}) written in terms of the previously mentioned transfer function for the gravitational potential. To completely determine the tensor power spectrum requires taking the necessary Wick contractions between the scalar fluctuations in the RHS of above. The definition for the dimensionless tensor power spectrum that we use here is as follows: 
\bea
\langle h_{k}(\tau)h_{k'}(\tau)\rangle = \frac{2\pi^{2}}{k^3}\Delta^{2}_{h}(k,\tau)\;\delta^3\left(\mbf{k+k'}\right).
\eea
Now, by using these above two relations we can write the primordial power spectrum of the induced GW as follows:
\bea \label{tensorpspec}
\Delta^{2}_{h}(k,\tau)&=& 4\int_{0}^{\infty}\;dv\int_{|1-v|}^{1+v}\;du\;\bigg(\frac{4v^2-(1+v^2-u^2)^2}{4uv}\bigg)^2\; {\cal I}^2 (u,v,x)\;\Delta^{2}_{\zeta}(ku)\;\Delta^{2}_{\zeta}(kv), \eea
where we have introduced to new variables $u$ and $v$, which are defined by the following expressions:
\bea v\equiv \frac{q}{k},  \quad \quad \quad u \equiv \frac{|\mbf{k-q}|}{k}.
\eea
The function ${\cal I}$ here represents the kernel which we are going to discuss in detail in the upcoming section. We present here the expression of the kernel in terms of the Green's and the source functions as follows.
\bea \label{kernel}
{\cal I}(u,v,x)= 2\bigg(\frac{3+3w}{5+3w}\bigg)^2 \int_{0}^{x} d\Tilde{x}\;{\cal G}(x,\Tilde{x})f(\Tilde{x},u,v),
\eea
where the Green's function ${\cal G}$ and the source function $f$ are found, after utilizing the eqn.(\ref{O1scalartensor}), simplifies to: 
\bea \label{simplG}
{\cal G}(x,\Tilde{x}) &=& \frac{\pi \Tilde{x}^{s}}{2x^{t}}[J_{t}(\Tilde{x})Y_{t}(x)-Y_{t}(\Tilde{x})J_{t}(x)], \\
\label{sourcef}
f(\Tilde{x},u,v) &=& \frac{4^{s}}{6s}\Gamma^2
[1+s]\frac{1+3w}{1+w}(uvc^{2}_{s}\Tilde{x}^2)^{1-s} \nonumber \\
&&\quad \quad \quad \quad \quad \quad \quad \quad \times \Bigg[J_{s-1}(uc_{s}\Tilde{x})J_{s-1}(vc_{s}\Tilde{x})+ \frac{3(1+w)}{2}J_{s+1}(uc_{s}\Tilde{x})J_{s+1}(vc_{s}\Tilde{x})\bigg].\eea
Here the following relation for the Bessel functions is used: $J_{s-1}(x)+J_{s+1}(x) = (2s/x)J_{s}(x)$ to write $J_{s}$ in terms of $J_{s\pm 1}$. This simplified source term is crucial to derive the analytical expression of the kernel for a general $w$ which is presented in the following section.

\subsection{Semi-Analytical computation of the transfer function}

The kernel from eqn.(\ref{kernel}) present in the power spectrum eqn.(\ref{tensorpspec}) can be evaluated numerically, but we can simplify the calculation by obtaining an analytic expression for the kernel (or transfer) function. We have followed the approach performed in \cite{Domenech:2021ztg}. Now, substituting in the relations from eqs.(\ref{simplG},\ref{sourcef}) into eqn.(\ref{kernel}), we get the following simplified expression of the kernel:
\bea \label{simplekern}
{\cal I}(u,v,x)= 4^{t}\frac{3\pi}{2s^{3}}\frac{1+w}{1+3w}\Gamma^{2}(t+2)(c^{2}_{s}uvx)^{-t}\{Y_{t}(x)I_{J}^{x}(u,v,w)-J_{t}(x)I_{Y}^{x}(u,v,w)\}
\eea
where the other new integrals in the RHS are defined as follows:
\bea \label{besselprod}
I^{x}_{B}(u,v,w) = \int^{x}_{0}d\tilde{x}\;\tilde{x}^{1-t}B_{t}(\tilde{x})\left\{J_{t}(uc_{s}\tilde{x})J_{t}(vc_{s}\tilde{x}) + \frac{3(1+w)}{2}J_{t+2}(uc_{s}\tilde{x})J_{t+2}(vc_{s}\tilde{x}) \right\} 
\eea
where $B$ labels the use of the two Bessel functions: $B=J,\;Y$ both of order $t$ and $t+2$ present. The above integral is not possible to give analytic results for arbitrary values of $x$; however, in the limit $x \gg 1$, this integral remains analytically solvable, which also corresponds to the scales deep inside the horizon, and hence, we work within this regime for our future analysis. For the integral, this means pushing the upper limit to $x\rightarrow \infty$ where we can gather the leading order effects that suffice for our purpose of studying the induced GWs. Such integrals have already been examined to give analytic results in terms of the Legendre and associated Legendre polynomials by Gervois and Navelet \cite{gervois1985integrals}. We now mention the resulting form of the kernel with the form of the two integrals written in the sub-Horizon regime ($x\gg 1$):
\bea
I_{J}(u,v,w,c_{s}) &=& \cos{\left(x-\frac{b\pi}{2}\right)}\bigg(P_{t-1/2}^{-t+1/2}(y) + \frac{3(1+w)}{2}P_{t+3/2}^{-t+1/2}(y)\bigg)\Theta(c_{s}(u+v)-1), \nonumber\\
I_{Y}(u,v,w,c_{s}) &=& \frac{2}{\pi}\sin{\left(x-\frac{b\pi}{2}\right)}\bigg[\bigg\{Q_{t-1/2}^{-t+1/2}(y) + \frac{3(1+w)}{2}Q_{t+3/2}^{-t+1/2}(y)\bigg\}\Theta(c_{s}(u+v)-1)\\
&& \quad\quad\quad\quad\quad\quad -\bigg\{ {\cal Q}_{t-1/2}^{-t+1/2}(-y) + 3(1+w){\cal Q}_{t+3/2}^{-t+1/2}(-y)\bigg\}\Theta(1-c_{s}(u+v)) \bigg],\\
{\cal I}(u,v,w,c_{s},x\gg 1) &=& x^{-(t+1/2)}\frac{4^{t}s}{3uvc^{2}_{s}}\frac{1+3w}{1+w}\Gamma^{2}(t+1)\bigg(\frac{Z}{2uv}\bigg)^{t-1/2}\times\bigg(I_{J}(u,v,w,c_{s})+I_{Y}(u,v,w,c_{s})\bigg).
\eea
where $2t = 2b+1$ and we have introduced two new variables, $Z$, and $y$, which are defined by the following expressions: 
\bea 
Z^{2}=4u^{2}v^{2}(1-y^{2}), \quad\quad y = -1+\frac{c_{s}^{2}(u+v)^{2}-1}{2c_{s}^{2}uv},
\eea 
and expansion of the Bessel functions for large arguments is used. The above results have some crucial properties that we may now discuss. For such sub-Horizon integrals, $P_{\alpha}^{\beta}(y)$  and $Q_{\alpha}^{\beta}(y)$ are the Ferrer's functions of the first and the second kind valid for particular values of $|y|<1$, while ${\cal Q}_{\alpha}^{\beta}$ is the Olver's function, which is an associated Legendre polynomial of the second kind, and has solutions for $|y|>1$. There exists a resonant condition from the argument of the Heaviside theta, when $c_{s}(u+v)=1$, which also corresponds to the case where the wavenumbers for the tensor mode equals the sum of two scalar modes. The Heaviside Theta, $\Theta(c_{s}(u+v)-1)$, helps us to separate the cases different from the resonance.
Neglecting any non-Gaussian contributions for now, we directly present the final result of the kernel after assuming mostly Gaussian fluctuations and this requires taking the square and oscillation averaged value from the integrals which ultimately gives us:
\bea \label{kernelavg}
\overline{{\cal I}^{2}(u,v,w,c_{s},x)} &=& x^{-(2t+1)}\frac{4^{2t}s^{2}}{9u^{2}v^{2}c_{s}^{4}}\frac{1+3w}{1+w}\Gamma^{4}(t+1)\bigg(\frac{Z}{2uv}\bigg)^{2t-1}\nonumber\\
&& \quad\quad\quad\quad \times\bigg\{\bigg(P_{t-1/2}^{-t+1/2}(y) + \frac{3(1+w)}{2}P_{t+3/2}^{-t+1/2}(y)\bigg)^{2}\Theta(c_{s}(u+v)-1) \nonumber\\
&& \quad\quad\quad\quad + \frac{4}{\pi^{2}}\bigg(Q_{t-1/2}^{-t+1/2}(y) + \frac{3(1+w)}{2}Q_{t+3/2}^{-t+1/2}(y)\bigg)^{2}\Theta(c_{s}(u+v)-1) \nonumber\\
&& \quad\quad\quad\quad + \frac{4}{\pi^{2}}\bigg({\cal Q}_{t-1/2}^{-t+1/2}(-y) + 3(1+w){\cal Q}_{t+3/2}^{-t+1/2}(-y)\bigg)^{2}\Theta(1-c_{s}(u+v)) \bigg\}
\eea
This expression is of main importance for our purposes to calculate the induced GW spectrum for a general EoS background.

We emphasize the fact about notation $c_{s}$, introduced after Sec.\ref{s7} and during the derivation of the transfer function, represents the speed of propagation in the hypothetical cosmological fluid with EoS $w$ where the metric fluctuations propagate and eventually take part in the evolution of the tensor modes. During this, the previous Galileon degrees of freedom collapse to the component(RD, MD, etc.), which has the new effective $c_{s}$. The same notation in the context of discussion present for the Galileon in the first half of this work, till Sec.\ref{s5}, represents the effective sound speed $c_{s}$ for the Galileon field dominating before the collapse occurs.

\subsubsection{Special Case I: Radiation Domination}

For the radiation-dominated epoch, we have $w=1/3$ which implies the value of $b=0$. Putting this in eqn.(\ref{kernelavg}) and taking appropriate approximations we obtain the following transfer function for radiation epoch :
\bea \label{rdtransfer}
{\cal T}_{\rm RD} (u,v, w=1/3,c_{s})&=&
\bigg[\frac{4v^2 - (1-u^2 +v^2)^2}{4u^2v^2}\bigg]^{2}\times\overline{{\cal I}^{2}_{\rm RD} (u,v, w=1/3,c_{s})} \nonumber\\
&=& \frac{y^2}{3c_s ^4}\bigg[\frac{4v^2 - (1-u^2 +v^2)^2}{4u^2v^2}\bigg]^2 \times \bigg[\frac{\pi ^2 y^2}{4}\Theta[c_s(u+v)-1]+\bigg(1-\frac{1}{2}y \ln \bigg|\frac{1+y}{1-y}\bigg|\bigg)^2\bigg].
\eea
this result gets used in eqn.(\ref{tensorpspec}) to provide the averaged tensor power spectrum and therefore the density of GWs. This result correctly matches with the kernel obtained in \cite{Kohri:2018awv}.

\subsubsection{Special Case II: Matter Domination}

For the matter-dominated epoch, we have $w=0$ which implies $b=1$. This is also known to describe a pressure-less fluid background and a scalar field oscillating coherently around the bottom of a potential can be categorized with this condition. The kernel for this case after using the mentioned values for $b,w$ in eqn.(\ref{kernelavg}) reduces to give:
\bea
{\cal T}_{\rm MD}(u,v,w=0,c_{s})&=& 
\bigg[\frac{4v^2 - (1-u^2 +v^2)^2}{4u^2v^2}\bigg]^{2}\times\overline{{\cal I}^{2}_{\rm MD} (u,v, w=0,c_{s})} \nonumber\\
&=&\frac{3^3 5^2}{2^{14}c_s ^4}\bigg[\frac{4v^2-(1-u^2+v^2)^2}{4u^2v^2}\bigg]^2  \times \bigg[\frac{\pi ^2}{4}(1-y^2)^2 (1+3y^2)^2 \Theta[c_s (u+v)-1]\nonumber \\
&& \quad \quad \quad \quad +\bigg(y(1-3y^2)-1/2(1+2y^2-3y^4)\ln{\abs{\frac{1+y}{1-y}}}\bigg)^2 \bigg].
\eea

\subsubsection{Special Case III: Kinetic Domination}

For the kinetic domination epoch, we have $w=1$ which gives us $b=0$. This corresponds to a scenario where the scalar field follows a steep potential leading to the kinetic energy dominating the potential energy. Using this values in eqn.(\ref{kernelavg}) the final form of the kernel looks like:
\bea
{\cal T}_{\rm KD}(u,v,w=1,c_s)&=&
\bigg[\frac{4v^2 - (1-u^2 +v^2)^2}{4u^2v^2}\bigg]^{2}\times\overline{{\cal I}^{2}_{\rm KD} (u,v, w=1,c_{s})} \nonumber\\
&=&\frac{4}{3 \pi c_s^4 \abs{1-y^2}}\bigg[\frac{4v^2 - (1-u^2 +v^2 )^2}{4u^2v^2}\bigg]^2 \nonumber \\
&& \times \bigg[(1+3y^2)\; \Theta[c_s(u+v)-1]+\big(1-3y^2+3y\sqrt{\abs{1-y^2}}\big)^2\; \Theta[1-c_s(u+v)]\bigg].
\eea

\subsubsection{Special Case IV: Soft Fluid Domination}

For the soft fluid scenario, we have $w=1/9$ which gives us $b=1/2$. The kernel in eqn.(\ref{kernelavg}) gives us the reduced form:
\bea
{\cal T}_{\rm SFD}(u,v,w=1/9,c_{s})&=&
\bigg[\frac{4v^2 - (1-u^2 +v^2)^2}{4u^2v^2}\bigg]^{2}\times\overline{{\cal I}^{2}_{\rm SFD} (u,v, w=1/9,c_{s})} \nonumber\\
&=&\frac{2^8}{3^8 \pi c_s^4}\bigg[\frac{4v^2 - (1-u^2+v^2)^2}{4u^2v^2}\bigg] \times \bigg[(4+45y^2)\; \Theta [c_s(u+v)-1]\nonumber \\
&& \quad \quad \quad \quad +\big(y(3-10y^2)+(2+10y^2)\sqrt{\abs{1-y^2}}\big)^2 \Theta[1-c_s(u+v)]\bigg].
\eea

\subsubsection{Special Case V:  Negative EoS Fluid Domination}

The negative EoS, $w<0$, condition is also a possible scenario for consideration in view of the induced GW production. We mention a specific example related to this case where $w=-1/9$ which gives $b=2$ and these values in eqn.(\ref{kernelavg}) reduces the kernel to the readable form:
\bea
{\cal T}_{\rm NEoS}(u,v,w=-1/9,c_{s}) &=&
\bigg[\frac{4v^2 - (1-u^2 +v^2)^2}{4u^2v^2}\bigg]^{2}\times\overline{{\cal I}^{2}_{\rm NEoS} (u,v, w=-1/9,c_{s})} \nonumber\\
&=&\frac{5^2 7^2}{2^8 3^9 c_s ^4}\bigg(\frac{4v^2 -(1-u^2+v^2)^2}{4u^2 v^2}\bigg)^2 \times \bigg[\frac{225 \pi^2}{4}(1-y^2)^2 (1+y^2-2y^4)^2 \;\Theta [c_s(u+v)-1] \nonumber \\ 
&& \quad \quad \quad \quad \quad \quad + \bigg(y(9+35y^2 - 30y^4)+\frac{15}{2}(1-3y^4+2y^6)\ln{\abs{\frac{1+y}{1-y}}\bigg)^2}\bigg].
\eea
We clarify here that we intend to refrain from explicitly mentioning the results for this case since we found it did not provide any relevant information when confronted with the NANOGrav15 signal. Thus, we choose to ignore these results and only mention them here for the sake of completeness.

\subsection{Computing $w$-SIGW abundance}

In this section, we use the general expressions for the kernel(or transfer) function mentioned in the previous sections to study the production of SIGWs by examining their resulting spectrum. The $w$-SIGW phenomenon corresponds to an arbitrary background EoS $w$, which is assumed during the last stages of inflation before the Universe ultimately recovers the hot Big Bang scenario. We use this situation for our Galileon inflation setup to visualize the induced GW spectrum via sourcing from the scalar modes sub-Horizon during such a general cosmological background. On that note, the spectrum of the induced GW for a general $w$ background is written as follows:
\bea
\label{GWdensity}
\Omega_{\rm{GW},0}h^2 = 1.62 \times 10^{-5}\;\bigg(\frac{\Omega_{r,0}h^2}{4.18 \times 10^{-5}}\bigg) \bigg(\frac{g_{*}(T_c)}{106.75}\bigg)\bigg(\frac{g_{*,s}(T_c)}{106.75}\bigg)^{-4/3}\Omega_{\rm GW,c},
\eea
where $\Omega_{r,0}h^2$ represents the radiation energy density as observed today, and $g_{*},g_{*,s}$ are the energy and entropy effective degree of freedom. The quantity $\Omega_{\rm GW,c}$ represents the GW energy density fraction during the radiation-dominated epoch where, the moment denoted here by ``c'', such induced GW behave as freely propagating GWs.
\textcolor{black}{Both the functions $g_{*}(T_c)$ and $g_{*,s}(T_c)$ change with temperature and can affect the induced GWs, which re-enter the Horizon at a particular moment in time before BBN. The most drastic changes in these degrees of freedom occur around $T\gtrsim 0.1{\rm GeV}$ due to the QCD crossover experienced by the standard model radiation bath. This regime is also crucial in the present context since the corresponding GWs generated lie within the interval, $f\sim 10^{-9}-10^{-8}{\rm Hz}$, that coincides with the low-frequency, or infrared, tail of the signal probed by PTAs. To precisely estimate the functions $g_{*},g_{*,s}$ during the crossover requires techniques of lattice QCD simulations \cite{Saikawa:2018rcs}. In section \ref{s9}, we show how including such effects can alter the nHz tail of the obtained induced GWs. }

We further express the energy density $\Omega_{\rm GW,c}$ using the kernel functions for the modes, with scales satisfying $k \geq k_{*}$, in the manner as:
\bea
\label{omegac}
\Omega_{\rm {GW},c}&=& \frac{k^{2}}{12a^{2}H^{2}}\times\overline{\Delta^{2}_{h}(k,\tau)}  \nonumber\\
&=& \bigg(\frac{k}{k_{*}}\bigg)^{-2b}\int_{0}^{\infty}dv \int_{|{1-v}|}^{1+v} du \; {\cal T}(u,v,w,c_s) \;\;\Big[\Delta^{2}_{\zeta}(ku)\Big]_{\bf Total} \times \Big[\Delta^{2}_{\zeta}(kv)\Big]_{\bf Total},
\eea
with the transfer function for a constant EoS background and having a constant speed of propagation $c_{s}$. The value $k_{*}$ refers to the pivot scale in our setup using which we conduct our upcoming analysis of the GW spectrum. For sake of clarity, we mention the final version of the transfer function used above, ${\cal T}(u,v,w,c_{s})$, which results from the use of eqn.(\ref{kernelavg}) and adjusting for other multiplicative factors present in eqn.(\ref{tensorpspec}):
\bea \label{transfer}
{\cal T}(u,v,w,c_{s}) &=& (b+1)^{-2(b+1)}\frac{4^{2b}}{3c^{4}_{s}}\bigg[\frac{3(1+w)}{1+3w}\bigg]^{2}\Gamma^{4}(b+3/2)\bigg[\frac{4v^2 - (1-u^2 +v^2 )^2}{4u^2v^2}\bigg]^{2}\bigg(\frac{Z}{2uv}\bigg)^{2} \nonumber\\
&& \quad\quad\quad\quad \times\bigg\{\bigg(P_{-b}^{b}(y) + \frac{3(1+w)}{2}P_{b+2}^{-b}(y)\bigg)^{2}\Theta(c_{s}(u+v)-1) \nonumber\\
&& \quad\quad\quad\quad + \frac{4}{\pi^{2}}\bigg(Q_{-b}^{b}(y) + \frac{3(1+w)}{2}Q_{-b}^{b+2}(y)\bigg)^{2}\Theta(c_{s}(u+v)-1) \nonumber\\
&& \quad\quad\quad\quad + \frac{4}{\pi^{2}}\bigg({\cal Q}^{-b}_{b}(-y) + 3(1+w){\cal Q}^{-b}_{b+2}(-y)\bigg)^{2}\Theta(1-c_{s}(u+v)) \bigg\}.
\eea
We use the above transfer function with various cases of $b$ or $w$ values to evaluate the GW density and analyze the behaviour of the spectrum for various EoS scenarios. Before discussing further results, we mention an important point regarding the applicability of this theory of generating induced GWs from a general $w$ background within the present Galileon EFT setup. In the beginning of Sec.\ref{s3}, we mentioned the importance of working with the decoupling limit for studying the self-interactions in Galileon theory. As a consequence of this limit, we can successfully neglect the mixing effects from the gravity sector, which complicates the overall analysis, and that would demand an entirely different formulation of the evolution of tensor modes in induced GWs. Hence, the decoupling limit saves us by focusing on the Galileon self-interactions independently rather than including couplings with the gravity sector, thus validating the above analysis for the metric perturbations and its application with the Galileon EFT.  

\section{PBH overproduction issue and its resolutions} \label{s8}

In the previous sections, we have attempted to discuss the generation of induced GWs within a general EoS $w$ setting and identified its energy density spectrum. The significant enhancements due to the couplings of scalar modes in the very early Universe can be linked to having a PBH counterpart. The enhanced power spectrum sensitive to the scales of NANOGrav15 can also lead to the production of near solar-mass PBH; thus, the detection of SIGW can also signal possible PBH production in the very early Universe. 

\subsection{What is PBH overproduction?}

Recently, there has been much development towards the problem concerning PBH overproduction. The SGWB signal released by the PTA collaborations was examined by many for its possible astronomical or cosmological origins. From the various possible scenarios, the SIGW model is one of those that fit well with the PTA data, but then soon, it was also suggested that this scenario may suffer from the problem of overproduction. The explanation and resolutions for this issue have been actively pursued by many; consider refs.\cite{Ferrante:2023bgz,Franciolini:2023pbf,Gow:2023zzp,Gorji:2023sil,Firouzjahi:2023xke} for respective literature. The crux of this problem concerns the fact that achieving a sizeable abundance of near solar-mass PBH that corresponds to the frequencies probed by the NANOGrav15 signal leads to the ultimate disagreement of the SIGW interpretation of the very SGWB signal confirmed by the PTA collaborations. Forcing greater statistical agreement with PTA will directly lead to a breakdown of perturbation theory due to the enhanced amplitude becoming $\sim {\cal O}(1)$ during the regime associated with PBH production. The above makes the issue more troublesome from the perspective of matching closely with the data. It calls for either an alternative that can better approximate the SIGW interpretation of the data or provide an alternative interpretation of the signal itself, ultimately evading the overproduction. To make the claims about overproduction more robust, the authors in \cite{Franciolini:2023pbf,Ferrante:2022mui} include the impact of non-Gaussianities and the inclusion of non-linearities in the density contrast present when in the super-Horizon regime. Such theoretically necessary factors and a thorough analysis between multiple models make the issue more rigid and demand further attention. Other fundamental questions exist regarding the exact theoretical calculations for the PBH abundance, including the proper threshold range of density contrast and higher-order estimates of abundance in the super-Horizon. However, these still need to be concretely established and are expected not significantly to alter the existing analysis and their results.

\subsection{Possible resolutions}

As mentioned above, many authors have recently sought to provide attractive resolutions to the overproduction issue for PBHs. These resolutions include a possible curvaton scenario that exists as a spectator field in the very early Universe and is responsible for generating the curvature perturbations at PBH-relevant scales. Also, there is the possibility of an extra tensor spectator field inducing the observed signal. The presence of non-attractor features in the theory of concern is also crucial for having abundant PBH production by the involvement of non-Gaussianities, which can become highly sensitive to the scalar power spectrum amplitude. The non-Gaussian features of the primordial density perturbations are of significant importance when explaining PBH production which we emphasize in this section. Also, since we are dealing with an arbitrary EoS parameter, the motivation for choosing this scenario, given the overproduction issue, will be discussed in this section.

\subsubsection{Using non-Gaussianity}

One of the typical scenarios
includes the addition of an ultra-slow roll (USR) phase to inflation, which brings with it large quantum fluctuations generating the primordial curvature perturbations that later  collapse to become PBHs, or in some cases, give rise to induced GWs. Although, for simplicity purposes, we take the power spectrum profile for the curvature perturbations to be Gaussian in nature, including non-Gaussianity provides for a detailed and comprehensive analysis. For instance, non-Gaussianity has a considerable impact on the tail of the probability distribution function (PDF) \cite{Kawaguchi:2023mgk,DeLuca:2022rfz,Taoso:2021uvl,Atal:2018neu,Young:2013oia,Byrnes:2012yx,Bullock:1996at} 
which is required to compute the PBHs abundance. From Ref.\cite{Pi:2022ysn}, a general analysis on the PDFs of single-field inflation reveals a logarithmic relation appearing for the fluctuations in e-folds, $\delta N$, in the presence of a non-attractor regime during inflation and dominance of such logarithms in $\delta N$ can quickly make the PDF develop exponential tails which can significantly affect the PBH mass fraction. In Ref.\cite{Choudhury:2023fwk}, within the framework of Galileon theory, we  worked with the threshold statistics of the compaction function in linear cosmological perturbation theory to address the non-linearity and non-Gaussianity associated with the curvature perturbations. We considered the large negative local non-Gaussianity associated with the primordial perturbations. Our findings suggest that PBH masses up to $M_{\rm PBH} \sim {\cal O}(10^{-3}-10^{-2} M_{\odot})$ could avoid PBH overproduction while considering $f_{\rm NL} \sim -6$.

\subsubsection{Using Equation of State Parameter (EoS)}

This provides an exciting ground for analysis whereby incorporating various constant EoS parameters allows us to navigate through possible scenarios of PBH formation. In this work, we have examined cases with discrete values of the EoS parameter $w$, namely $w\in (-5/9, 1/3)$, which results from working within the linear regime of the perturbation theory. Also, we focus our discussions on the scenarios where for the speed of propagation in eqn.(\ref{transfer}) we keep $c_{s}^{2}=1$. The choice for $c_{s}$ was made clear at the beginning of Sec.\ref{s6}. Until now, the case of an RD era has commonly been chosen given the well-known PBH formation mechanism coming from critical collapse theory. However, we make use of the flexibility of not knowing about the EoS in the very early Universe to investigate the supposed effects of the large fluctuations entering into such an era on the induced GWs and corresponding PBH production. By confronting the results of our analysis with the recent data, we can get a more robust picture of the cosmological background where many cosmological processes can take place.  Through our work with Galileon, we can make the case for $w=1/3$ more robust compared to the effects of other values of $w$. The significance of this analysis is directly felt on the SIGW spectrum and mass fraction for the near solar-mass PBH.

\begin{figure*}[htb!]
    	\centering
    \subfigure[]{
      	\includegraphics[width=8.5cm,height=7.5cm]{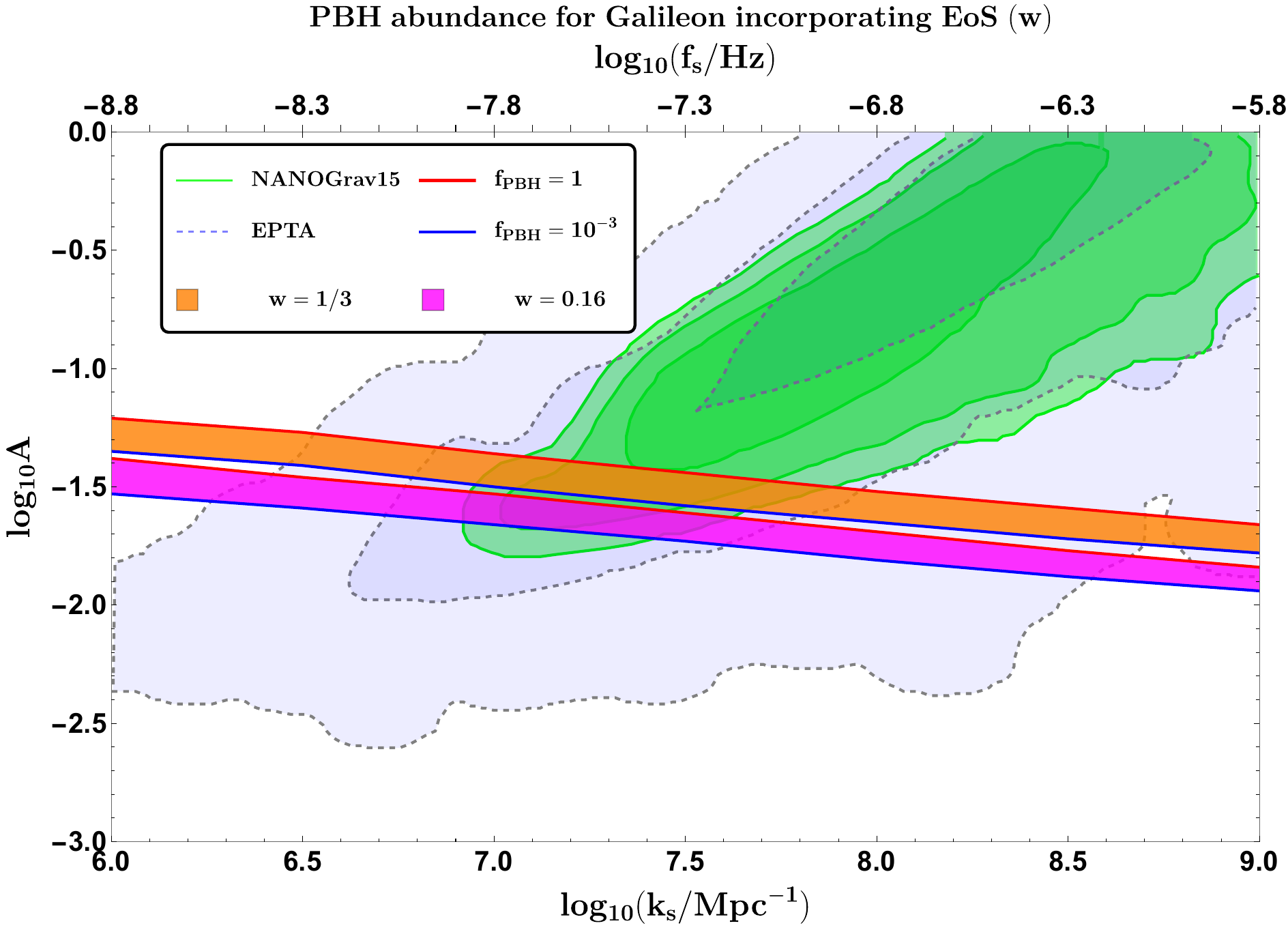}
        \label{wgalfpbh}
    }
    \subfigure[]{
       \includegraphics[width=8.5cm,height=7.5cm]{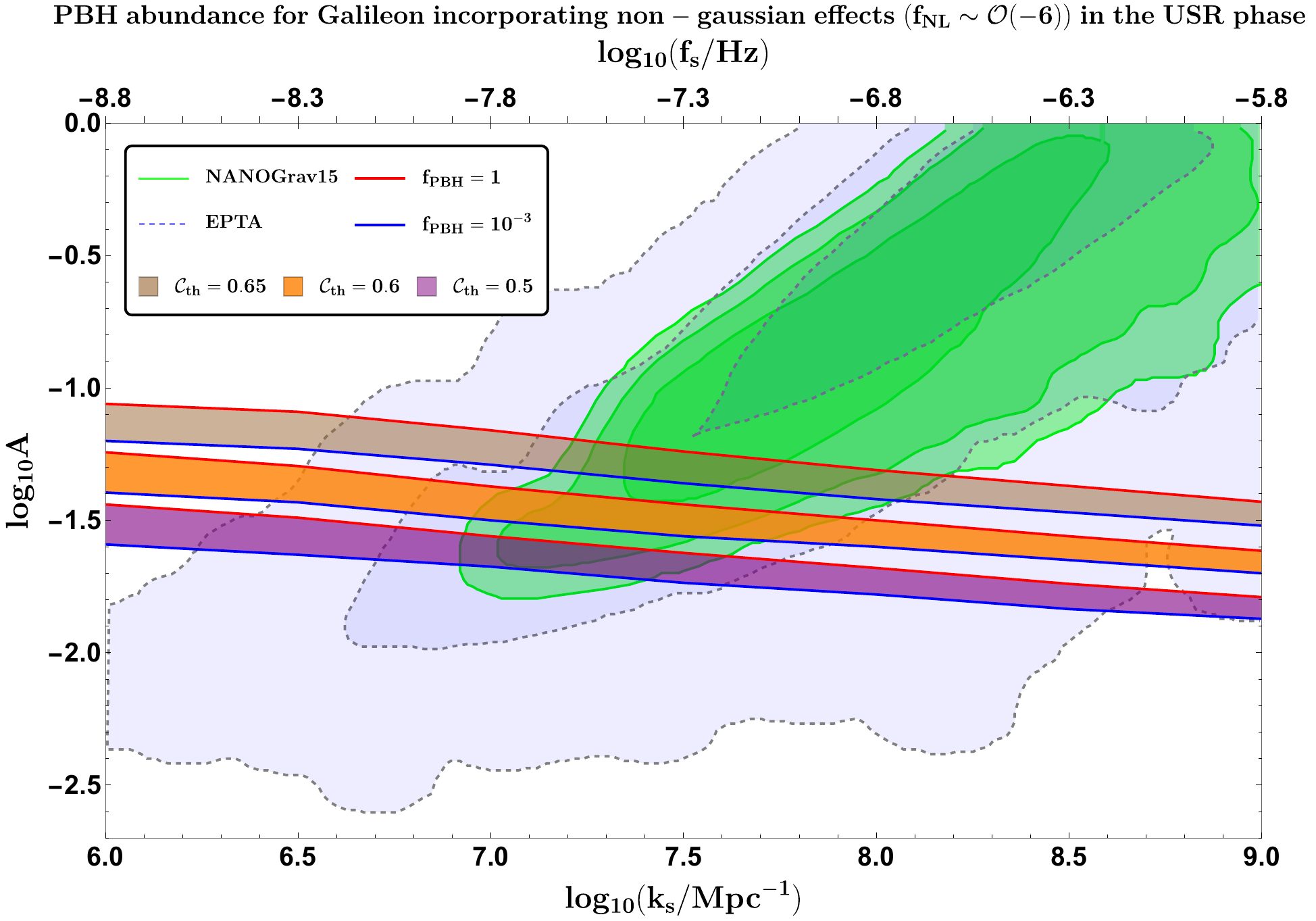}
        \label{overprod}
    }
    	\caption[Optional caption for list of figures]{Comparison between using non-Gaussianity and implementing EoS parameter to resolve PBH overproduction. \textit{Left panel}: The orange and magenta coloured bands represent regions of sizeable PBH abundance, $f_{\rm PBH} \in (10^{-3},1)$, for EoS parameters $w=1/3$ and $w=0.16$, respectively. \textit{Right panel}: Non-gaussian effects generated in the USR along with various values of the compaction function threshold, ${\cal C}_{\rm th} \in \{0.65,0.6,0.5\}$, are included to achieve $f_{\rm PBH} \in (10^{-3},1)$. The red and blue lines refer to $f_{\rm PBH}=1$ and $f_{\rm PBH}=10^{-3}$, respectively. The background green and light blue filled contours, taken from \cite{Franciolini:2023pbf}, represent the sensitivity curves for the NANOGrav15 and EPTA signals, respectively. } 
    	\label{compare1}
    \end{figure*}

To make a comparative analysis between the two resolution methods discussed above we present the fig.(\ref{compare1}). The left panel presents the results of this work, where we consider different EoS values to analyze the amplitude of the total scalar power spectrum related to a sizeable abundance. We find the EoS parameter $w$ to lie within $w \sim (0.16,1/3)$ such that we achieve $f_{\rm PBH} \in (10^{-3},1)$ and the overproduction avoiding scenarios lie within $1\sigma-2\sigma$ when confronted with NANOGrav15. For $w$ values lower than the above range, we do not obtain sufficient abundance for PBH masses corresponding to the scales of the NANOGrav15 signal. We use the results from \cite{Choudhury:2023fwk} in the right panel for a comparative study with the present analysis. For the results in the right, we include the effects of non-Gaussianity and implement the compaction function approach to estimate the PBH abundance. There, the two factors, namely the compaction threshold and the non-Gaussianity amplitude, are crucial to evaluating $f_{\rm PBH}$. We notice that the scalar power spectrum amplitude is quite sensitive to small changes in the threshold, and the extreme value of ${\cal C}_{\rm th}=0.65$ provides the best agreement with the NANOGrav15 data. Using both methods for the resolution, we find that regions avoiding overproduction are not removed beyond $2\sigma$ while including large negative non-gaussian effects can provide a better way to interpret the observed signal and simultaneously avoid overproduction.

We want to highlight the important correspondence between the EoS $w$ and the compaction threshold ${\cal C}_{\rm th}$. Recall from the relation, $\delta_{\rm th} = 3(1+w)/(5+w)$, the connection between the parameter $w$ and the threshold $\delta_{\rm th}$ of the density contrast. The analysis we perform with different values of $w$ also considers values for the density contrast. For example, $w=1/3$ corresponds to $\delta_{\rm th}=2/3$, $w=0.25$ corresponds to $\delta_{\rm th}=0.652$, $w=0.16$ corresponds to $\delta_{\rm th}=0.635$, and so on. On the other hand, when dealing with non-linearities and the compaction function formalism, we require different values of the compaction threshold, ${\cal C}_{\rm th}$, such as ${\cal C}_{\rm th} = \{0.65,\;0.6,\;0.5\}$ as chosen in the right panel of fig.(\ref{compare1}). Hence, effectively speaking, we are working with changes in two different threshold variables: (a) the density contrast threshold when considering the linear regime approximation on super-Hubble scales, see eqn.(\ref{deltalinear}), where non-Gaussianities are absent, and (b) the compaction threshold when including the non-linearities present in the super-Hubble and the non-Gaussianities, $f_{\rm NL}$, coming as a part of this in the comoving curvature perturbation. Also crucial is to notice that the equality between the compaction threshold and the density contrast threshold can only be established once the PBH forming modes re-enter into the Horizon, and we can, therefore, use the same interval found for the density contrast under linear approximations with the compaction function. These observations culminate in the fact that there exists a one-to-one correspondence between the EoS parameter $w$ and the compaction threshold ${\cal C}_{\rm th}$.

\begin{figure*}[htb!]
    	\centering
    {
        \includegraphics[width=17cm,height=11.5cm]{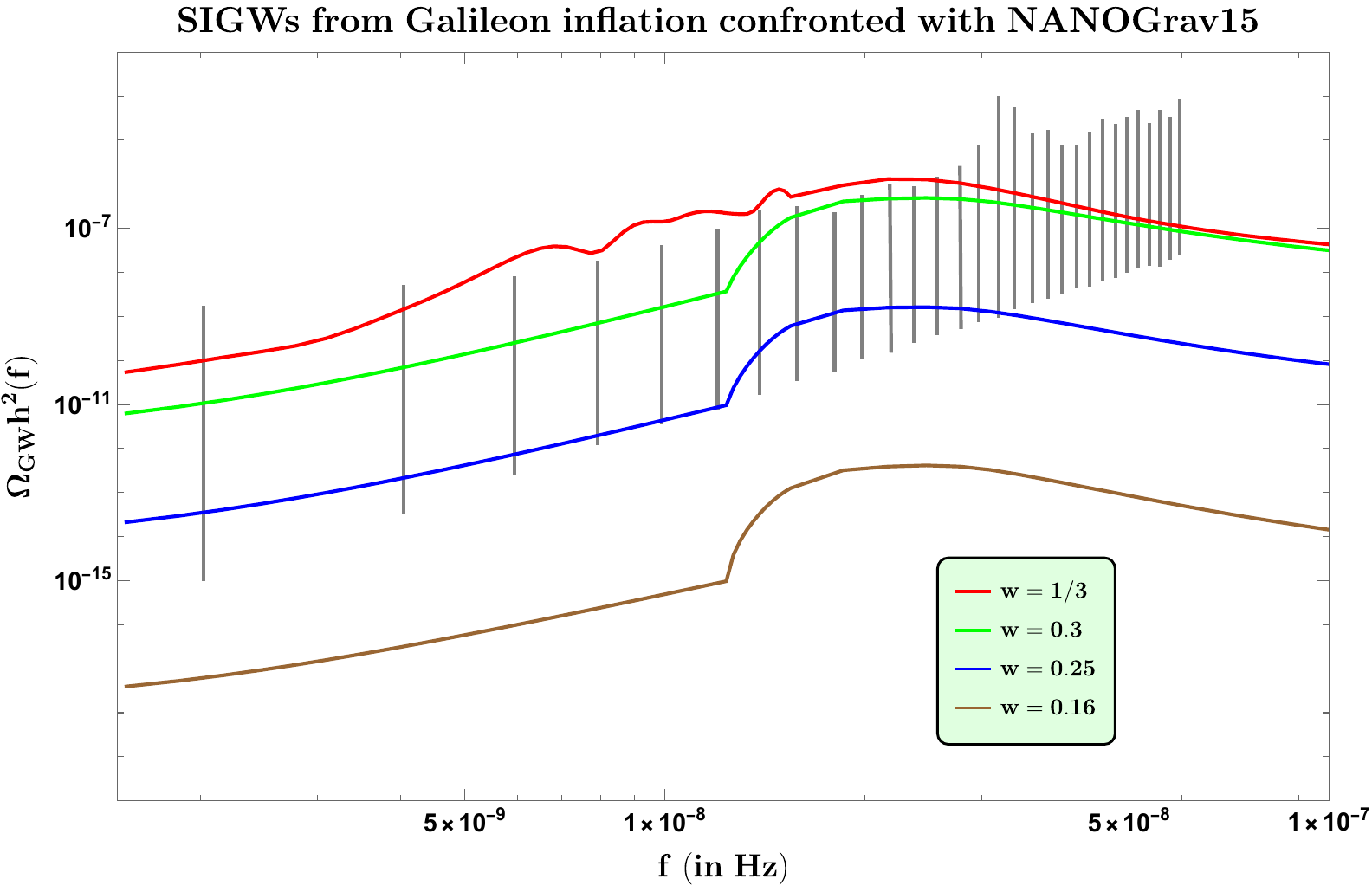}
        \label{wGalnano}
    }
    \caption[Optional caption for list of figures]{Spectrum of GW energy density as a function of the frequency for various EoS $w$ values and confronted with the NANOGrav15 signal. The spectra is plotted for values $w=1/3$ (red), $w=0.3$ (green), $w=0.25$ (blue), and $w=0.16$ (brown), respectively. For $w=1/3$, the transfer function from eqn.(\ref{rdtransfer}) is used, while for general $w$ transfer function from eqn.(\ref{transfer}) is used.  
    }
\label{wGalNANO}
    \end{figure*}

\begin{figure*}[htb!]
    	\centering
    {
        \includegraphics[width=17cm,height=11.5cm]{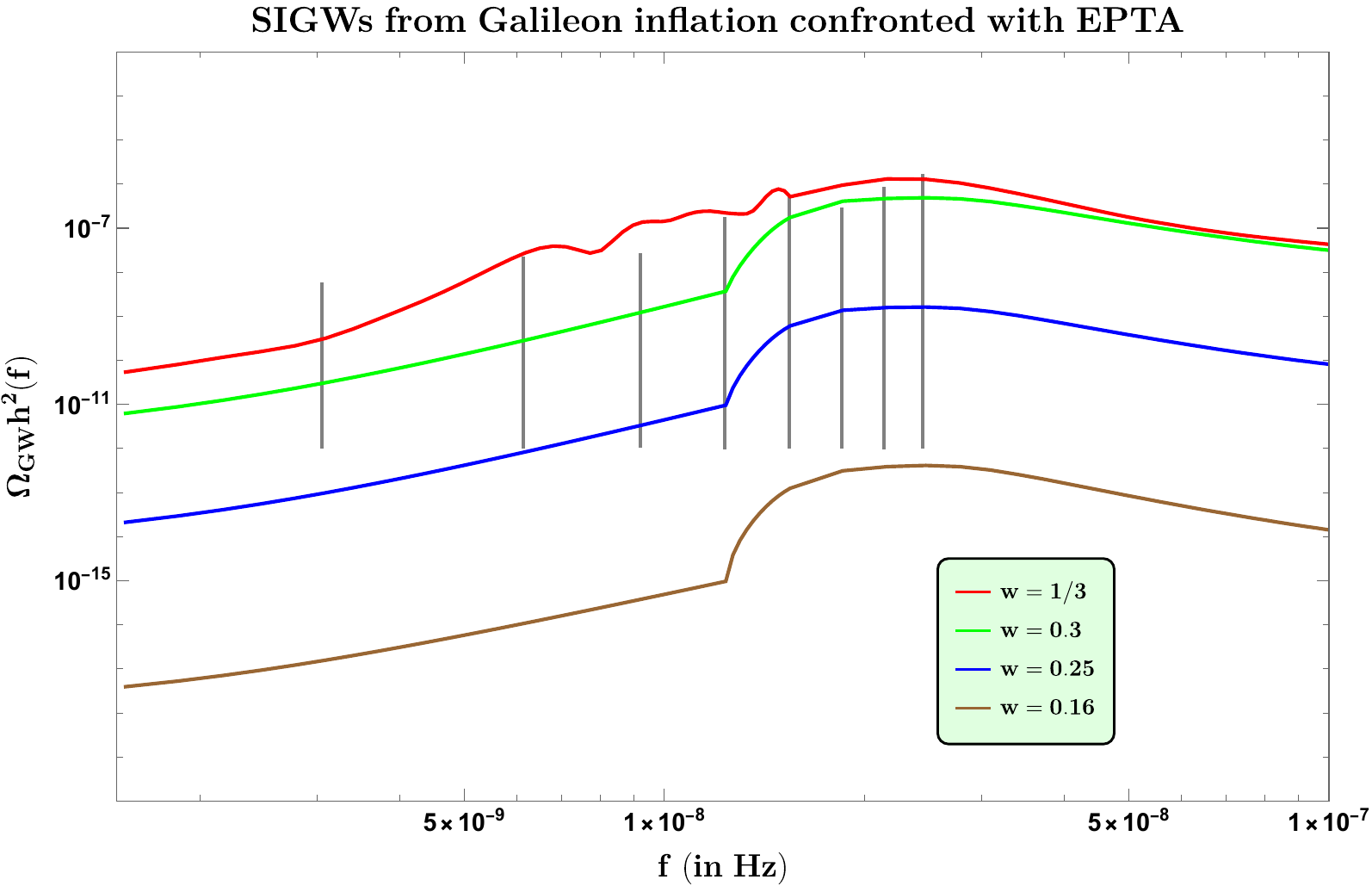}
        \label{wGalepta}
    }
    \caption[Optional caption for list of figures]{Spectrum of GW energy density as a function of the frequency for various EoS $w$ values and confronted with the EPTA signal. The spectra is plotted for values $w=1/3$ (red), $w=0.3$ (green), $w=0.25$ (blue), and $w=0.16$ (brown), respectively. For $w=1/3$, the transfer function from eqn.(\ref{rdtransfer}) is used, while for general $w$ transfer function from eqn.(\ref{transfer}) is used. 
    }
\label{wGalEPTA}
    \end{figure*}

We again emphasise that  our analysis  holds only when we consider working within the linear regime for the density contrast; see eqn.(\ref{deltalinear}). During this assumption, the Press-Schechter formalism modified with the EoS $w$ predicts $w\simeq 1/3$ as the best candidate to resolve the overproduction issue, see figs.(\ref{compare1})(left panel) $\&$ (\ref{wGalNANO},\ref{wGalEPTA}).
The inclusion of non-linear statistics of the density contrast and incorporating non-Gaussianities, $f_{\rm NL}$, into the study of the PBH mass fraction, makes the current analysis with the EoS insufficient. Generally, the curvature perturbations during PBH formation do not obey Gaussian statistics due to slow-roll violation, and to account for this a local perturbative expansion of the following form is usually chosen,
\bea
\zeta({\bf x}) = \zeta_{G}({\bf x}) + \frac{3}{5}f_{\rm NL}(\zeta_{G}^{2}({\bf x}) - \langle\zeta_{G}^{2}({\bf x})\rangle) + \cdots,
\eea 
such consideration of non-Gaussianity can sufficiently impact the PBH formation.
Before concluding this discussion, we highlight the particular non-linear relationship mentioned above,  
\bea \label{NLdensity}
\delta({\bf x},t) &=& -\frac{4}{9}\left(\frac{1}{aH}\right)^{2}e^{-2\zeta({\bf x})}\bigg[\nabla^{2}\zeta({\bf x}) + \frac{1}{2}\partial_{i}\zeta({\bf x})\partial_{i}\zeta({\bf x})\bigg],
\eea
which forces us to consider the non-Gaussian statistics for $\delta({\bf x},t)$ and when coupled with the local non-Gaussianity in the curvature perturbation, $f_{\rm NL}$, can significantly alter the analysis for the PBH mass fraction \cite{Choudhury:2023fwk, Young:2019yug, DeLuca:2019qsy}. 

\begin{figure*}[htb!]
    	\centering
    {
        \includegraphics[width=17cm,height=11.5cm]{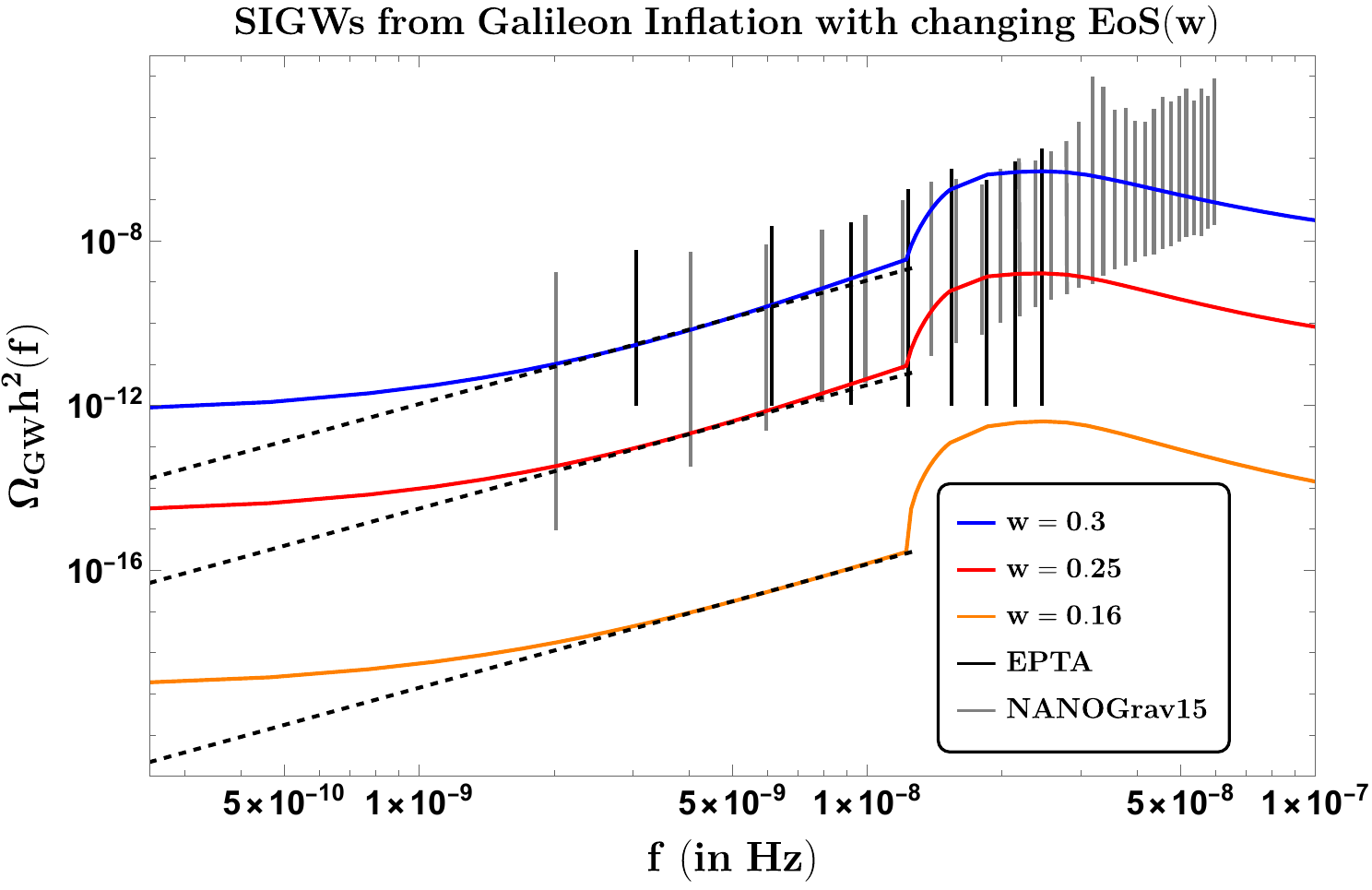}
        \label{IRsigw}
    }
    \caption[Optional caption for list of figures]{Spectrum of GW energy density as a function of the frequency for various EoS $w$ values and confronted with the NANOGrav15 and EPTA signal. The dashed-black line describes the IR tail of the relation, $\Omega_{\rm GW}h{^2}\propto k^{3}$. The spectra is plotted for values $w=0.3$ (blue), $w=0.25$ (red), $w=0.16$ (orange), and  the PTA (NANOGrav5 and EPTA) signals present in the background in gray and black solid lines, respectively. 
    }
\label{GalIR}
    \end{figure*}

\begin{figure*}[htb!]
    	\centering
    {
        \includegraphics[width=18cm,height=12cm]{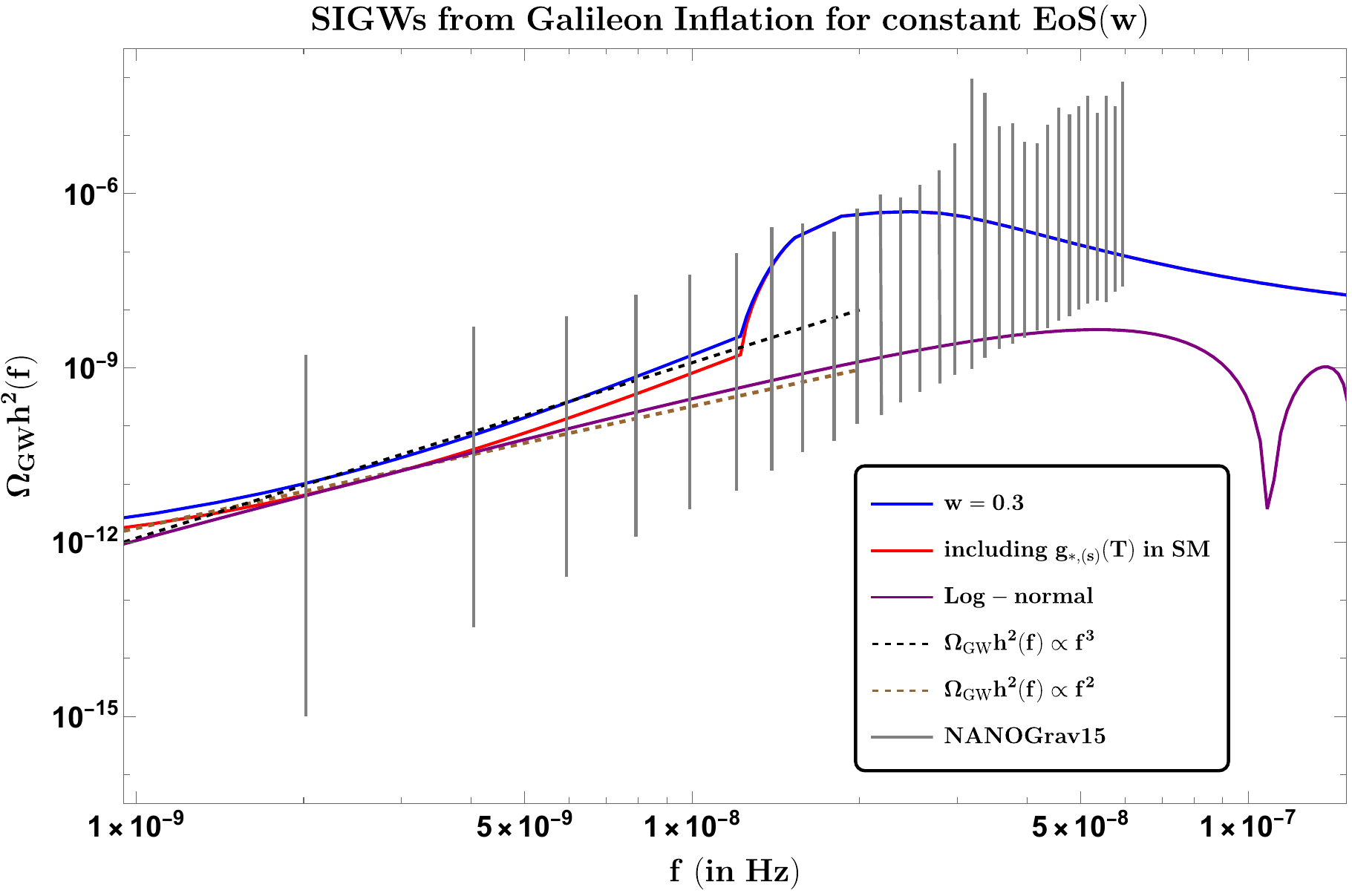}
        \label{qcdsigw}
    } 
    \caption[Optional caption for list of figures]{Spectrum of induced GWs as a function of their frequency and including the effects from QCD crossover. We compare the GW spectrum for a constant EoS $w=0.3$ in the two cases, without QCD effects (in blue), and including impact of $g_{*}(T),g_{*,s}(T)$ (in red). The dashed-black and dashed-brown lines show the approximate $f^{3}$ and $f^{2}$ scaling, respectively. The gray lines in the background represent the NANOGrav15 signal. }
\label{qcdgalileon}
    \end{figure*}

\section{Numerical outcomes and Discussions} 
\label{s9}

This section focuses on the results for the energy density spectrum of GWs obtained within the present framework of Galileon inflation after considering a varying background EoS $w$. We use the eqn.(\ref{transfer}) for this process and examine our results.

\begin{figure*}[htb!]
    	\centering
    \subfigure[]{
      	\includegraphics[width=8.6cm,height=7.5cm]{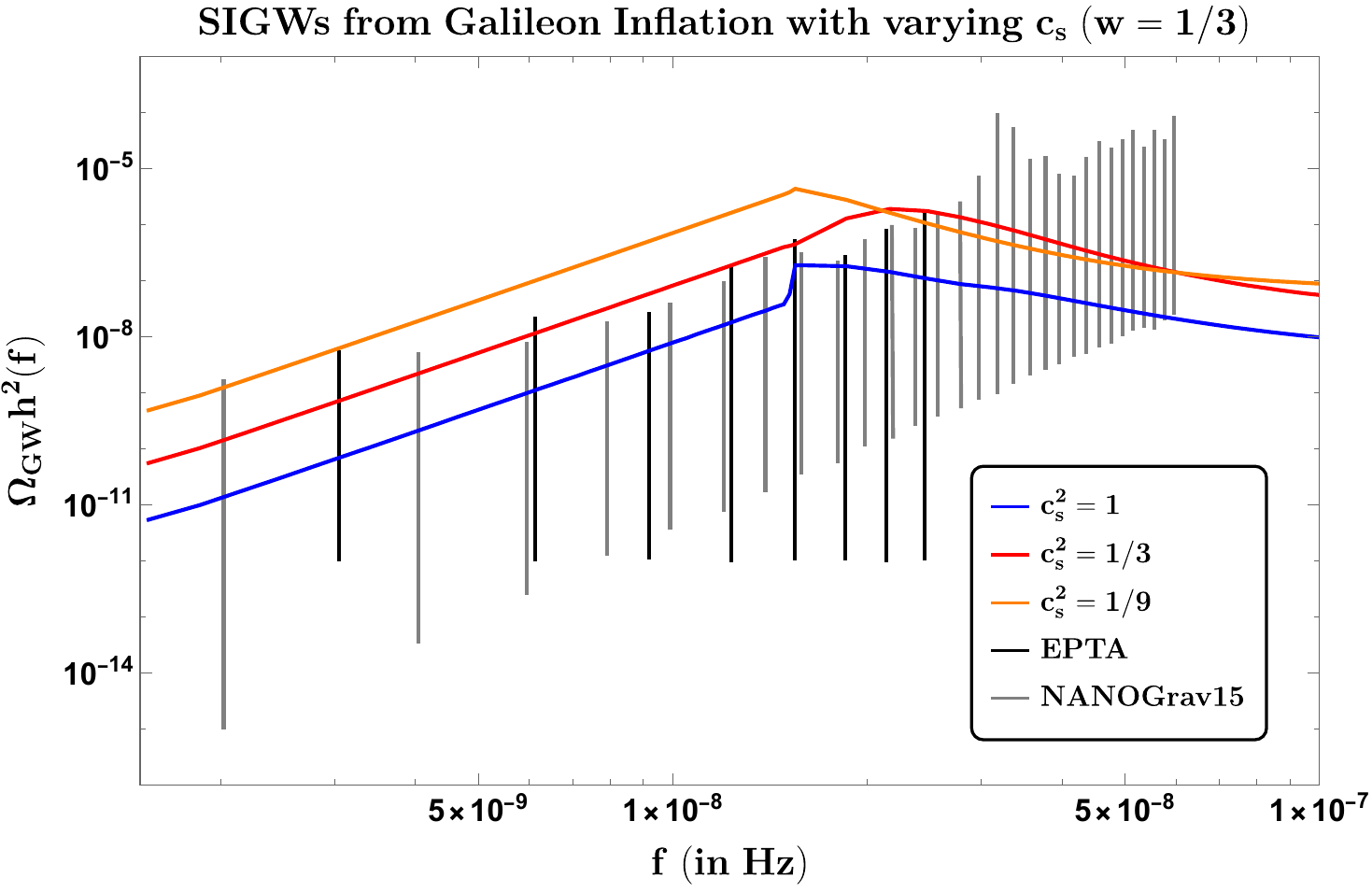}
        \label{SIGWcs1}
    }
    \subfigure[]{
       \includegraphics[width=8.6cm,height=7.5cm]{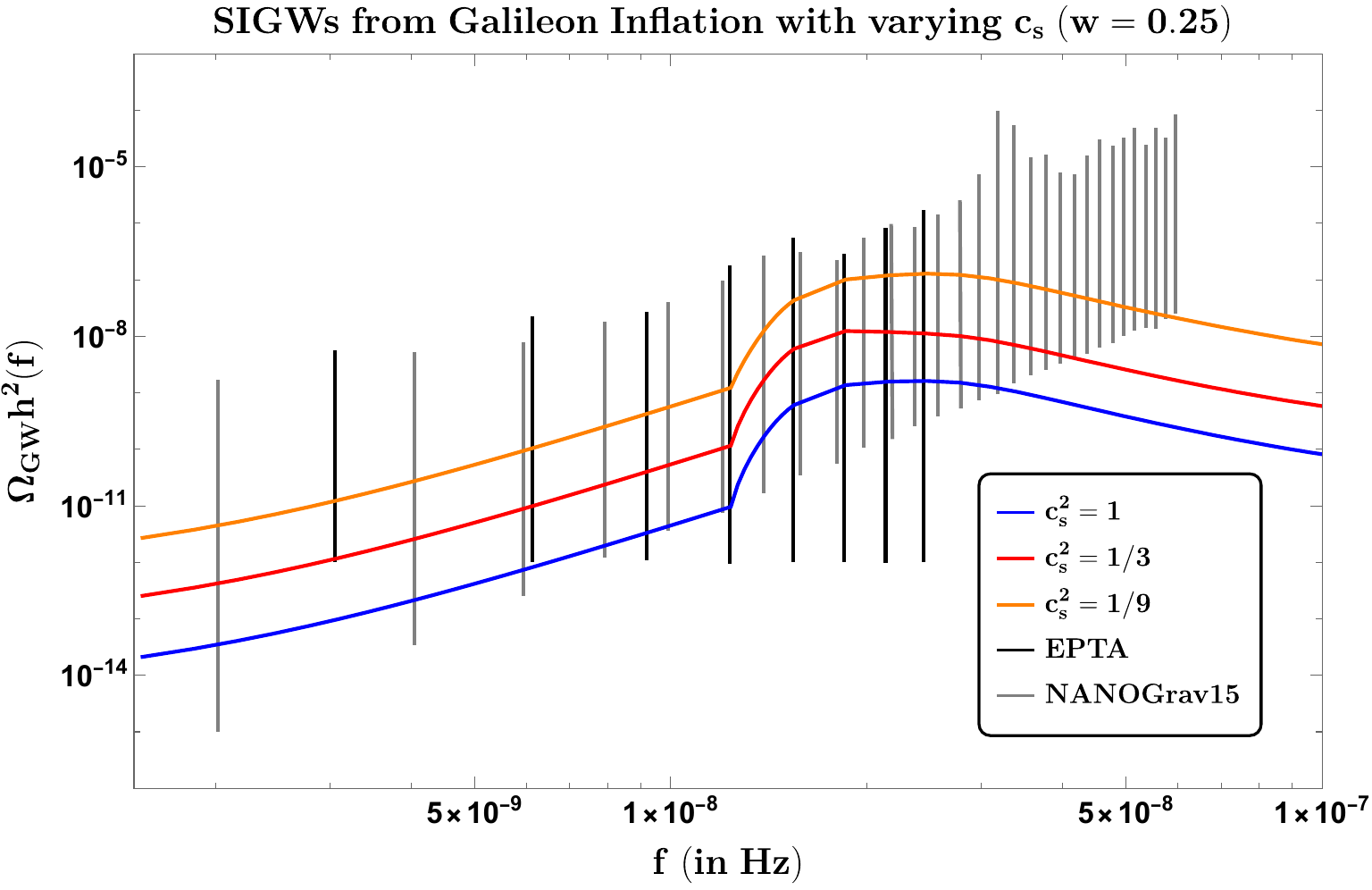}
        \label{SIGWcs2}
    }
    	\caption[Optional caption for list of figures]{Plots of SIGW spectrum with varying values of the propagation speed $c_{s}$ once for EoS $w=1/3$ in the left and $w=0.25$ in the right. The values of $c_{s}$ are chosen with $c_{s}^2=1$ (blue), $c_{s}^2=1/3$ (red), and $c_{s}^2=1/9$ (orange), for both EoS cases and with the PTA (NANOGrav15 and EPTA) signals present in the background in gray and black solid lines, respectively. We note differences in the high-frequency tail for $w=1/3$ with changing values of $c_{s}$ while not much significant change in $w=0.25$ except increase in SIGW spectrum amplitude for both as $c_{s}$ keeps decreasing. } 
    	\label{SIGWcs}
    \end{figure*}

\begin{figure*}[htb!]
    	\centering
    {
        \includegraphics[width=17cm,height=12cm]{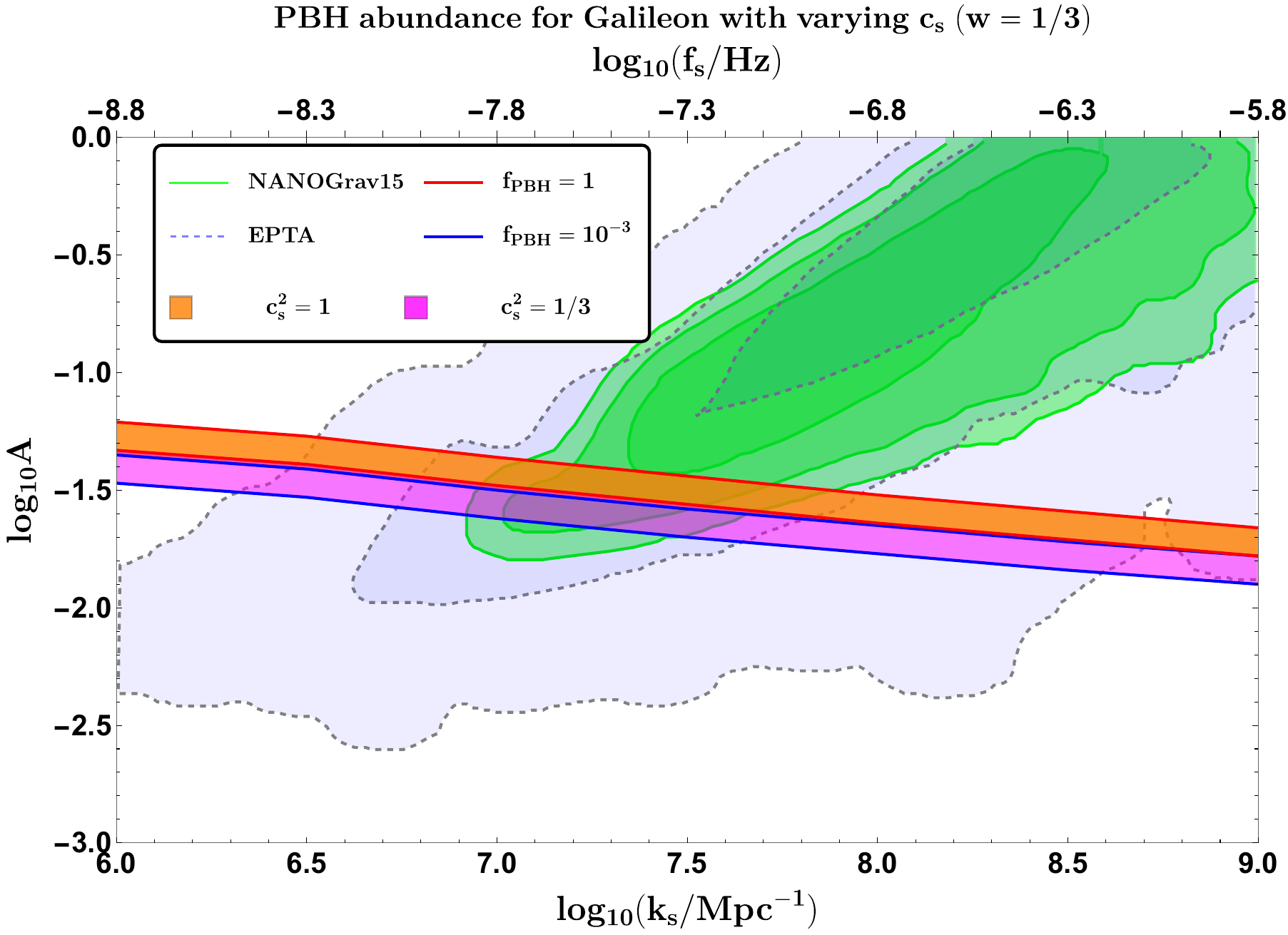}
        \label{pbhabundcs}
    } 
    \caption[Optional caption for list of figures]{Amplitude of the total scalar power spectrum $A$ as function of the transition wavenumber $k_{s}$ when the EoS is fixed at $w=1/3$. The orange band corresponds to the region of sizeable abundance, $f_{\rm PBH}\in (10^{-3},1)$ when $c_{s}^2=1$ is chosen and the magenta band corresponds to the sizeable abundance region when $c_{s}^2=1/3$ is chosen. The amplitude $A$ increases when $c_{s}^2$ is decreased keeping the highlighted regions of abundance within the  2$\sigma$ contour of the PTA signal. The background green and light blue filled contours, taken from \cite{Franciolini:2023pbf}, represent the sensitivity curves for the NANOGrav15 and EPTA signals, respectively. }
\label{abundancecs}
    \end{figure*}

\begin{figure*}[htb!]
    	\centering
    {
        \includegraphics[width=17cm,height=12cm]{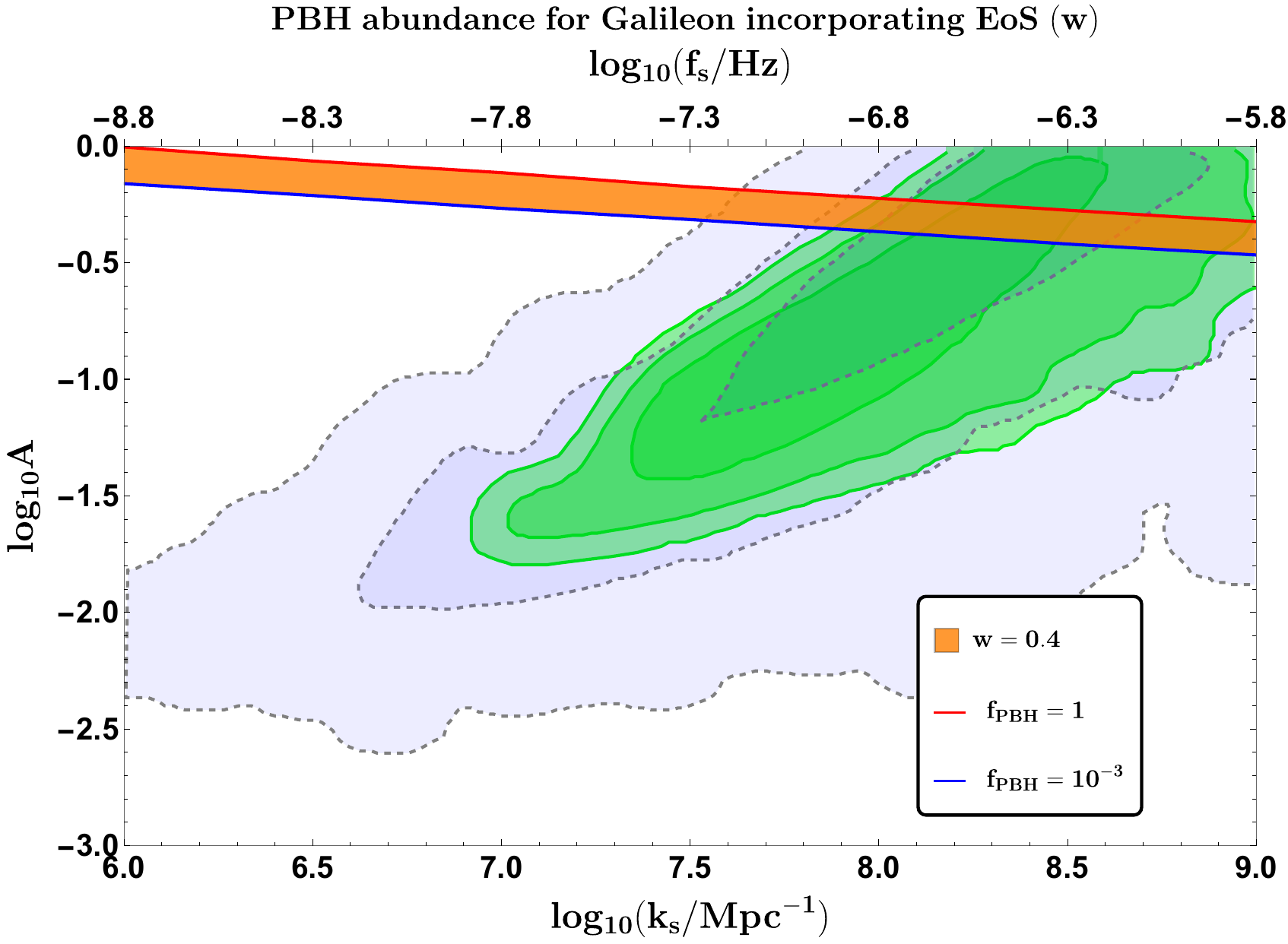}
        \label{wGal2}
    } 
    \caption[Optional caption for list of figures]{ Amplitude of the total scalar power spectrum $A$ as function of the transition wavenumber $k_{s}$ for the USR. The EoS parameter $w=0.4$ is fixed for this. The background green and light blue filled contours, taken from \cite{Franciolini:2023pbf}, are the sensitivity curves for the NANOGrav15 and EPTA signals, respectively.  }
\label{wGal2}
    \end{figure*}

\begin{figure*}[htb!]
    	\centering
    {
        \includegraphics[width=19cm,height=12cm]{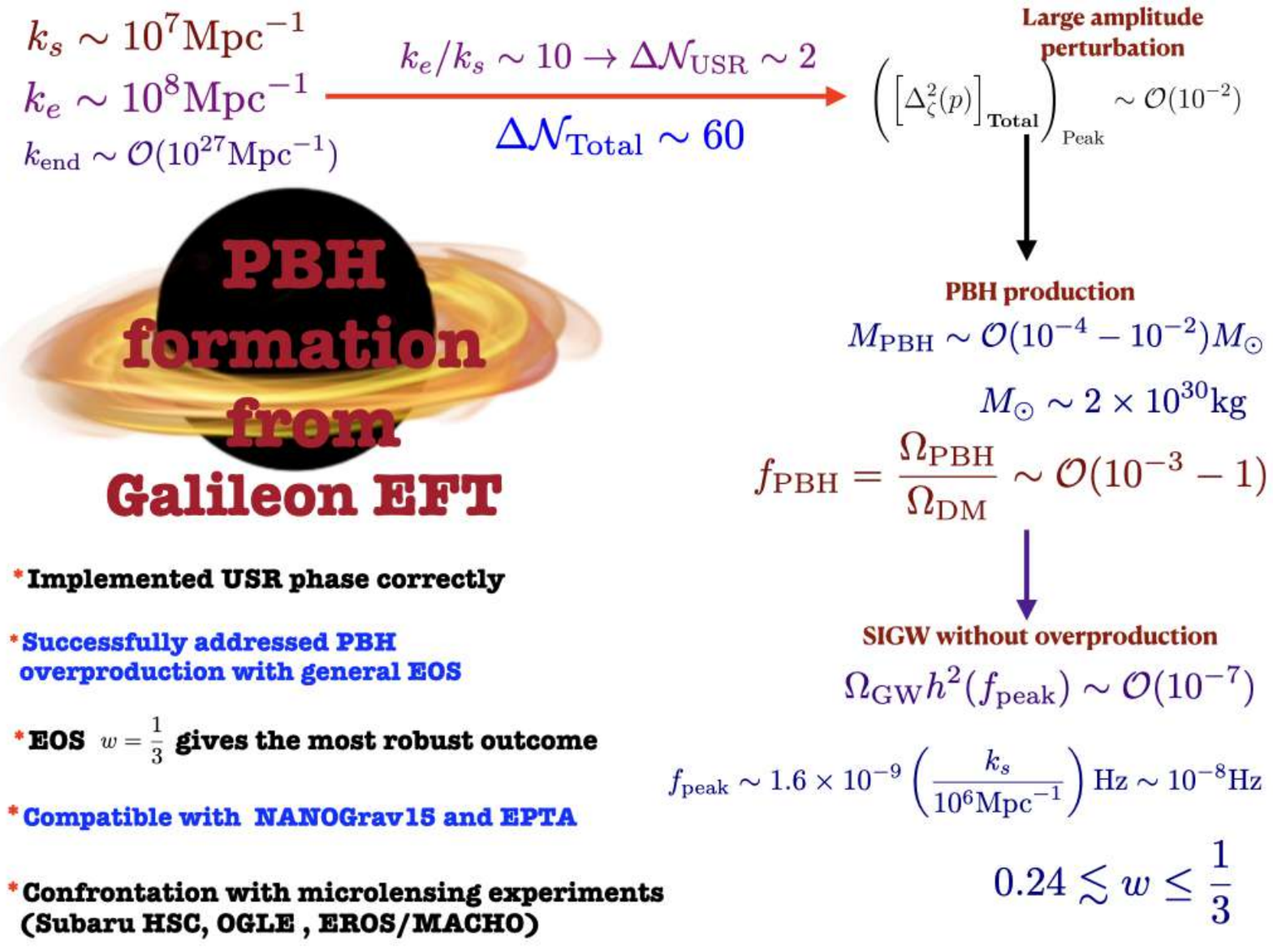}
        \label{wgalileon}
    } 
    \caption[Optional caption for list of figures]{Representative diagram depicting the key findings from the present work of integrating EoS parameter $w$ with the Galileon EFT framework and analysing the PBH formation and SIGW generation. }
\label{wgalileon}
    \end{figure*}

The plots in fig.(\ref{wGalNANO},\ref{wGalEPTA}) shows the SIGW spectra for the case of multiple constant EoS values. The grey vertical bands in the background correspond to the recently observed NANOGrav15 and EPTA signal, respectively. For the RD era, the obtained spectrum in red results from setting $w=1/3$ in the kernel or in eqn.(\ref{rdtransfer}). The RD era remains consistent with the observational signal with the highest possible spectrum amplitude. As we descend in $w$ values, the amplitude of the GW spectra falls quickly. Up to $w \gtrsim 0.24$, the region above the blue spectrum remains accessible from the NANOGrav15 signal. Going further below this value makes the spectrum move even lower than the signal amplitudes, as seen clearly from using $w=0.16$ and its spectrum in brown colour. In fig.(\ref{wgalfpbh}), we already observed that for $w=0.16$, the magenta band avoiding overproduction remains removed by near $2\sigma$ from the NANOGrav15 posteriors. From the above plot, we can thus conclude that going below $w \sim 0.24$ disagrees with the PTA signal; hence, only the region lying between $w \in (1/3,0.24)$ remains a potentially viable interval to understand the cosmological background in the very early Universe and the generation of SIGWs.

\textcolor{black}{Another crucial characteristic of the SIGWs generated in a universe filled with a fluid of arbitrary EoS $w$ is the behaviour of their spectrum in the infrared (IR) regime. In fig.(\ref{GalIR}), we highlight this specific behaviour for multiple values of the EoS and with $c_{s}^{2}=1$ fixed. The dashed-black line there represents the approximate $k^{3}$ scaling of the GW spectrum. In \cite{Cai:2019cdl}, a universal IR scaling relation, $\Omega_{\rm GW}\propto k^{3}$, was derived for the tensor modes that re-enter the universe in a radiation-dominated era. Here we aim to visualise the IR scaling behaviour of the SIGW spectrum from Galileon inflation and changing EoS values and how it compares to the scaling relation derived for the radiation era. Due to the highly complex structure of the scalar power spectrum in the present Galileon theory, an analytic treatment of the GW spectrum in the IR scenario, with the wavenumbers $k$ much less than the peak scale, becomes impossible here. We notice from fig.(\ref{GalIR}), that the scaling along the wavenumbers less than the peak position of the GW spectrum is approximately $k^{3}$ till $f\sim 2\times 10^{-9}{\rm Hz}$. For frequencies $f\lesssim 10^{-9}{\rm Hz}$ (nano-Hz), the GW spectrum does not fall and begins to become flatter in nature the lower we go. We also note that for all the $w$ values considered in this analysis, the scaling nature is almost the same till we descend below the nano-Hz frequency range.Even below the nano-Hz scale, the spectra tend to remain almost parallel, with the only difference present in their amplitude, thus signaling that the GWs from the different EoS lose their strengths similarly in the IR.  }

\textcolor{black}{Earlier in fig.(\ref{GalIR}), we showed how the GW spectrum showcased an IR tail with the scaling, $\Omega_{\rm GW}h^{2}\propto k^{3}$, for different EoS $w\leq 1/3$. However, it has been recently pointed out that an IR tail with such features can be greatly improved once we account for the effects from QCD crossover in the energy and entropy degrees of freedom \cite{Franciolini:2023wjm} which coincides with the nHz regime of the PTAs. While the use of eqn.(\ref{omegac}) takes into account the dilution of the GW energy density, when modes become sub-horizon, relative to the general EoS background, the factors present outside the eqn.(\ref{GWdensity}) for the induced GWs spectrum provide a damping effect on their amplitude proportional to the factor $g_{*}(T_c)*(g_{*,s}(T_c))^{-4/3}$. We use the data for effective degrees of freedom as a function of temperature provided in \cite{Saikawa:2018rcs} to implement this damping. In fig.(\ref{qcdgalileon}), we now highlight the case of the GW spectrum corresponding to a constant EoS $w=0.3$ and how the presence of QCD effects in the nHz regime can alter the IR tail of the signal. The red line is visible as lying between the $k^{3}$ (dashed-black) and $k^{2}$ (dashed-brown) scalings. Also, we provide a benchmark signal (in purple) generated by having a sharp log-normal power spectrum for comparison. The spectrum after QCD effects (in red) exhibit a scaling of $\Omega_{\rm GW}h^{2}\propto k^{2}$ in the IR inside the frequency interval, $f\sim (1-5){\rm nHz}$. Up to this, the spectrum also coincides closely with the GW spectrum resulting from having a sharp log-normal power spectrum. Further inside the NANOGrav frequency bins, the spectrum for $w=0.3$ changes its scaling, and, due to the absence of the QCD damping effects, the spectrum in blue breaks away from an approximate $k^{2}$ scaling before the red signal. We see in the present case of induced GWs from Galileon inflation that the impact of QCD effects is not sufficient enough to alter the IR scaling greatly when close to the peak of the spectrum, but it shows to improve the scaling near the initial frequency bins of the NANOGrav15 signal, and it is also in this region where it approximates closely to a GW spectrum coming from the log-normal case.}

\textcolor{black}{We consider the implications an arbitrary propagation speed $c_{s}$ can have on the generated SIGWs for a fixed value of the EoS $w$. The figure fig.(\ref{SIGWcs}) highlights changes in the generated SIGWs for $w=1/3$, on the left, and $w=0.25$, on the right, where $c_{s}^{2}\in \{1,1/3,1/9\}$ is considered for both the scenarios. A similar change having different $c_{s}$ brings is shifting of the SIGW spectrum to larger amplitudes as $c_{s}$ decreases from the case of $c_{s}^{2}=1$. For $w=1/3$, the curve in red shows SIGWs generated when $c_{s}^2=1/3$ for which the transfer function in eqn.(\ref{rdtransfer}) gets considered and, in this limit of sound speed, the result matches with the spectrum of radiation-dominated. It is interesting to notice that for the $w=1/3$ case, the tails all look similar in the low-frequency region of the signal, with only a difference in the amplitude. Notable changes only occur past the frequency corresponding to the transition wavenumber $k_{s}$, after which we see the differences in the behaviour of the high-frequency tails brought by varying $c_{s}$, with tail for $c_{s}^2=1/9$ looking to cross over the tail of $c_{s}^2=1/3$. Thus, in the case of a background EoS $ w=1/3 $ and having arbitrary $c_{s}$ can affect the SIGW spectrum in a manner that may become important for observations at the higher frequencies tails. In the case of a different $w=0.25$ scenario, the effects from varying $c_{s}$ are primarily to induce changes in the spectrum amplitude and not in the tail regions in the vicinity of the PTA signal. The overall impact on the amplitude looks similar throughout the frequency range.
}

\textcolor{black}{The impact of having varying $c_{s}$ is also examined in the context of overproduction of PBHs when the EoS is kept fixed to $w=1/3$ in fig.(\ref{abundancecs}). The region of significant abundance, or $f_{\rm PBH}\in (10^{-3},1)$ is highlighted using orange and magenta bands with the red and blue lines marking the upper and lower limits of this $f_{\rm PBH}$ interval. The orange bands corresponds to the previously studied case where the propagation speed $c_{s}^2=1$ is fixed. The case becomes interesting when we focus on having $c_{s}^2=w=1/3$ in which case the criteria of using the threshold for studying PBH abundance, from eqn.(\ref{deltath}), does not hold good here. We require a sensible choice of the density contrast threshold value to provide sufficient abundance for the conditions of interest. The analysis of the threshold is sensitive to the shape of the primordial spectrum which directly impacts the density contrast as we witness from the linear regime relation in eqn.(\ref{deltalinear}). In the present context, the peak in the total primordial power spectrum comes from the USR phase with an amplitude of $A\sim {\cal O}(10^{-2})$, see fig.(\ref{s5d}). As the PBH mass fraction is exponentially sensitive to the amplitude of the primordial power spectrum, we focus on the behavior of the small scales in the USR phase to give us the most significant contributions towards PBH abundance. The shape remains sharply peaked in the USR but has a small, non-vanishing width during the interval and for such reasons, we approximate the spectrum behaviour close to a log-normal power spectrum peaked in the limit of a finite but small width.} 

\textcolor{black}{As a consequence of the above arguments, we choose to use the value of $\delta_{\rm th}=0.59$ found for a power spectrum with such features in the condition of $c_{s}^2=w=1/3$ \cite{Musco:2020jjb} giving us a good approximate estimate to conduct the abundance analysis with these parameters. The larger value of $c_{s}^{2}=1$ is seen to prefer a higher amplitude of the scalar power spectrum amplitude compared to the other case of having $c_{s}^2=1/3$ with the new corresponding threshold value. }

\textcolor{black}{In light of the discussions till now, we clarify an essential choice in our analysis of using the NANOGrav and EPTA posteriors; see background contours in fig.(\ref{compare1}). The choice is motivated by a few assumptions based on our power spectrum and the resulting GW spectrum. As elaborated in the above discussions, from fig.(\ref{s5d}), our power spectrum in the USR closely approximates a log-normal power spectrum, giving a sharp peak with a finite width. Since the region of spectrum contributing most significantly to the PBH abundance and the peak amplitude in the GW spectrum comes from the USR region, we approximate features of our power spectrum for such cases as being similar to a log-normal spectrum. Utilising the same, we have also shown in fig.(\ref{GalIR}) that the GW spectrum for various EoS $(w)$ values exhibit a causality tail, $\Omega_{\rm GW}\propto k^{3}$, in the IR till we remain within the NANOGrav signal frequency range. Such a tail feature is also a characteristic result in the IR limit from use of a log-normal spectrum. These above approximations enable us to take advantage of the posteriors developed for a log-normal power spectrum. Thus, we do not believe there is a need for an explicit likelihood analysis to develop the posteriors in this work. In the future, we aim to develop this analysis more rigorously via thorough numerical modelling and studying the PBH overproduction issue further. }

Fig.(\ref{wGal2}) focuses on EoS larger than the RD era $w=1/3$. We take $w=0.4$ to demonstrate the effect increasing EoS beyond the RD era has on the amplitude of the total scalar power spectrum. We notice that slightly exceeding the bound on $w$ can drastically change the amplitude. The total amplitude gets close to breaking the perturbation theory assumptions within the wavenumber interval for the NANOGrav15. The orange colour region highlights the conditions for $f_{\rm PBH} \in (10^{-3},1)$. From this plot, we conclude that cases with $w > 1/3$ damage perturbativity approximations and, hence, are unsuitable for examining other related events, including SIGW and PBH formation.  

Before concluding this section, we mention the essential features of fig.(\ref{compare1}). Giving significance to the non-gaussian nature of density fluctuations provides a thorough analysis in calculating the mass fraction of PBHs. The right panel, in fig.(\ref{overprod}), uses the threshold statistics on compaction function as a method to accurately estimate PBH abundance and, after taking into account $f_{\rm NL} \sim -6$, gives a relatively close agreement with the NANOGrav15 signal than what we see using the $w-$Press-Schechter used in fig.(\ref{wgalfpbh}). Considering the statistical factors involved, the $w-$Press-Schechter formalism is still a suitable method to evaluate the PBH mass fraction and, therefore, provides insights into the nature of $w$ and its effects on processes that have empirical support.  

\section{Conclusion} 
\label{s10}

In this paper, we have demonstrated the explicit realization of a USR phase within the framework of the effective field theory of single-field Galileon inflation. In this setup, we have studied the formation of PBHs and the generation of SIGWs. We also examined the recently encountered PBH overproduction issue in detail without explicitly using the concept of primordial non-Gaussianity.  We have shown that the set-up under consideration results in the generation of SIGWs well consistent with the PTA signal along with obtaining a comfortable abundance of near-solar-mass black holes, $10^{-3} \lesssim f_{\rm PBH} \lesssim 1$.

In our earlier work related to Galileon EFT \cite{Choudhury:2023hvf,Choudhury:2023kdb,Choudhury:2023hfm,Choudhury:2023fwk}, we had assumed the presence of a USR phase followed by its implications for the PBH formation and generation of the SIGWs. Our present analysis, in particular,  focuses on the implementation of a USR phase based upon a detailed analysis of the Galileon EFT coefficients for the USR and the SR phases.
The said background is then used to discuss the PBH formation in the Galileon EFT framework targeting their overproduction issue by incorporating the general equation of state parameter $(w)$ for a cosmological background. 
We found $w=1/3$ to be the most robust scenario consistent with the present PTA observations. 
This is 
also consistent with our previous investigations performed using the compaction function approach to consider the non-linearities in the comoving curvature perturbations. The current analysis parallels our previous work \cite{Choudhury:2023fwk} dealing with the effects of non-linearities and non-gaussianities in Galileon EFT inflation. We have presented a comparative analysis of the two approaches. We have also dedicated a section specifically to introduce the overproduction problem and outline the possible resolutions.
The formation of PBHs during inflation requires a transition into a phase that enables the necessary enhancement of the primordial curvature perturbations. The inclusion of the ultra-slow roll phase in our setup puts conditions on the Galileon EFT coefficients. We have analyzed the suitable ranges of these coefficients that can facilitate the specific construction of an SRI, a sharp transition into the USR, followed by another sharp transition into the final SRII phase. The necessary constraints during this analysis came from the behaviour of the slow-roll parameters, $\epsilon$ and $\eta$, during each phase, the chosen parameterization of the effective sound speed $c_{s}$, and the amplitude of the total scalar power spectrum for each phase. The mild symmetry breaking, introduced through the term $c_{1}$ with the linear term in the action, is found to decrease in magnitude as we progress through each phase, while the coefficients for the higher-derivative interaction terms, especially the terms with $c_{4},c_{5}$, increase significantly as we reach SRII. However, $c_{4},c_{5}$ remain highly suppressed by large powers of the cut-off scale, $\Lambda$, which is necessary for our EFT description to remain valid.
After acquiring the suitable values for the EFT coefficients, we moved towards briefly explaining the derivation of the scalar power spectrum. Here, we highlighted the essential non-renormalization theorem and included the cubic order Galileon action responsible for one-loop corrections to the amplitude of the scalar power spectrum. 
 
We then outlined the PBH formation procedure by incorporating the EoS into the Press-Schechter formalism and evaluating the PBH mass fraction and present-day abundance. Further, we discussed the derivation of the GW energy density spectrum using the evolution of the tensor modes sourced by scalar modes at second order in the presence of an arbitrary but constant EoS background and constant speed of propagation $c_{s}$.  

We found that in the linear regime for the density contrast, the use of a general EoS, $w$, can provide a window around $(w=1/3)$ that can avoid overproduction after working with a SIGW interpretation of the recent PTA signal. However, the current analysis breaks down as one attempts to include the intrinsic non-linearities between $\delta_{\rm th}$ and the comoving curvature perturbation $\zeta({\bf x})$. In such a scenario, the effects of the non-Gaussianities should be accounted for to avoid overproduction. 
The case of $w=1/3$ predicts the highest amplitude of the scalar power spectrum to allow a sizeable PBH abundance, $10^{-3}\leq f_{\rm PBH} \leq 1$ consistent with NANOGrav15 data at the $1\sigma$ level. Subsequently, the values near $w \sim 0.16$ further lower the scalar power spectrum amplitude until the resulting case gets removed by $2\sigma$ from the NANOGrav15 data. It is interesting that the peak value of amplitude is achieved by the equation of state, $w=1/3$, which gives the robust outcome in the present context of solving overproduction issue. 
\textcolor{black}{We examined the infrared (IR) tail of the GW spectrum for the other values of the EoS, $w\ne 1/3$, considered in our analysis. After comparing the IR scaling with the universal scaling relation existing during the radiation era, $\Omega_{\rm GW}\propto k^{3}$, we found that the same scaling relation followed in the IR for other $w$ values until we got lower than $f\sim 10^{-9}{\rm Hz}$ (nano-Hz) regime. We also looked into the changes occurring in the GW spectrum upon considering arbitrary values of the sound speed $c_{s}$. We observed that decreasing $c_{s}$ led to an overall increase in the amplitude of the generated spectrum. For $w=1/3$, the large wavenumber tail suffered noticeable changes while the IR tail remained the same with changing $c_{s}$. This observation contrasted for a different EoS, $w=0.25$, where the behaviour along both tails remained the same for the different $c_{s}$ considered. The impact of $c_{s}$ on the PBH abundance was then analysed further. We found that, for $w=1/3$, a larger value of $c_{s}^2=1$ needed larger amplitude to produce the same amount of PBH than for $c_{s}^2=1/3$, though the amount by which it got enhanced was not quite significant.}
We conclude that the EoS approach to resolving overproduction works well after working with a modified $w$-Press-Schechter formalism to give $w=1/3$ as the best option for the cosmological background producing the SIGW. The interval $0.24 \lesssim w \leq 1/3$ is shown to generate enough induced GW density that remains consistent with the PTA signal. 

The fig.(\ref{wgalileon}) provides a representative version of the overall findings from the analysis done in this work. The major conclusions are summarized in the respective diagram.


\section*{Acknowledgement}
SC would like to thank The National Academy of Sciences (NASI), Prayagraj, India for being elected as a member of the academy.
SP is supported by the INSA Senior scientist position at NISER, Bhubaneswar through the Grant number INSA/SP/SS/2023. MS is supported by Science
and Engineering Research Board (SERB), DST, Government of India under the Grant Agreement number CRG/2022/004120 (Core Research Grant). MS is
also partially supported by the Ministry of Education
and Science of the Republic of Kazakhstan, Grant No.
0118RK00935, and CAS President’s International Fellowship Initiative (PIFI).

\newpage

\appendix
\section*{Appendix} 

\section{Useful integrals in the Superhorizon regime}   \label{app:A}

In this appendix, we consider the limiting versions of the integrals in eqn.(\ref{besselprod}), where we have a product of Bessel functions to examine them in the super horizon regime. We follow the analysis as presented in \cite{Domenech:2021ztg}. In this case, the variable $x=k\tau$ goes as $x \ll 1$. During this regime, there exist modes that are super-Horizon relative to the pivot scale, as chosen for our case. Although we have not considered the details of the super-Horizon kernel for our analysis, we feel it relevant to provide a brief understanding of the solutions in this regime. Another essential assumption concerns the scale contributing to the peak of the scalar power spectrum, which is assumed to enter way before the pivot scale, i.e., $k_{s} \gg k_{*}$. Such an assumption helps to avoid worrying about significant changes occurring due to sourcing during or after the re-entry of the modes at a pivot scale.

Based on the above assumptions we can consider for the variable, $v \sim k_{s}/k \gg 1$ which then leads to having $u\sim v \gg 1$. Thus incorporating these approximations into the Bessel function products and integrating under the conditions $u \sim v$, requires us to implement the following relations:
\bea
\int dx \;x\;J_{p}^{2}(ax) &=& \frac{x^{2}}{2}(J_{p}^{2}(ax)-J_{p-1}(ax)J_{p+1}(ax)),\\
\int dx\; x^{-2p+1}\;J_{p}^{2}(ax) &=& \frac{x^{-2p+2}}{2(1-2p)}(J^{2}_{p}(ax) + J^{2}_{p-1}(ax))
\eea
Using these relations, the integrals present in eqn.(\ref{besselprod}) can be reduced to give us:
\bea
I_{J} &\approx & \frac{3+2b}{1+b}\frac{2^{-b-1/2}}{\Gamma(b+3/2)}\frac{x}{\pi v},\\
I_{Y}  &\approx & \frac{3+2b}{b(1+b)c_{s}\pi}\bigg(2^{b-1/2}\Gamma(b+1/2)\frac{x^{-2b}}{\pi v} - 2^{-b-3/2}\frac{c_{s}^{2b}}{\Gamma(b+3/2)}\frac{1+b+b^{2}}{1+b}\bigg)
\eea

\section{Useful integrals in the Subhorizon regime}   \label{app:B}

In this appendix, we discuss the integrals involving products of Bessel functions and useful approximations under the limiting case of the sub-Horizon regime. The variable $x=k\tau$ in this regime goes as $x \gg 1$. The kernel in eqn.(\ref{simplekern}) contains the integral in eqn.(\ref{besselprod}) which involves product of three Bessel function and, as we mentioned before, is not analytically doable for general $x$. On taking the limit $x \gg 1$, there does exist analytical results and here mention those relevant integrals, also found in \cite{gervois1985integrals}. The results presented are written using a few new variables, we label them as $p$, $q$, and $r$, and the eqn.(\ref{besselprod}) under the condition $|p-q| < r < p+q$ reads as: 
\bea
I_{B}^{x \gg 1} &=& \int_{0}^{\infty}d\tilde{x}\;\tilde{x}^{1-a}\;B_{a}(r\tilde{x})J_{b}(p\tilde{x})J_{b}(q\tilde{x}), \nonumber\\
&=& \frac{\sqrt{2}}{\pi\sqrt{\pi}}\frac{(pq)^{a-1}}{r^{a}}(\sin{\theta})^{a-1/2}\times \left\{
	\begin{array}{ll}
		\displaystyle \frac{\pi}{2}P^{-a+1/2}_{b-1/2}  \quad\quad & \mbox{when}\quad  B_{a}(rx)=J_{a}(rx) \\ 
			\displaystyle 
			\displaystyle -Q^{-a+1/2}_{b-1/2} \quad\quad & \mbox{when }\quad  B_{a}(rx)=Y_{a}(rx)
	\end{array}
\right.
\eea
which involves the relations: 
\bea
16\Delta^{2} \equiv (r^{2}-(p-q)^{2})((p+q)^{2}-r^{2}), \quad\quad \cos{\theta} = \frac{p^{2}+q^{2}-r^{2}}{2pq}, \quad\quad \sin{\theta} = \frac{2\Delta}{pq}.
\eea
Contrary to the above, when $r > p+q$ is satisfied we have from the integrals the following:
\bea
I_{B}^{x \gg 1} &=& \int_{0}^{\infty}d\tilde{x}\;\tilde{x}^{1-a}\;B_{a}(r\tilde{x})J_{b}(p\tilde{x})J_{b}(q\tilde{x}), \nonumber\\
&=& \frac{\sqrt{2}}{\pi\sqrt{\pi}}\frac{(pq)^{a-1}}{r^{a}}(\sinh{\theta})^{a-1/2}\Gamma(b-a+1){\cal Q}^{-a+1/2}_{b-1/2}(\cosh{\theta})\times \left\{
	\begin{array}{ll}
		\displaystyle -\sin{(b-a)\pi}  \quad\quad & \mbox{when}\quad  B_{a}(rx)=J_{a}(rx) \\ 
			\displaystyle 
			\displaystyle \cos{(b-a)\pi} \quad\quad & \mbox{when }\quad  B_{a}(rx)=Y_{a}(rx)
	\end{array}
\right.
\eea
which involves the new relations: 
\bea
16\tilde{\Delta}^{2} \equiv (r^{2}-(p-q)^{2})(r^{2}-(p+q)^{2}), \quad\quad \cosh{\theta} = \frac{r^{2}-(p^{2}+q^{2})}{2pq}, \quad\quad \sinh{\theta} = \frac{2\tilde{\Delta}}{pq}.
\eea

For the purpose of the integrals done in this text, the following re-labelling of variables must be done:
\bea
r=1,\quad\quad p=c_{s}v, \quad\quad q=c_{s}u
\eea
Based on the above, for the variables in eqn.(\ref{omegac}) the range $|1-v| < u < 1+v$ can be separated into $1 > c_{s}(u+v) (r > p+q)$ and $1 < c_{s}(u+v) (r < p+q)$.

\section{Notes on Associate Legendre functions} \label{app:C}
Here we will present some of the important formulas associated with Legendre's polynomials that have been used in our analyses. Firstly, we start by presenting the definition of the hypergeometric function for which a quadratic transformation exists:
\bea
\textbf{F}(l,m;n;x) = \frac{1}{\Gamma[n]}{\cal F}(l,m;n;x),
\eea
where $l,m,n$ are rational parameters, and ${\cal F}(l,m;n;x)$ represents the Gauss's Hypergeometric function. With this, let's write the expression for Ferrer's and Olver's functions.
\bea
P_{\nu}^{\mu}(x)&=& \bigg[\frac{1+x}{1-x}\bigg]^{\mu /2}\; \textbf{F}(\nu+1,-\nu;1-\mu;1/2-1/2x), \\
Q_{\nu}^{\mu}(x)&=&\frac{\pi}{2\sin{(\mu \pi)}}\bigg(\cos{(\mu \pi)}\bigg[\frac{1-x}{1+x}\bigg]^{\mu/2}\; \textbf{F}(\nu+1,-\nu;1-\mu;1/2-1/2x) \nonumber \\ 
&& \quad \quad \quad \quad - \frac{\Gamma(\nu +\mu +1)}{\Gamma(\nu - \mu +1)}\bigg[\frac{1-x}{1+x}\bigg]^{\mu/2}\; \textbf{F}(\nu+1,-\nu;1+\mu;1/2-1/2x)\bigg), \\
{\cal Q}_{\nu}^{\mu}(x) &=& \frac{\pi}{2\sin{(\mu \pi)} \Gamma(\nu+\mu+1)}\bigg(\bigg[\frac{x+1}{x-1}\bigg]^{\mu/2}\; \textbf{F}(\nu+1,-\nu;1-\mu;1/2-1/2x) \nonumber \\
&& \quad \quad \quad \quad - \frac{\Gamma(\nu +\mu +1)}{\Gamma(\nu - \mu +1)}\bigg[\frac{x-1}{1+x}\bigg]^{\mu/2}\; \textbf{F}(\nu+1,-\nu;1+\mu;1/2-1/2x)\bigg),
\eea
Now these $P_{\nu}^{\mu}(x)$ and $Q_{\nu}^{\mu}(x)$ are Legendre's functions on the cut or Ferrer's functions which are defined only for $|x|<1$. However, ${\cal Q}_{\nu}^{\mu}(x)$ is defined only for $\abs{x}>1$, and is known as Olver's functions. It is sometimes more useful to work with a real-valued version of ${\cal Q}_{\nu}^{\mu}(x)$ given by:
\bea
{\cal Q}_{\nu}^{\mu}(x) \equiv e^{-\mu \pi i}\frac{{\cal Q}_{\nu}^{\mu}(x)}{\Gamma[\mu +\nu +1]}
\eea
Here are a few useful interconnections between the Legendre's functions of the first and second kind.
\bea
{\cal Q}_{\nu}^{\mu}(x)&=& \sqrt{\frac{\pi}{2}}\frac{1}{(x^2-1)^{1/4}}P^{-\nu - 1/2}_{\mu- 1/2}\bigg(x(x^2 -1)^{-1/2}\bigg), \\
P_{\nu}^{\mu}(-x) &=& -(2/\pi)\sin{((\nu+\mu)\pi)}\;Q+{\nu}^{\mu}(x) +  \cos{((\nu + \mu)\pi)}\; P_{\nu}^{\mu}(x) \nonumber \\
Q_{\nu}^{\mu}(-x) &=& \frac{-1}{2}\pi \sin{((\nu +\mu)\pi)}\; P_{\nu}^{\mu}(x) - \cos{((\nu+\mu)\pi)}\;Q_{\nu}^{\mu}(x)
\eea
\subsection{Asymptotic approximation}

This section of the appendix is indebted to the necessary asymptotic behaviour of the above-mentioned associated Legendre functions near the singular points $x \sim 1$, and $x \to \infty$. 

\begin{itemize}
    \item \underline{\textbf{Case I}} : Limit $x \to 1^{-}$ \\
This is the case where $\abs{x}<1$, and so only concerns the Ferrer's functions. The Ferrer's function of the first kind converges to:
\bea
\lim_{x\to 1^{-}} P_{\nu}^{\mu}(x) \sim \frac{1}{\Gamma(1-\mu)}\bigg(\frac{2}{1-x}\bigg)^{\mu /2}
\eea
While Ferrer's function of the second kind goes to:
\bea
\lim_{x \to 1^{-}}Q_{\nu}^{\mu}(x) &=& \frac{\Gamma(\mu)\Gamma(\nu - \mu +1)}{2 \Gamma (\nu+\mu+1)} \bigg(\frac{2}{1-x}\bigg)^{\mu /2} \; \; \text{for $\mu > 0$}\\
\lim_{x \to 1^{-}}Q_{\nu}^{\mu}(x) &=& \frac{\cos{(\mu \pi)}}{2} \Gamma(\mu) \bigg(\frac{2}{1-x}\bigg)^{\mu /2} \; \; \text{for $\mu < 0, \mu \ne 1/2$}
\eea

    \item \underline{\textbf{Case II}} : Limit $x \to 1^{+}$ \\
    This case is concerned with Olver's functions since $\abs{x} > 1$. Here we have the limiting behaviour as,
\bea
\lim_{x \to 1^{+}}{\cal Q}_{\nu}^{\mu}(x) = \frac{\Gamma(\mu)}{2 \Gamma (\nu +\mu +1)}\bigg(\frac{2}{x-1}\bigg)^{\mu /2}
\eea 

    \item \underline{\textbf{Case III}} :Limit $x \to \infty$ \\
This case is also concerned with Olver's function of the second kind. The asymptotic limit gives:
\bea
\lim_{x \to \infty}{\cal Q}_{\nu}^{\mu}(x) = \frac{\sqrt{\pi}}{\Gamma(\nu +3/2)(2x)^{\nu+1}}
\eea
\end{itemize}

\subsection{Resonance approximation}

From the previous section, notice that the integrals under the asymptotic limits have a divergent nature when their argument follows $x \sim 1$, when $\mu \ne 0$. To treat the functions, appearing in the eqn.(\ref{besselprod}), near this regime we must make use of their limiting form when $x\sim 1$. The term ``resonance'' for this limit can be understood based on the singular nature of the associated Legendre functions discussed in the previous section and from the specific form of the argument for these functions as mentioned before eqn.(\ref{kernelavg}) which coincides with the behaviour of the Heaviside Theta argument. We now present the resonance approximations vital to evaluate the integrals near such resonant conditions.

Starting with the limit $y \rightarrow -1^{+}$, we have:
\bea
P_{b}^{-b}(y) + \frac{b+2}{b+1}P_{b+2}^{-b}(y) \sim \frac{3+2b}{1+b}\frac{1}{\Gamma[1+b]}\bigg(\frac{2}{1+y}\bigg)^{-b/2}
\eea

\bea
Q_{b}^{-b}(y) + \frac{b+2}{b+1}Q_{b+2}^{-b}(y) \sim - \frac{3+2b}{1+b} \begin{cases}
(1+b+b^2)\displaystyle{\frac{\Gamma[b]}{\Gamma[2b+3]}\bigg(\frac{2}{1+y}\bigg)^{b/2}}, & \text{for } b>0  \\
\displaystyle{\frac{\cos{(b \pi)}}{2}}\Gamma[-b]\bigg(\frac{2}{1+y}\bigg)^{-b/2}, & \text{for } b<0
\end{cases}
\eea
and for the case $y \rightarrow 1^{+}$, we get:
\bea
{\cal Q}_{b}^{-b}(y) + 2 \frac{b+2}{b+1}{\cal Q}_{b+2}^{-b}(y) \sim \frac{3+2b}{1+b} \begin{cases}
(1+b+b^2)\displaystyle{\frac{\Gamma[b]}{\Gamma[2b+3]}\bigg(\frac{2}{1-y}\bigg)^{b/2}}, & \text{for } b>0  \\
\displaystyle{\frac{\Gamma[-b]}{2}\bigg(\frac{2}{1-y}\bigg)^{-b/2}}, & \text{for } b<0
\end{cases}
\eea

\newpage

\bibliography{Refs}
\bibliographystyle{utphys}

\end{document}